M. Douarre,† V. Martí-Centelles,† C. Rossy, A. Tron, I. Pianet, N. D McClenaghan (†Equal contribution)

Macrocyclic Hamilton receptor-shuttling dynamics in [2]rotaxanes

# Macrocyclic Hamilton receptor-shuttling dynamics in [2]rotaxanes

Maxime Douarre,[a,‡] Vicente Martí-Centelles,[a,‡] Cybille Rossy,[a] Arnaud Tron,[a] Isabelle Pianet,*[b] and Nathan D. McClenaghan*[a]

[a]*Institut des Sciences Moléculaires, CNRS (UMR 5255), University of Bordeaux, Talence, France;* [b]*Université Bordeaux Montaigne, IRAMAT (UMR 5060), Maison de l'Archéologie, Pessac, France*

‡ These authors contributed equally. Corresponding Author *E-mail: nathan.mcclenaghan@u-bordeaux.fr (N.McC.), isabelle.pianet@u-bordeaux-montaigne.fr (I.P.)

Electronic Supplementary Information (ESI) available: experimental procedures, characterization data of **1**, **2**, and **4** ($^1$H, $^{13}$C, VT NMR) and molecular modeling structures of rotaxanes **1**, **7**, **8** and **9**.

# Macrocyclic Hamilton receptor-shuttling dynamics in [2]rotaxanes

Inherent movement of a macrocycle comprising a multi-site hydrogen bonding Hamilton-type receptor in a series of [2]rotaxanes was investigated by dynamic and VT NMR. In these rotaxanes the varying nature and number of hydrogen-bond motifs and stations on the molecular axles influenced ring dynamics. Triazole stations interact selectively by hydrogen bonding with the isophthalamide amide bonds present in the Hamilton macrocycle and fast shuttling (72 kHz at 25 °C) was observed in a two station variant.

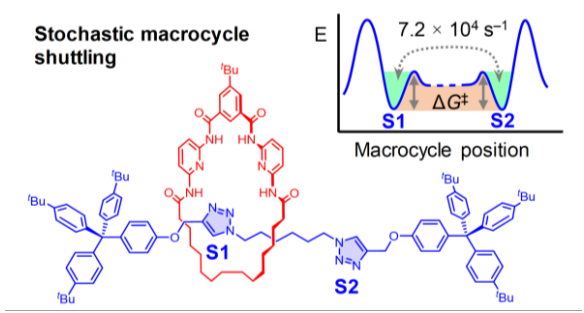

Keywords: molecular shuttle; rotaxane; hydrogen bonding; dynamic NMR; macrocycle

## Introduction

Artificial molecular machines based on interlocked molecules have attracted much attention in the last decades, often such sophisticated structures mimic complex functions of biological systems.[1] In particular, [2]rotaxane systems comprising mobile macrocycles have been developed as molecular shuttles,[2–4] and determining shutting parameters via dynamic NMR spectroscopy is an established approach.[1] This allowed determining the key role of ring movement in molecular machines and obtaining detailed kinetic analyses providing key structure-property information in a number of different molecular machines,[1, 5–7] including non-degenerate molecular shuttles used as switches in response to external stimulus,[2, 8, 9] and degenerate rotaxanes exhibiting spontaneous ring motion.[5, 10–16] Relevant works include Stoddart molecular shuttles [1, 17–19], hydrogen-bonded molecular machines by Leigh,[3, 20, 21] Loeb's ring-through ring shuttling rotaxanes and related systems,[2, 22] Hirose shuttling molecular machine studies on ring size and axle length,[5, 23], Coutrot's pioneering studies on shuttling effects in rotaxanes,[24] and Brouwer's shuttling dynamics studies showing the complexity of that macrocycle shuttling events.[6]

Among the different rotaxane formation methods, the Huisgen 1,3-dipolar cycloaddition involving reaction of alkyne and azide groups, also known as CuAAC "click" chemistry, has been extensively used as an efficient tool for the synthesis of interlocked molecules.[25, 26] In our previous studies, we employed this methodology to synthetize rotaxanes based on Hamilton-like macrocyclic receptor **3** on different axles (**7** and **9**) as well as Glaser coupling (**8**), however the ring shuttling dynamics in these systems were not explored.[27, 28] Herein, we enlarge the series to include double station rotaxane **1** and we focused on studying the stochastic molecular shuttling on the ensemble of the different axles via dynamic NMR spectroscopy to evaluate the effect of the different levels of ring–axle supramolecular interactions for rotaxanes **1**, **7**, **8** and **9** whose structures are shown in Figure 1. A schematic and qualitative potential energy surface representation of the macrocycle shuttling between two different energy stations is shown in Figure 1. The shuttling energy barrier depends only on the macrocycle binding affinity to the initial station and it is independent of the affinity to the arrival station, therefore the shuttling rate is slower if the macrocycle–station binding affinity to the initial station is higher.[27, 28]

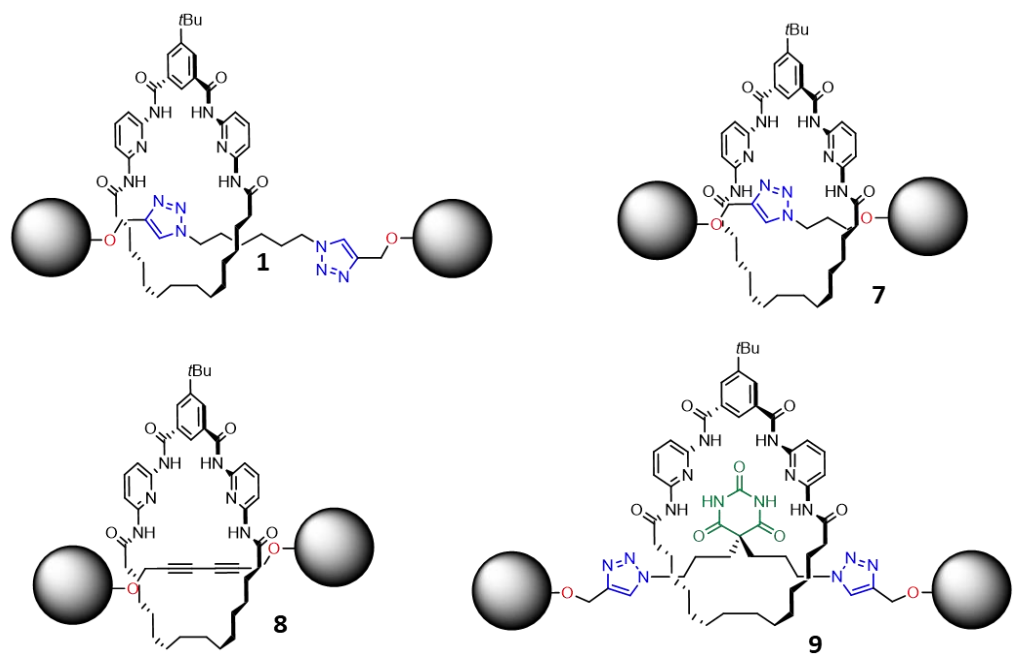

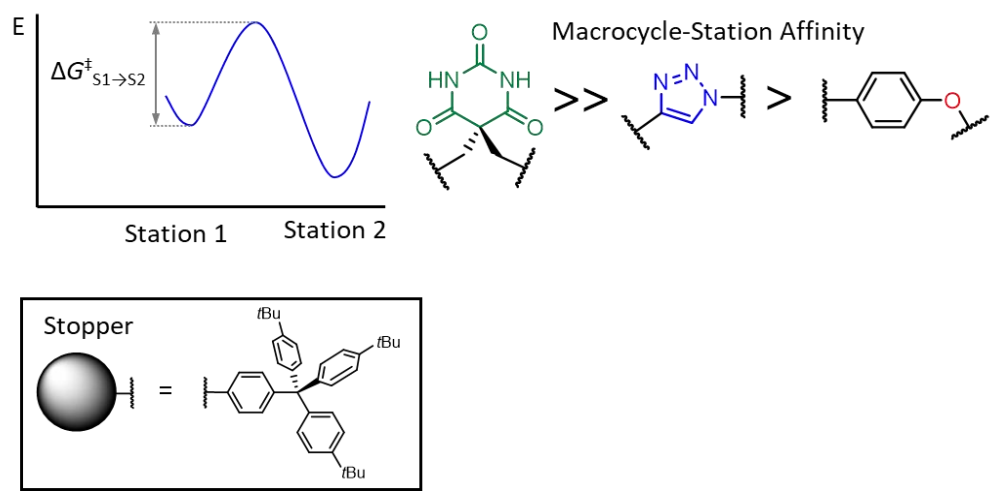

**Figure 1.** Rotaxanes under investigation incorporating H-bonding macrocycle **3**: **1** (synthesis developed in this work) and **7–9** (synthesis previously reported).[27, 28] Right) Qualitative potential energy surface as a function of ring position (not to scale) showing the relative energy of the different stations and the $\Delta G^{\ddagger}_{S1 \to S2}$ shuttling energy barrier. Energetics on non-bonding portions and possible alternative co-conformations involving the interaction with more than one station at the same time are not considered.

**Results and discussion**

The affinity of the macrocycle with the different stations is qualitatively represented in an idealized way in Figure 1. The macrocycle integrating a Hamilton receptor has a high affinity towards the barbiturate station, as evidenced by binding studies with 5,5-diethylbarbiturate, where an interaction energy of –6.0 kcal/mol ($K_{ass} = 23500$ M$^{-1}$ in CHCl$_3$) was determined.[28] In addition, the triazole ring can act as a hydrogen-bond acceptor station,[29, 30] and also the phenol-ether may act as an station with very weak affinity as the interaction of an amide with a phenol in CHCl$_3$ is an unfavourable interaction ($\Delta\Delta G_{\text{H-bond}} = +3.6$ kJ/mol).[31] Macrocycle shuttling in rotaxanes **1**, **7**, **8** and **9** was investigated by VT NMR to study the effect of macrocycle–axle hydrogen bonding interaction strength of the different stations.

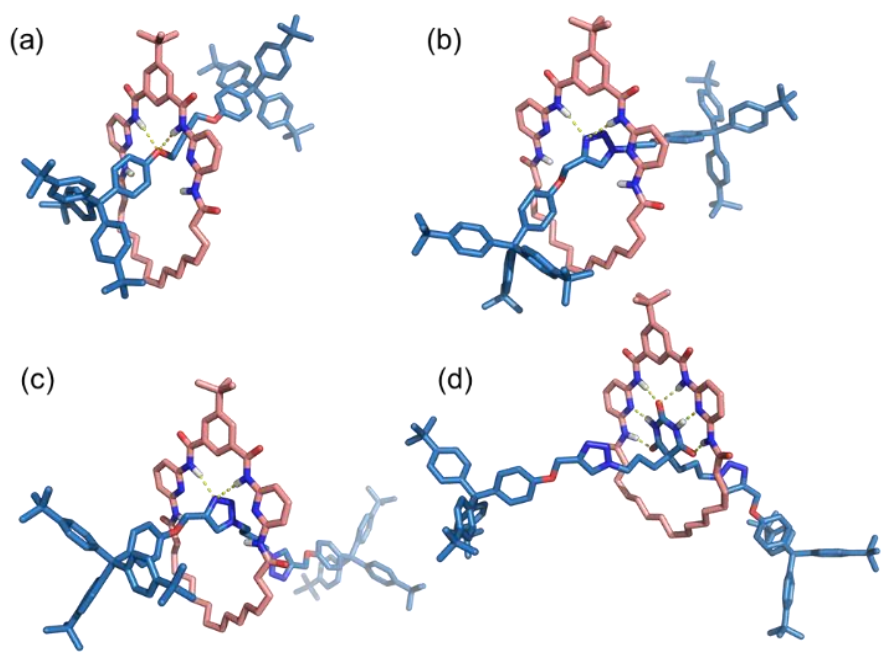

**Figure 2.** Representative molecular models of rotaxanes **1**, **7**–**9** showing the theoretical number of hydrogen bonds between axle and macrocycle. (a) **8**, (b) **7**, (c) **1**, (d) **9**. Note: Flexibility of both macrocyclic and axle components can result in alternative co-conformations (not shown) in this complex system.

Rotaxane **1** comprises the Hamilton-like macrocyclic receptor **3** [28] and a symmetrical axle integrating two triazole groups (Figure 3). Initial attempts to prepare rotaxane **1** from 1,6-diazidehexane **6** and stopper **5** involving two-click reactions in the presence of macrocycle **3** and Cu(I) yield a complex mixture and no evidence of formation of [2]rotaxane **1** (or the corresponding [3]rotaxane) presumably associated to the low rotaxane formation yield (see Figure 3). To facilitate the formation of the rotaxane with fewer possible parasite reactions, a different synthetic strategy was adopted. Synthesis involved a CuAAC "click" active template rotaxane formation involving alkyne stopper **5** and azide-terminated half-axle **4**, prepared from commercial 1,6-dibromohexane.[32] Using this approach one triazole moiety is already present in the azide half-axle **4** and the second triazole ring is created in the rotaxane formation step.

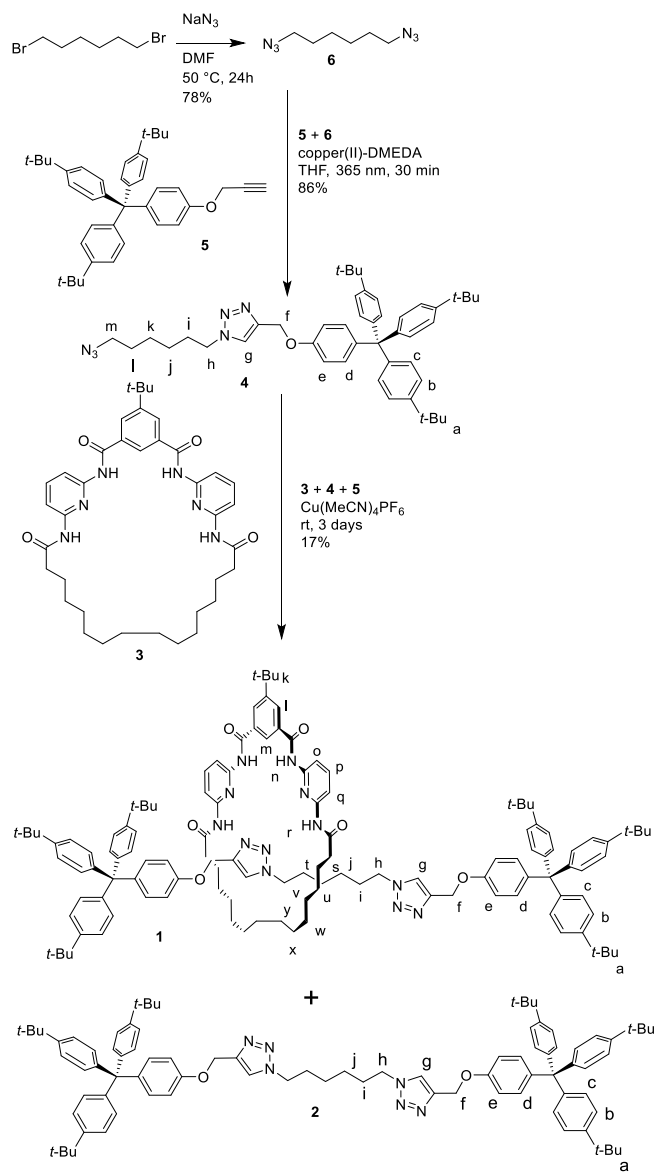

**Figure 3.** Synthesis of rotaxane **1** and free axle **2** side-product via a click reaction. Letters in rotaxane **1** structure refer to $^1$H NMR attributions.

Complexation of copper(I) by macrocycle **3** allows CuAAC "click" active template rotaxane formation.[27] Rotaxane **1** and the axle side-product (**3**) were isolated by column chromatography (SiO$_2$: CH$_2$Cl$_2$/ethyl acetate 10:0 to 8:2, v/v) in 17% and 58% yields, respectively. The obtained moderate yield for rotaxane formation was comparable to the obtained 20% yield for the homologue **7**, also involving a CuAAC

"click" active template rotaxane formation reaction.[27] We note in passing that, undoubtedly due to the relatively large macrocycle size and monodentate pyridine chelator, the obtained yield for these systems is low compared to certain other complex interlocked structures that have >90% rotaxane formation yield, for example based on small macrocycles and strong bipyridine chelators.[26, 33]

Rotaxane **1** was fully characterized by 1D- and 2D-NMR, and mass spectrometry (see electronic supporting information, ESI). The $^1$H NMR spectrum of rotaxane **1** (Figure 4) shows downfield shifts of isophthalamide NH ($\Delta\delta$ = 1.08 ppm, *n*) and CH ($\Delta\delta$ = 0.84 ppm, *m*) signals with respect to non-interlocked macrocycle **3**. This confirms the existence of a selective hydrogen-bonding interaction between the triazole nitrogen atom in axle **2** and the isophthalamide motif in macrocycle **3** (Figure 4). In contrast, the amide NH *r* signal shows an upfield shift of $\Delta\delta$ = –0.18 ppm, suggesting that instead of participating in a hydrogen bonding interaction the proton is shielded by the triazole ring, as observed in the molecular model of rotaxane **1** (see ESI). The cartoon representation in Figure 4 shows a possible co-conformation as flexibility of both macrocyclic and axle components can result in alternative conformations resulting in a complex system.

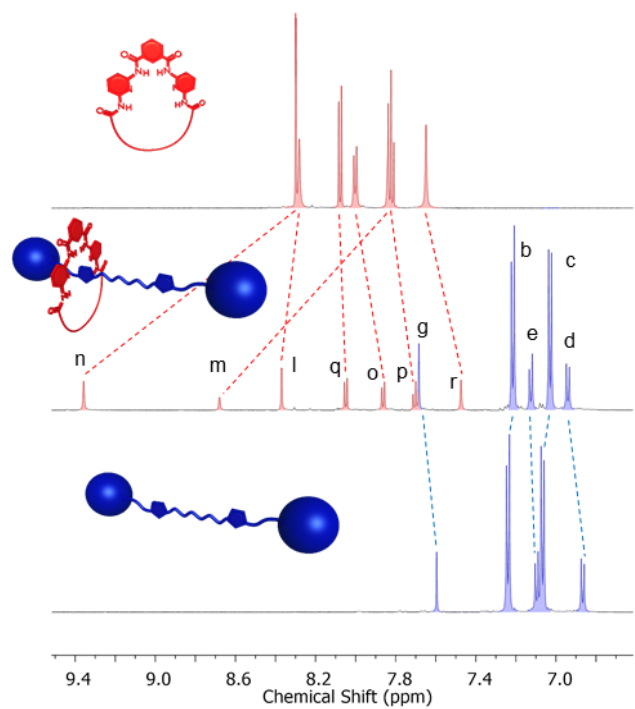

**Figure 4.** Partial $^1$H NMR (600 MHz, TCE-$d_2$, rt) spectra of axle **2** (bottom), rotaxane **1** (middle) and macrocycle **3** (top). Designation of the signals is described in Figure 3.

Additionally, DOSY NMR were performed in TCE-$d_2$ at 25 °C. Proton resonances corresponding to the constituent components in rotaxane **1** have the same diffusion coefficient of ($1.14 \times 10^{-6}$ cm$^2$/s, hydrodynamic radius = 10.4 Å), which is consistent with the interlocked nature of rotaxane **1** (see Figures S9-S12) and smaller compared to free macrocycle **3** ($1.82 \times 10^{-6}$ cm$^2$/s, hydrodynamic radius = 6.5 Å) and axle **2** ($1.20 \times 10^{-6}$ cm$^2$/s, hydrodynamic radius = 9.9 Å).

On cooling a solution of rotaxane **1** in CD$_2$Cl$_2$ from 25 °C to –40 °C (600 MHz spectrometer), only one set of signals was observed indicating fast exchange of the different co-conformation of the macrocycle and axle. A careful analysis of the shapes of the signals shows broadening of resonances, especially the methylene groups *f* and *h* next to the triazole moieties, suggesting rapid macrocycle shuttling between the two triazole groups that would act as stations. We repeated the same experiment in more viscous TCE-$d_2$ (1.84 cP for TCE-$d_2$ and 0.43 cP for CD$_2$Cl$_2$)[34] and a lower dielectric constant (8.20 for TCE-$d_2$ and 8.9 for CD$_2$Cl$_2$).[34] Cooling down to – 30 °C also results in one set of signals with enhanced signal broadening. The methylene group *f*, that is close to the triazole proton showed the most significant broadening (Table 1).

**Table 1.** Peak width at half height for protons *f* and *a* at different temperatures in CD$_2$Cl$_2$ and TCE-$d_2$.

| T (°C) | CD$_2$Cl$_2$ | | TCE-$d_2$ | |
|---|---|---|---|---|
| | *f* $v_{1/2}$ (Hz) | *a* $v_{1/2}$ (Hz) | *f* $v_{1/2}$ (Hz) | *a* $v_{1/2}$ (Hz) |
| 25 | 1.73 | 0.99 | 3.23 | 1.48 |
| 10 | 2.90 | 2.02 | 5.39 | 2.84 |
| 0 | 2.44 | 1.30 | 6.56 | 2.57 |
| −10 | 3.06 | 1.31 | 9.29 | 3.14 |
| −20 | 4.00 | 1.53 | 19.67 | 4.49 |
| −30 | 5.34 | 1.97 | 31.65 | 8.33 |
| −40 | 10.50 | 2.51 | — | — |

As fast exchange slows down on cooling, a significant broadening of signals is observed but the coalescence temperature is not reached. (Note: Technical limitations preclude further lowering the temperature). However, the observed broadening of signals in TCE-$d_2$ is adequate for determining the shuttling exchange rate constant. In this case, $k > v_{S1-S2}$ and the process is slow enough to contribute to its width. The exchange rate constant $k$ can be obtained using Equation 1.[35]

$$k \approx \frac{\pi\, v_{vS1-S2}^2}{2(\Delta v - \Delta v_{ref})} \tag{1}$$

In Equation 1, $\Delta v$ is the peak width at half-height, $\Delta v_{ref}$ is the peak width at half height of a non-exchanging reference. The *t*-Bu signal peak *a* was chosen as the non-exchanging. $v_{S1-S2}$ is the peak separation of *f* and *f'* after coalescence and can be estimated as $\nu_{S1\text{-}S2} = 2\times(\nu_{rotax} - \nu_{axle}) = 232$ Hz (0.386 ppm in the 600 MHz spectrometer) considering that the observed chemical shift in the rotaxane (5.348 ppm, $\nu_{rotax}$) is the average of "free" and "occupied" stations and the chemical shift of the free station has been estimated from the free axle observed chemical shift (5.16 ppm, $\nu_{axle}$). Signals were fitted to Lorentzian peaks to obtain the width at half height and the corresponding exchange rate constants $k$ obtained using equation 1 (Figure 5a). Thermodynamics of the macrocycle shuttling process can be obtained from the $k$ values at different temperatures by an Eyring plot (Figure 5b). From this plot, a value of $\Delta H^{\ddagger} = 7.4$ kcal/mol and $\Delta S^{\ddagger} = -11.5$ cal/mol K are obtained. These parameters allow extrapolating at 25 °C a value of $\Delta G^{\ddagger}_{25\,°C} = 10.8$ kcal/mol that corresponds to an exchange rate constant $k = 72000$ s$^{-1}$.

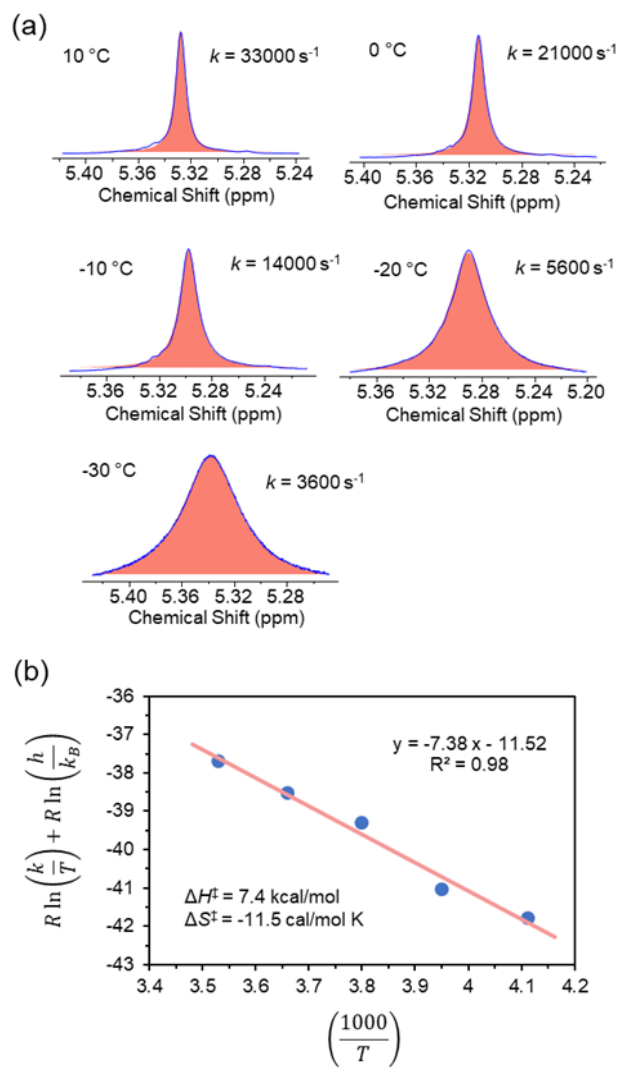

**Figure 5.** Determination of shuttling rate parameters (a) Fittings to Lorentzian peaks of $^1$H NMR signal *f* at different temperatures (600 MHz, TCE-$d_2$) and the corresponding exchange rate constants *k* Fittings displayed as a dashed area and experimental data in a solid line. (b) Eyring plot of the exchange rate constants *k*.

Efforts to obtain precise shuttling rate constants or macrocycle rocking rate constants by VT NMR spectroscopy for rotaxanes **7**, **8**, and **9** in different solvents (CDCl$_3$, CD$_2$Cl$_2$, TCE-$d_2$, DMSO-$d_6$, etc.) were unsuccessful as coalescence was not reached. As such, only qualitative shuttling information could be obtained for rotaxanes **8** and **9**. On cooling, $^1$H NMR signal broadening was observed for rotaxane **7**, which has only one triazole station, which could be associated with macrocycle rocking movement between the amide protons *n* and *r* (Figure S15). Weaker interaction of the isophthalamide with the phenol-ether stations in rotaxane **8** results in insignificant signal broadening on cooling **8** in CD$_2$Cl$_2$ (Figure S16), consistent with very fast macrocycle shuttling which is anticipated to be faster than in **1**. In the case of rotaxane **9**, a stronger interaction was observed between the barbiturate station and the Hamilton receptor motif. Indeed, even in a competitive solvent mixture (CDCl$_3$/MeOD/D$_2$O 45:45:10 v/v, Figure S17) or heating in a very strongly hydrogen bonding solvent (DMSO-*d$_6$*, Figure S18), which should weaken hydrogen bonding interactions between the two interlocked components, the macrocycle remains predominantly positioned at the barbiturate station. No specific evidence points to even a minor population on the triazole stations.[20] The corresponding thermodynamic equilibrium constant between the barbiturate station the triazole stations would be largely biased to towards the barbiturate station. As the shuttling energy barrier only depends on the starting station and not the end station,[1] the shuttling rate from the triazole stations to the barbiturate station is predicted to be equivalent to that determined for rotaxane **1** with occasionally shuttling from the barbiturate station the triazole stations. This rate would be much slower and estimated by molecular modeling to be in the one-shuttling event per second time scale (see Figure S19 in the Supporting Information).

**Conclusions**

In summary, herein we report the study of macrocycle shuttling in rotaxanes featuring the Hamilton-type hydrogen-bonding receptor (**1**, **8**, and **9**). The rotaxanes have different hydrogen bonding affinities between the axle stations and the macrocycle isophthalamide motifs. These different affinities result in dramatic changes in the macrocycle shuttling rates at 25 °C from the different stations in the following order: barbiturate (**9**) < triazole (**1**) < phenol (**8**). Information on macrocycle-axle interactions and resulting ring dynamics could prove useful in introducing predetermined macrocycle shuttling rates in 2-station rotaxanes.

**Experimental**

*Materials and methods*

All chemicals and solvents were obtained from commercial sources and used without further purification unless a specific procedure is described. Dry solvents were obtained using solvent purification system inert® PureSolv Model PS-MD-5. Mass spectra and NMR were performed in the CESAMO analytical facilities (Bordeaux, France).

*NMR*

$^1$H and $^{13}$C NMR spectra were recorded on a Bruker Avance I 300 MHz (300 MHz $^1$H, 75 MHz $^{13}$C), Bruker Avance II 400 MHz (400 MHz $^1$H, 100 MHz $^{13}$C); or Bruker Avance III 600 MHz (600 MHz $^1$H, 150 MHz $^{13}$C) spectrometer. Chemical shifts are reported in ppm and referenced to the solvent residual peaks. For the assignment of signals, the following abbreviations are used are s = singlet, d = doublet and t = triplet. DOSY NMR was performed in a Bruker Avance III 600 spectrometer using the standard Bruker TopSpin 2D Stimulated Echo experiment using bipolar gradients (stebpgp1s) for diffusion measurements at 25 °C. The diffusion parameters were optimized to obtain a >70% on signal-intensity decay. From the DOSY diffusion coefficient, the hydrodynamic radius was obtained by using the Stokes–Einstein equation $R = kT/(6\pi\eta D)$. VT NMR was performed in a Bruker Avance III 600

spectrometer, technical limitations preclude lowering the temperature below –40 °C.

*Mass spectrometry*

Electrospray spectra (ESI) were recorded on a Qexactive (Thermo) mass spectrometer. The instrument is equipped with an ESI source and spectra were recorded in positive mode. The spray voltage was maintained at 3200 V and capillary temperature set to 320 °C. Samples were introduced by injection through a 20 µL loop into a 300 µL/min flow of methanol from the LC pump.

*VT NMR and exchange rate constant calculation*

The $\Delta G^\ddagger$ at 25 °C was obtained to all systems by interpolation/extrapolation using Eyring plot using from the measured exchange rate constants (equations 2 and 3).[36]

$$R \ln\left(\frac{k_{ex}}{T}\right) + R \ln\left(\frac{h}{k_B}\right) = -\Delta H^\ddagger \left(\frac{1}{T}\right) + \Delta S^\ddagger \quad (2)$$

With $R = 1.99$ cal mol$^{-1}$ K$^{-1}$, $h = 6.63 \times 10^{-34}$ J s and $k_B = 1.23 \times 10^{-23}$ J K$^{-1}$ and a factor of 1000 in the enthalpic term to use the kcal.mol$^{-1}$ units in $\Delta H^\ddagger$, equation 2 converts to equation 3.

$$1.99 \ln\left(\frac{k_{ex}}{T}\right) - 47.18 \frac{\text{cal}}{\text{mol K}} = -\Delta H^\ddagger \left(\frac{1000}{T}\right) + \Delta S^\ddagger \quad (3)$$

The exchange rate constant $k$ can be obtained using equation 1.[35] An approximate value for $v_{S1-S2}$ has been obtained assuming that in the rotaxane the observed chemical shift is the average of "free" and "occupied" station and the chemical shift of the free station has been estimated from the free axle.

*Molecular Modelling*

Structures were minimized with the MMFF force field using the Spartan '18 software.[37] The isophthalamide motif in the macrocycle was constrained to be planar. The molecular models allowed determining the number of hydrogen bonds between the macrocycle and thread for the different rotaxanes described in this research (see ESI).

*Synthetic procedures*

1,6-Diazidehexane **6**, alkyne stopper **5**, Hamilton-like macrocyclic receptor **3** and rotaxanes **7**, **8** and **9** were prepared using literature procedures.[27, 28, 38, 39]

*Synthesis of rotaxane 1 and axle 2*

Macrocycle **3** (43 mg, 0.0657 mmol, 1.1 eq.) and Cu(MeCN)$_4$PF$_6$ (25 mg, 0.0671 mmol, 1.1 eq.) are dissolved in anhydrous chloroform (4 mL) previously degassed with gentle N$_2$ bubbling for 15 min. Solid azide **4** (43 mg, 0.0605 mmol, 1 eq.) and alkyne **5** (47.5 mg, 0.0875, 1.4 eq.) were simultaneously added to the mixture in one portion. The reaction mixture was stirred at room temperature for 3 days in the dark under N$_2$ atmosphere. The reaction mixture was quenched with 1 mL of a solution of EDTA (1M) in concentrated ammonium hydroxide (35%) and stirred at room temperature for 30 min. The mixture was diluted with dichloromethane (40 mL) and the organic layer was washed successively with water (2 × 50 mL) and brine (1 × 50 mL). The organic layer was dried over magnesium sulfate, filtered and concentrated *in vacuo*. Purification by silica gel column chromatography (DCM/EtOAc, 10:0 to 8:2, v/v) afforded double-station rotaxane **1** as a colorless solid (13 mg, 17% yield) and axle **2** as a colorless solid (44 mg, 58% yield).

Rotaxane **1**: **$^1$H NMR** (600 MHz, CD$_2$Cl$_2$) δ 9.34 (s, 2H, $H_n$), 8.60 (brs, 1H, $H_m$), 8.34

(d, *J* = 1.4 Hz, 2H, *H$_l$*), 8.02 (d, *J* = 8.0 Hz, 2H, *H$_q$*), 7.84 (d, *J* = 8.0 Hz, 2H, *H$_o$*), 7.69 (s, 2H, *H$_g$*), 7.67 (t, *J* = 8.0 Hz, 2H, *H$_p$*), 7.56 (brs, 2H, *H$_r$*), 7.23 (d, *J* = 8.5 Hz, 12H, *H$_b$*), 7.16 (d, *J* = 8.9 Hz, 4H, *H$_e$*), 7.10 (d, *J* = 8.5 Hz, 12H, *H$_c$*), 6.93 (d, *J* = 8.9 Hz, 4H, *H$_d$*), 5.32 (bs 4H, *H$_f$*), 4.17 (t, *J* = 7.1 Hz, 4H, *H$_h$*), 2.11 (t, *J* = 7.8 Hz, 4H, H$_s$), 1.60 (m, 8H, H$_t$ + H$_i$), 1.41 (s, 9H, H$_k$), 1.28 (s, 54H, *H$_a$*), 1.27 – 1.23 (m, 20H, *H$_{u-y}$*), 1,08 (m, 4H, *H$_j$*). **$^{13}$C NMR** (151 MHz, CD$_2$Cl$_2$) δ 171.9, 165.5, 156.6, 153.8, 150.8, 150.3, 149.0, 145.0, 144.8, 141.5, 141.0, 134.6, 132.7, 130.9, 130.5, 124.9, 123.4, 122.0, 114.2, 110.2, 109.7, 63.7, 63.1, 50.8, 38.0, 35.7, 34.8, 31.7, 31.5, 30.3, 29.3, 29.2, 29.1, 29.0, 28.7, 26.1, 25.5. **HRMS (ESI)**: calcd for C$_{124}$H$_{154}$N$_{12}$O$_6$+H$^+$ [M+H]$^+$ m/z = 1908.2187, found m/z = 1908.2166.

Axle **2**: **$^1$H NMR** (600 MHz, TCE-*d$_2$*) δ 7.60 (s, 2H, *H$_g$*), 7.25 (d, *J* = 8.6 Hz, 12H, *H$_b$*), 7.10 (d, *J* = 8.9 Hz, 4H, *H$_d$*), 7.07 (d, *J* = 8.6 Hz, *H$_c$*), 6.87 (d, *J* = 9.0 Hz, 4H, *H$_e$*), 5.16 (s, 4H, *H$_f$*), 4.32 (t, *J* = 7.1 Hz, 4H, *H$_h$*), 1.89 (m, 4H, *H$_i$*), 1.38 – 1.33 (m, 4H, *H$_j$*), 1.31 (s, 54H, *H$_a$*). **$^{13}$C NMR** (75 MHz, CDCl$_3$) δ 156.3, 148.5, 144.6, 144.2, 140.3, 132.5, 130.8, 124.2, 122.6, 113.4, 63.20, 62.18, 50.2, 34.4, 31.5, 30.1, 26.0. **HRMS (ESI)** calcd for C$_{86}$H$_{104}$N$_6$O$_2$+H$^+$ [M+H]$^+$ m/z = 1253.8294, found m/z = 1253.8278.

*Synthesis of azide 4*

Azide **4** was synthesized using a general copper-click procedure from Beniazza *et al.*[32] In a Schlenk tube, copper(II)-DMEDA complex (6.2 mg, 8.56 µmol, 1 mol%), alkyne **5** (465 mg, 0.856 mmol, 1 eq.) and azide **6** (1.44 g, 8.56 mmol, 10 eq.) were dissolved in THF (5 mL). The reaction mixture, protected from light by aluminium foil, was then degassed by gentle argon bubbling for 20 minutes. The reaction mixture was irradiated at 365 nm using a TLC lamp placed at ~ 1 cm from the tube and stirred for 30 min at room temperature. The reaction mixture was diluted with EtOAc (50 mL) and washed successively with water (2 × 50 mL) and brine (1 × 50 mL). The organic layer was dried over magnesium sulfate, filtered and concentrated *in vacuo*. Purification by silica gel column chromatography (cyclohexane/EtOAc 10:0 to 9:1, v/v) afforded azide **4** as a colorless solid (526 mg, 86% yield). **$^1$H NMR** (300 MHz, CDCl$_3$) δ 7.59 (m, 1H, $H_g$), 7.23 (m, 6H, $H_b$), 7.09 (m, 8H, $H_c + H_d$), 6.86 (m, 2H, $H_e$), 5.19 (s, 2H, $H_f$), 4.37 (t, $J$ = 7.2 Hz, 2H, $H_h$), 3.26 (t, $J$ = 6.7 Hz, 2H, $H_m$), 1.94 (p, $J$ = 7.3 Hz, 1H, $H_i$), 1.65 – 1.53 (m, 2H, $H_l$), 1.49 – 1.35 (m, 4H, $H_k + H_j$), 1.30 (s, 27H, $H_a$). **$^{13}$C NMR** (100 MHz, CDCl$_3$) 156.2, 148.5, 144.5, 144.2, 140.3, 132.5, 130.8, 124.2, 122.6, 113.4, 63.2, 62.2, 51.3, 50.4, 34.4, 31.5, 30.2, 28.7, 26.2, 26.2. **HRMS (ESI)** calcd for C$_{46}$H$_{58}$N$_6$O+H$^+$ [M+H]$^+$ m/z = 711.4745, found m/z = 711.4731.

## Conflicts of interest

There are no conflicts to declare.

## Acknowledgements

We thank the analytical facilities CESAMO (NMR and HRMS) of the Institut des Sciences Moléculaires, University of Bordeaux. This project has received funding from the European Union's Horizon 2020 research and innovation programme under the Marie Skłodowska-Curie grant agreement No. 796612. Equally, financial support from the Agence Nationale de la Recherche (project ANR-16-CE29-0011) and CNRS is gratefully acknowledged.

## Notes and references

# Supporting information

# NMR spectra

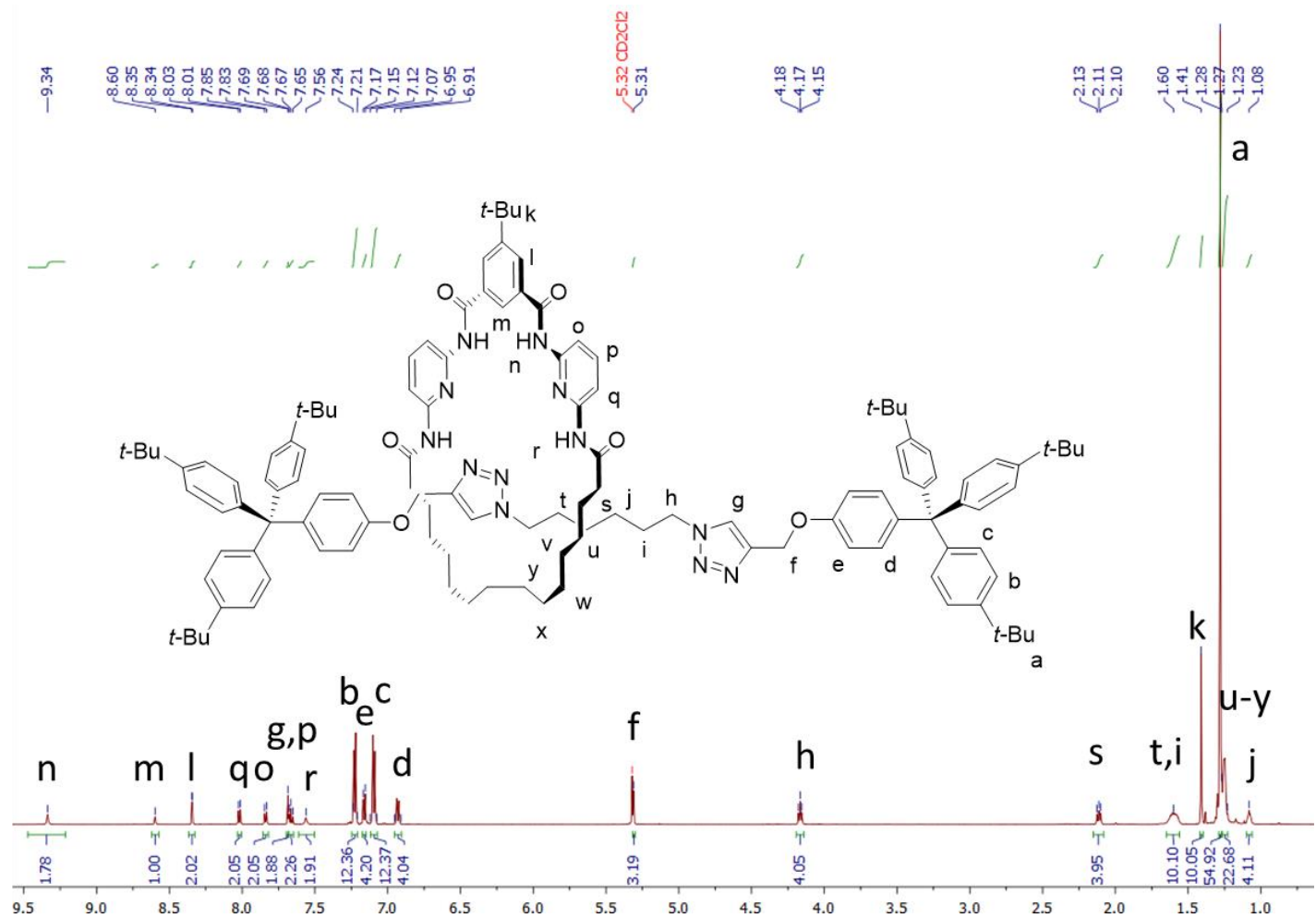

**Figure S1**. $^1$H-NMR of rotaxane **1** (600 MHz in CD$_2$Cl$_2$).

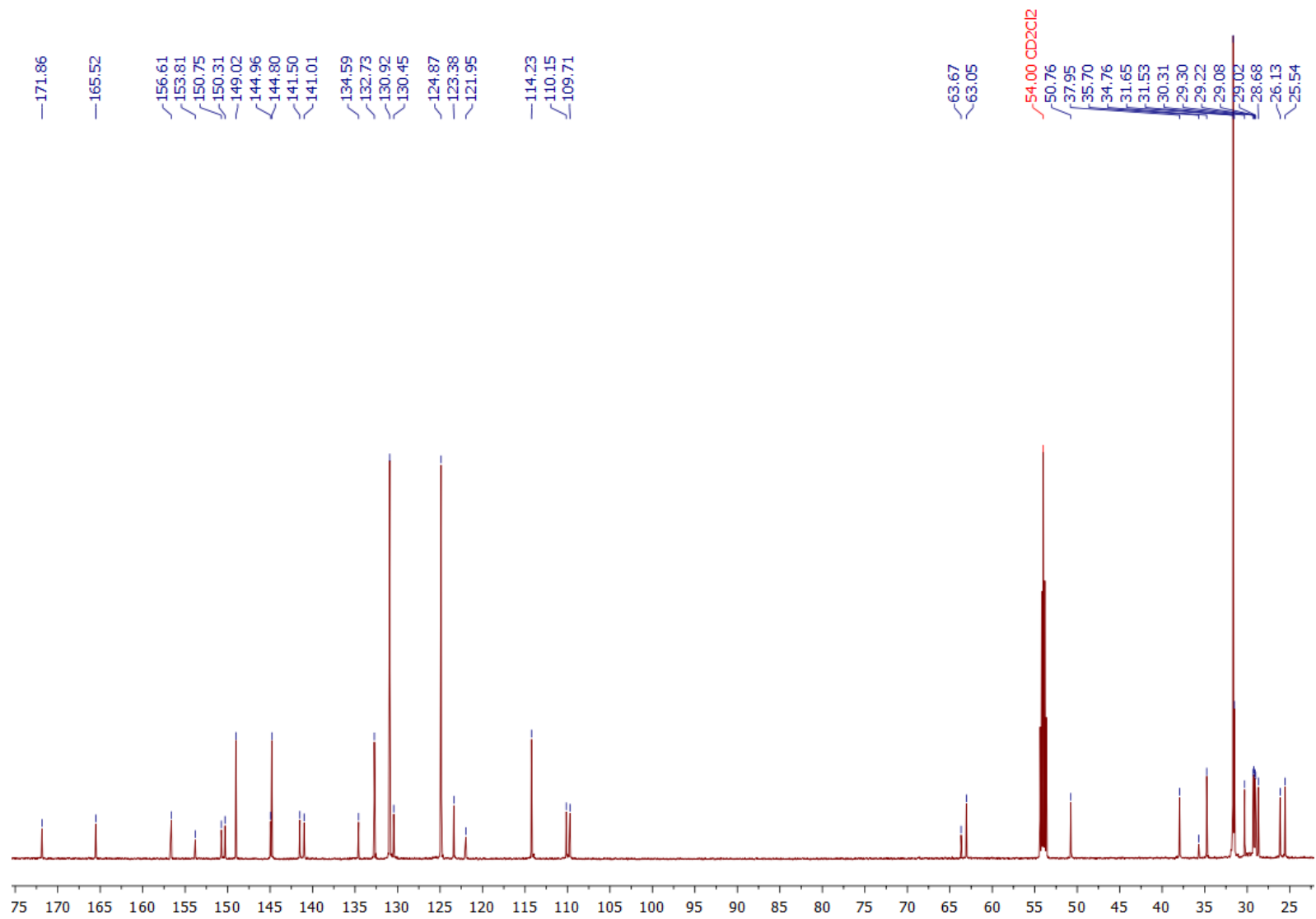

**Figure S2**. $^{13}$C-NMR of rotaxane **1** (151 MHz in $CD_2Cl_2$).

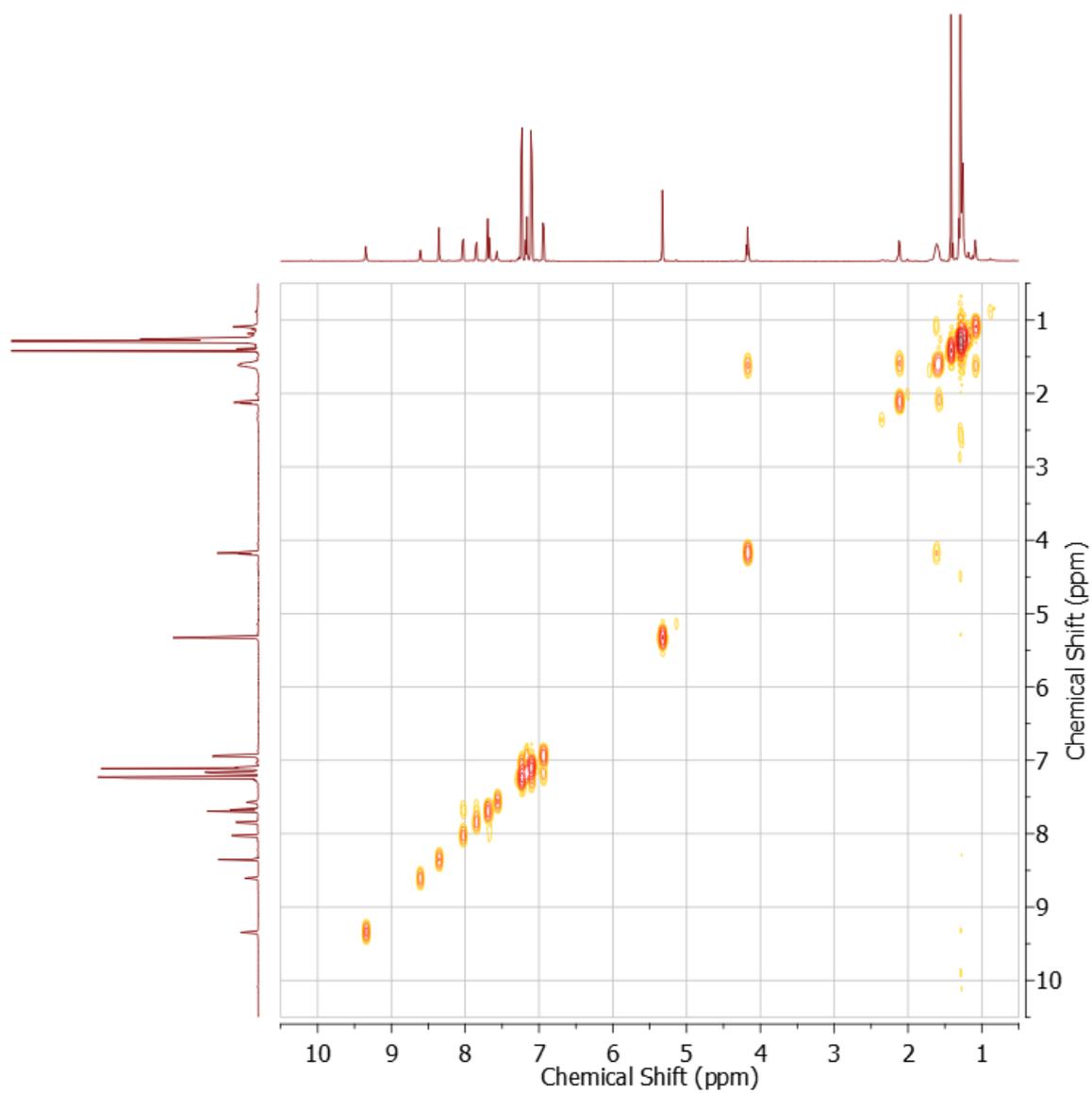

**Figure S3**. $^1$H–$^1$H COSY-NMR of rotaxane **1** (600 MHz in CD$_2$Cl$_2$).

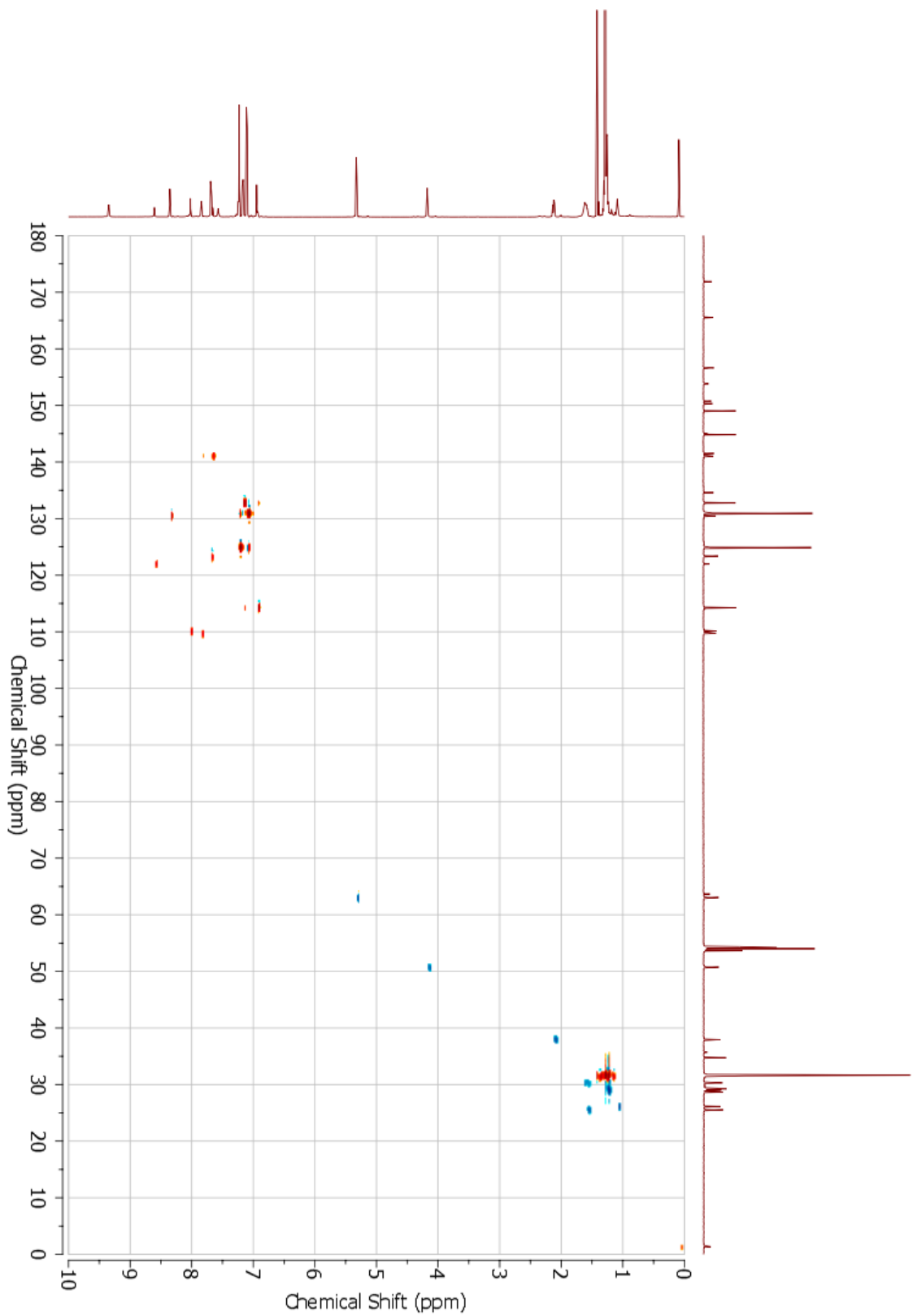

**Figure S4**. ($^1$H,$^{13}$C)-HSQC NMR of rotaxane **1** (600 MHz in CD$_2$Cl$_2$).

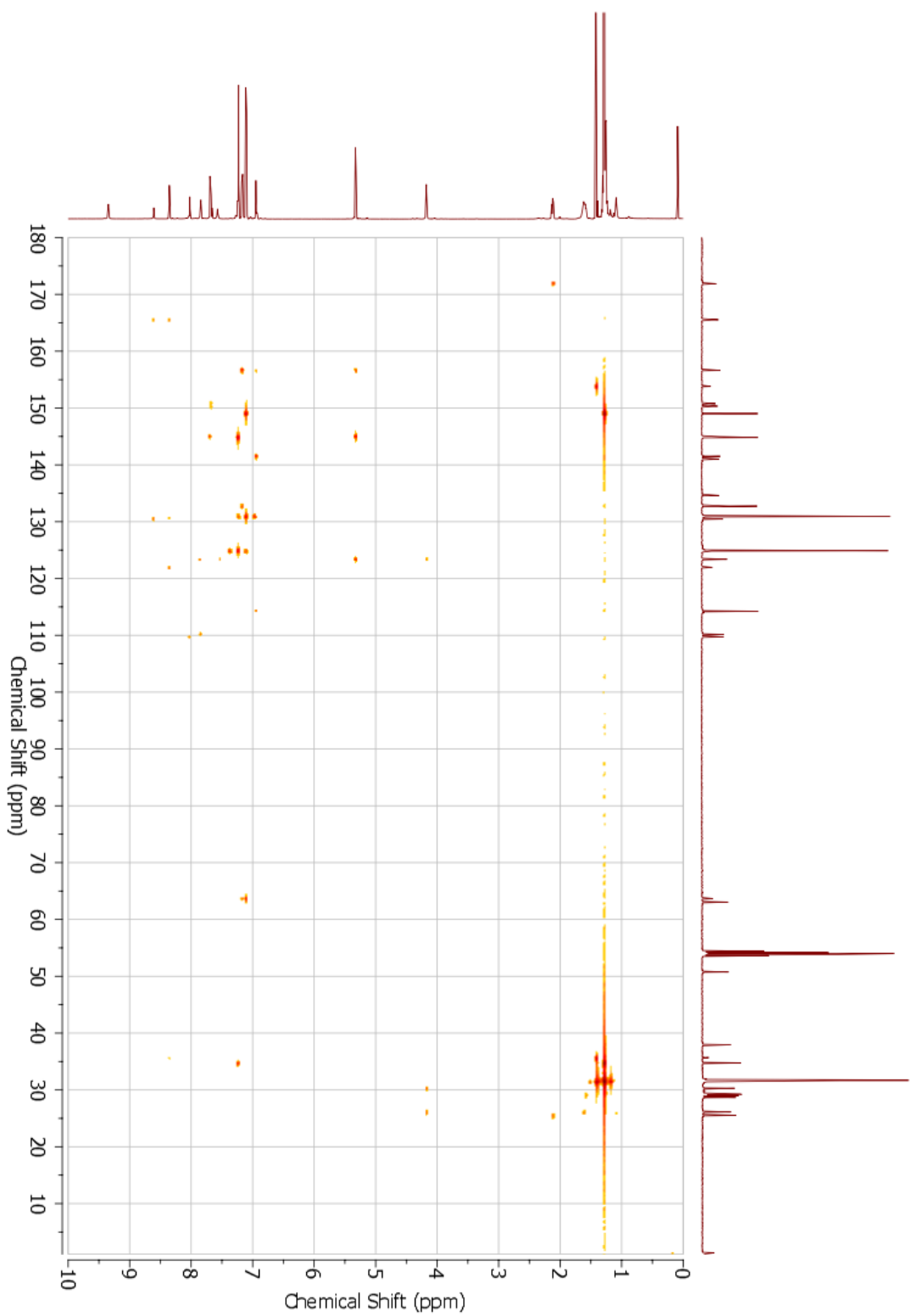

**Figure S5**. ($^1$H,$^{13}$C)-HMBC NMR of rotaxane **1** (600 MHz in CD$_2$Cl$_2$).

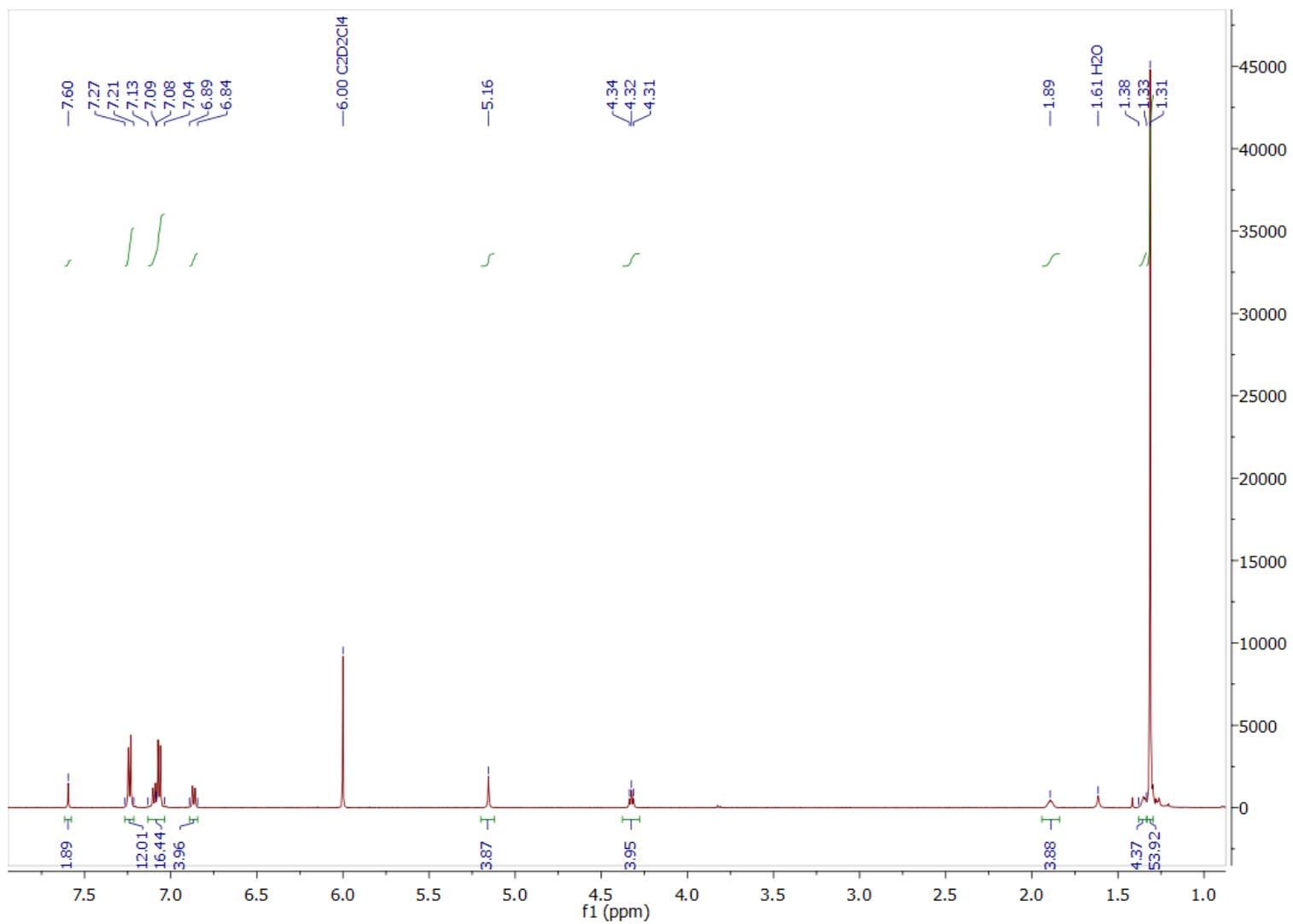

**Figure S6.** $^1$H-NMR of thread **2** (600 MHz in TCE-$d_2$).

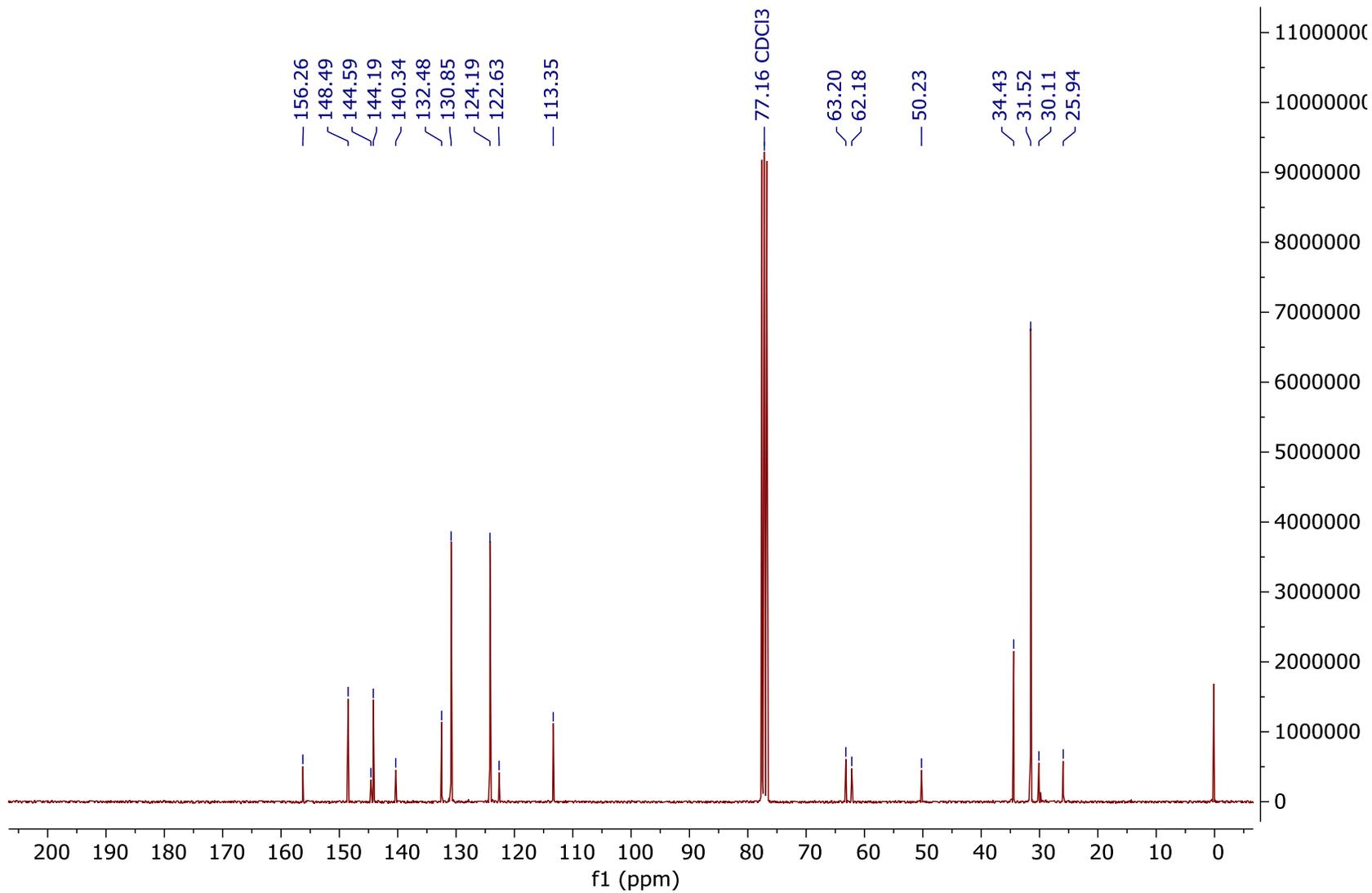

**Figure S7.** $^{13}$C-NMR of thread **2** (75 MHz in CDCl$_3$).

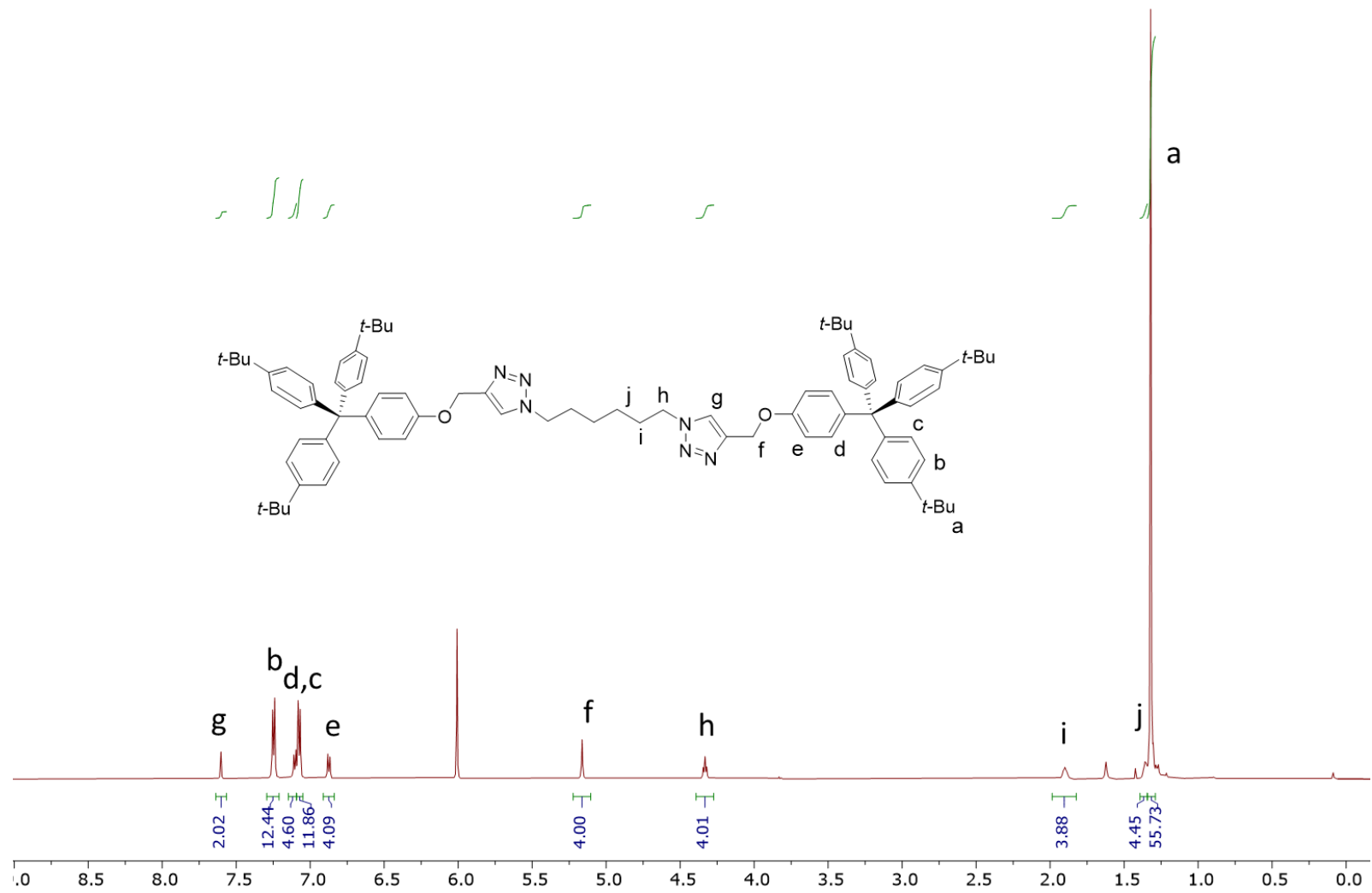

**Figure S8**. $^1$H-NMR of azide **4** (300 MHz in TCE-$d_2$).

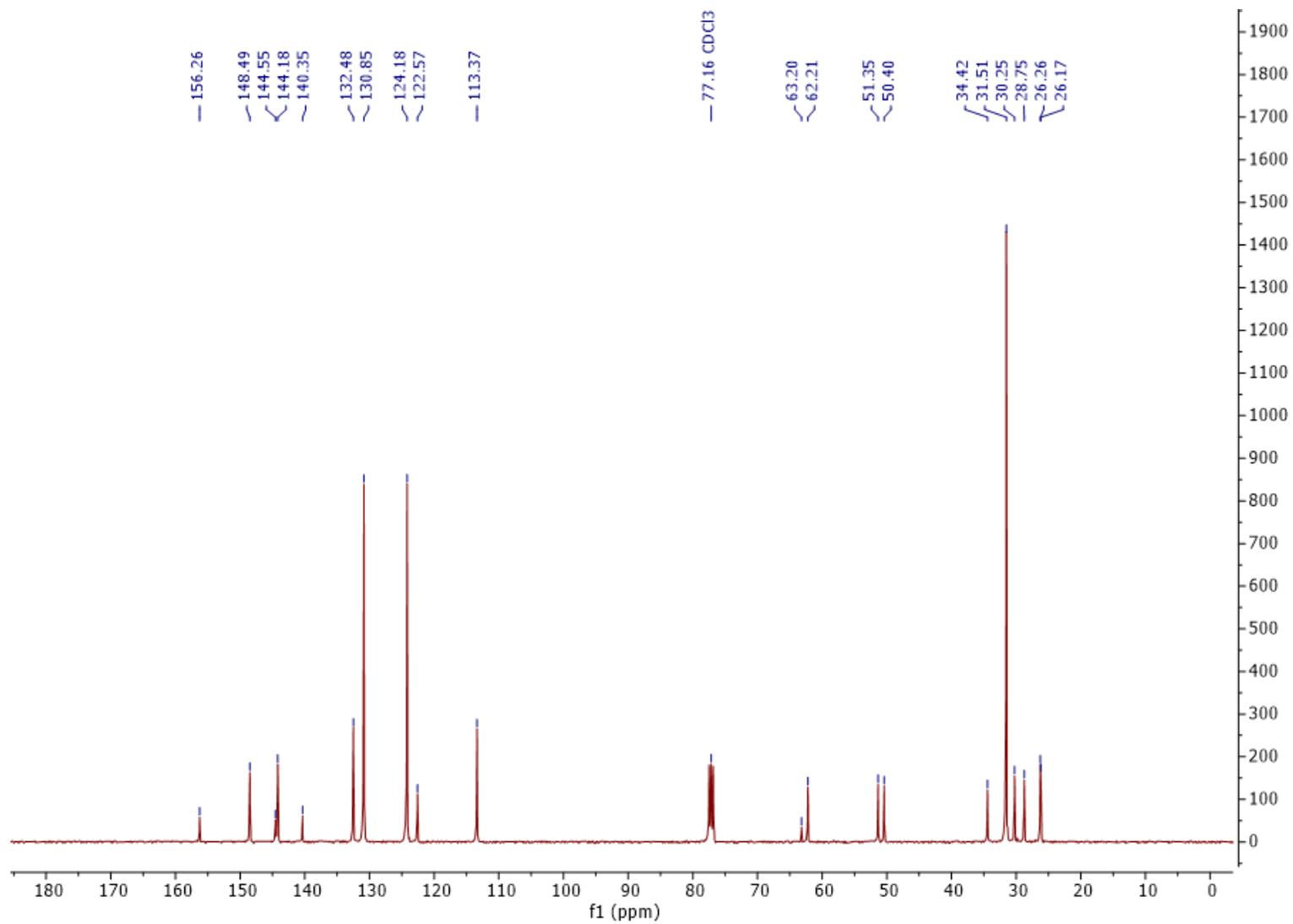

**Figure S9**. $^{13}$C-NMR of azide **4** (100 MHz in CDCl$_3$).

## NMR diffusion studies (DOSY)

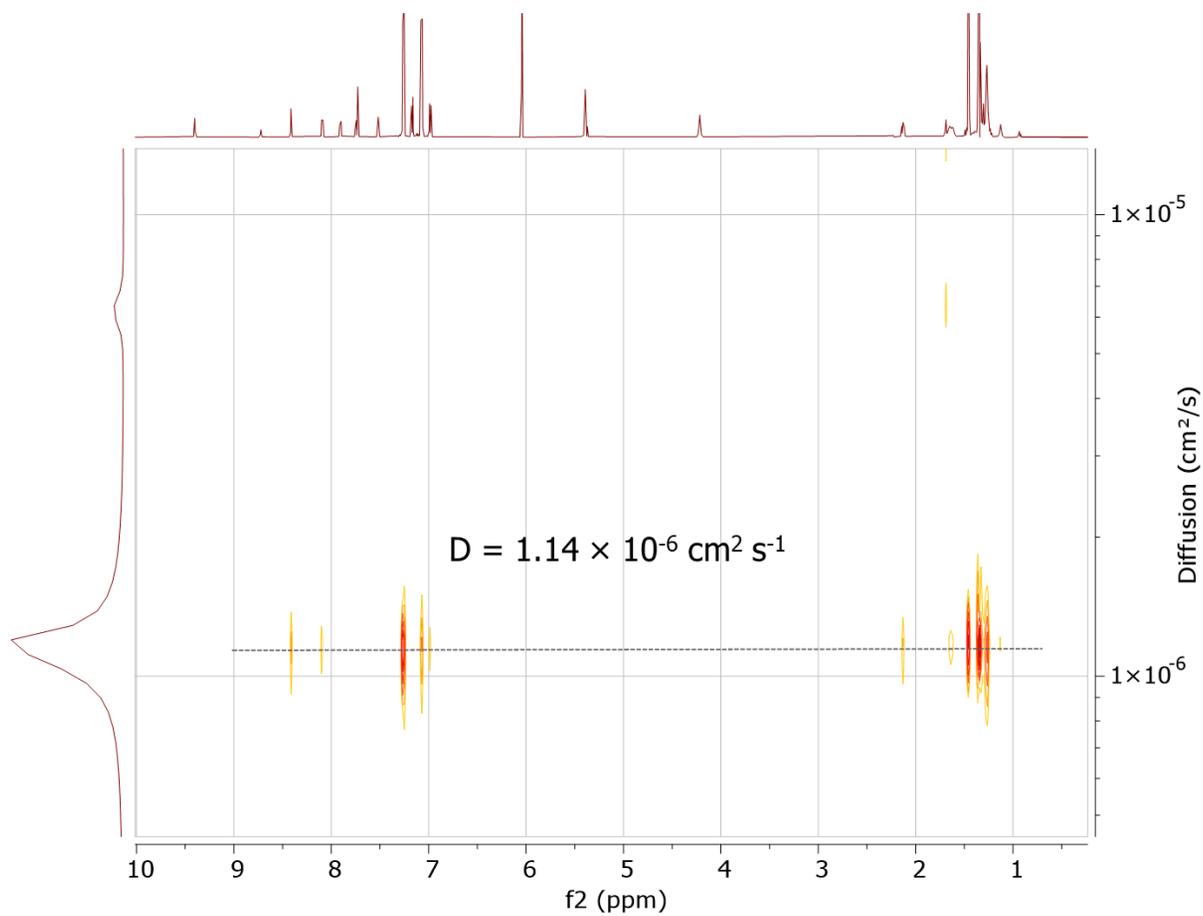

**Figure S10**. DOSY NMR of rotaxane **1** (600 MHz in TCE-$d_2$).

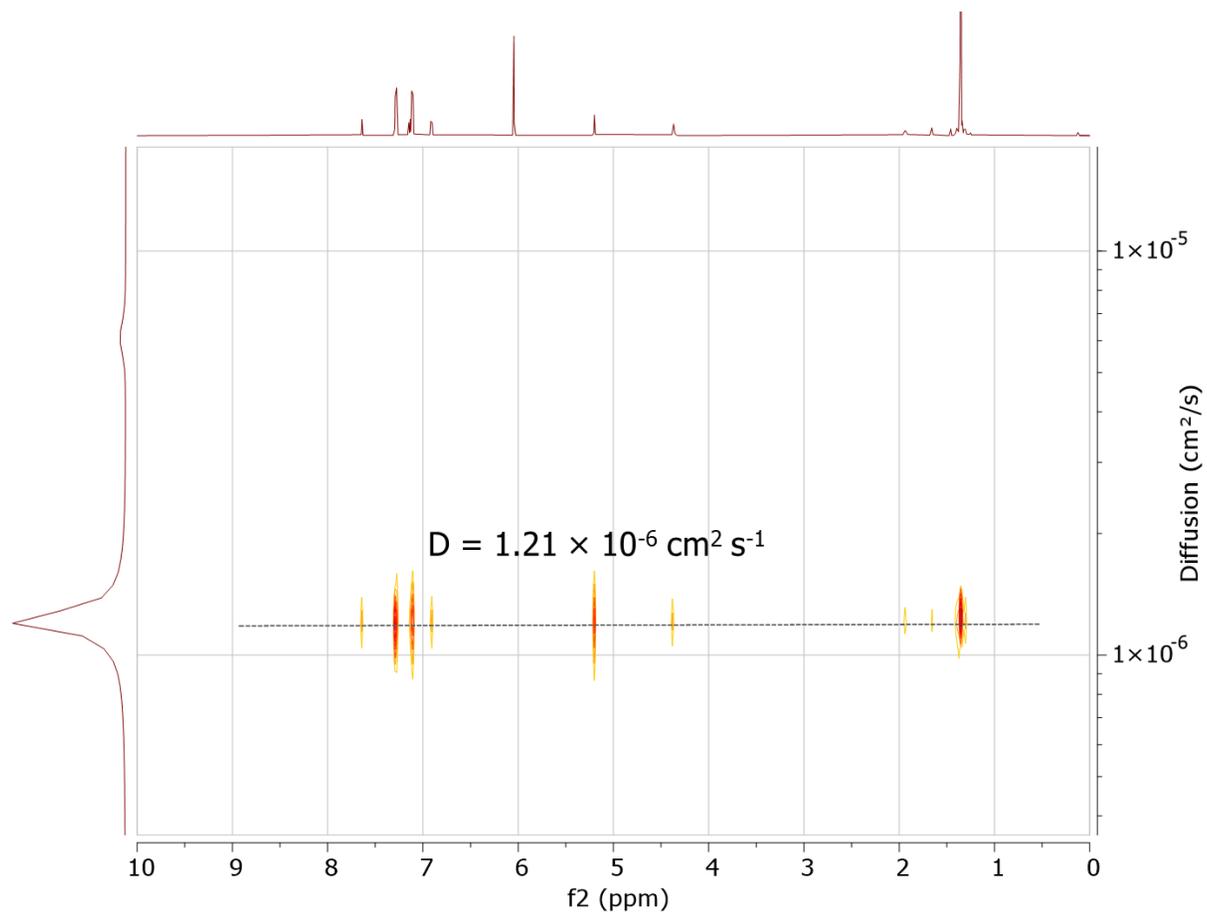

**Figure S11**. DOSY NMR of thread **2** (600 MHz in TCE-$d_2$).

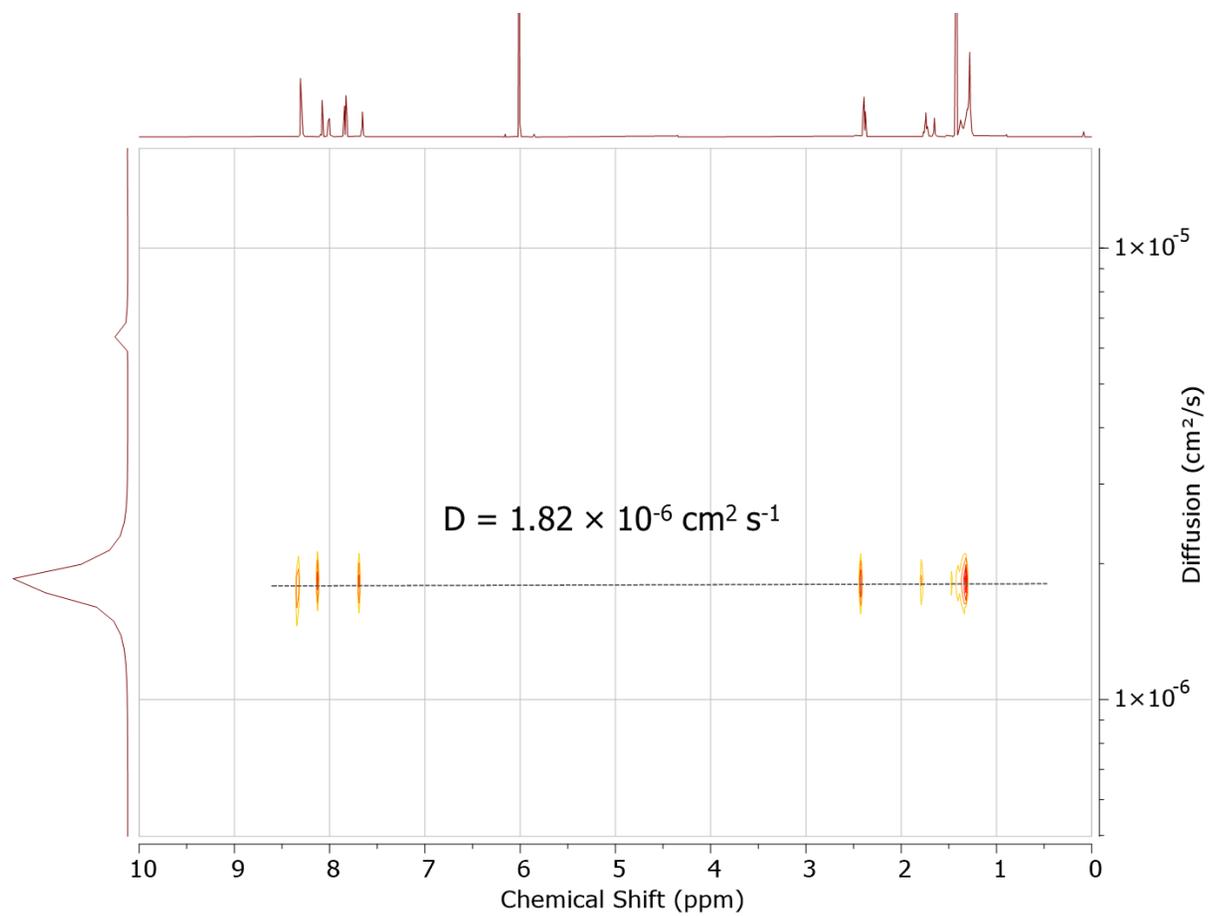

**Figure S12**. DOSY NMR of macrocycle **3** (600 MHz in TCE-$d_2$).

# VT NMR and exchange rate constant calculation

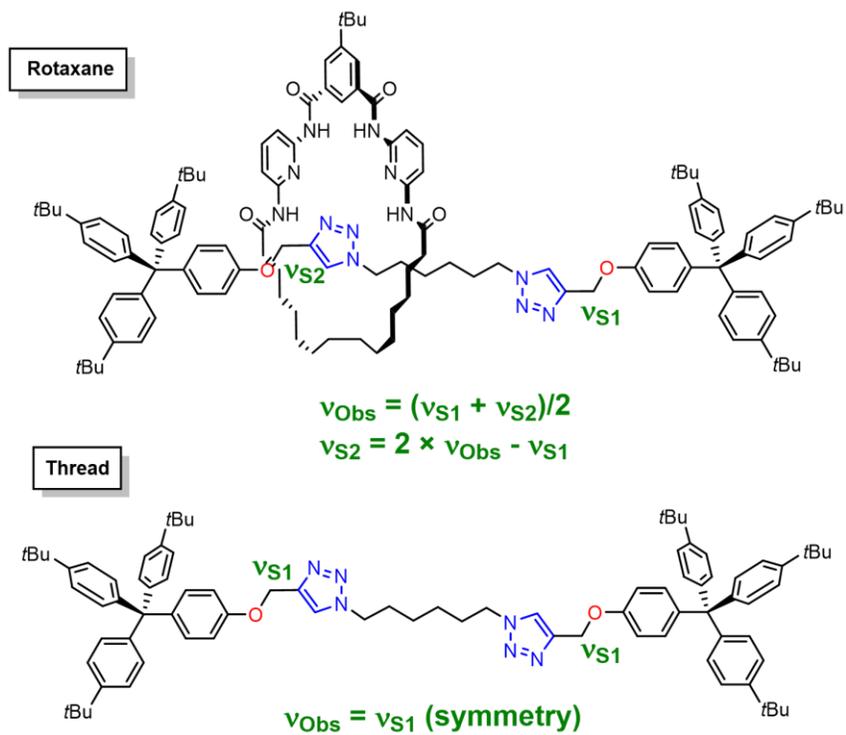

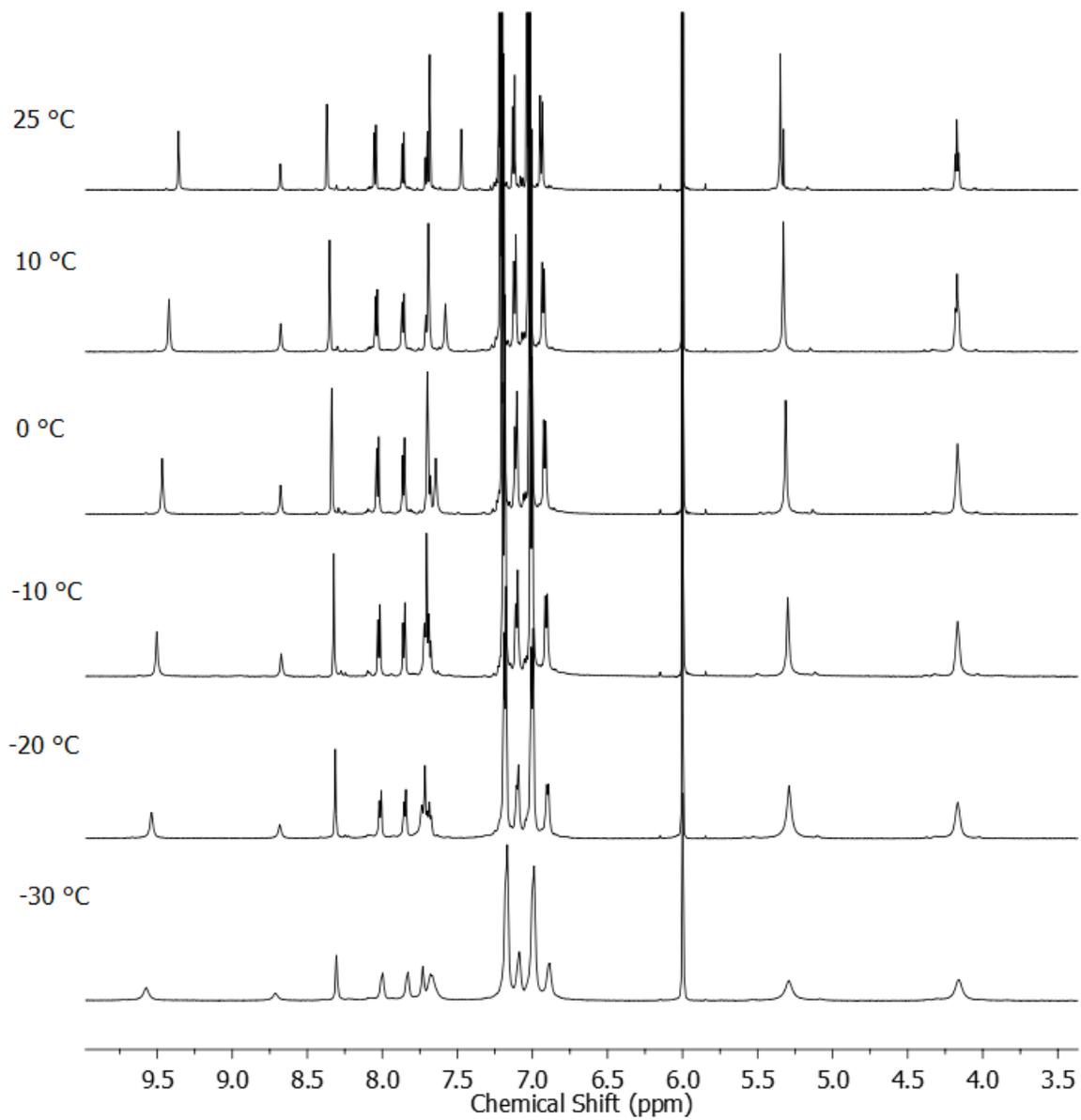

**Figure S13**. $^1$H-NMR of rotaxane **1** at different temperatures (600 MHz in TCE-$d_2$).

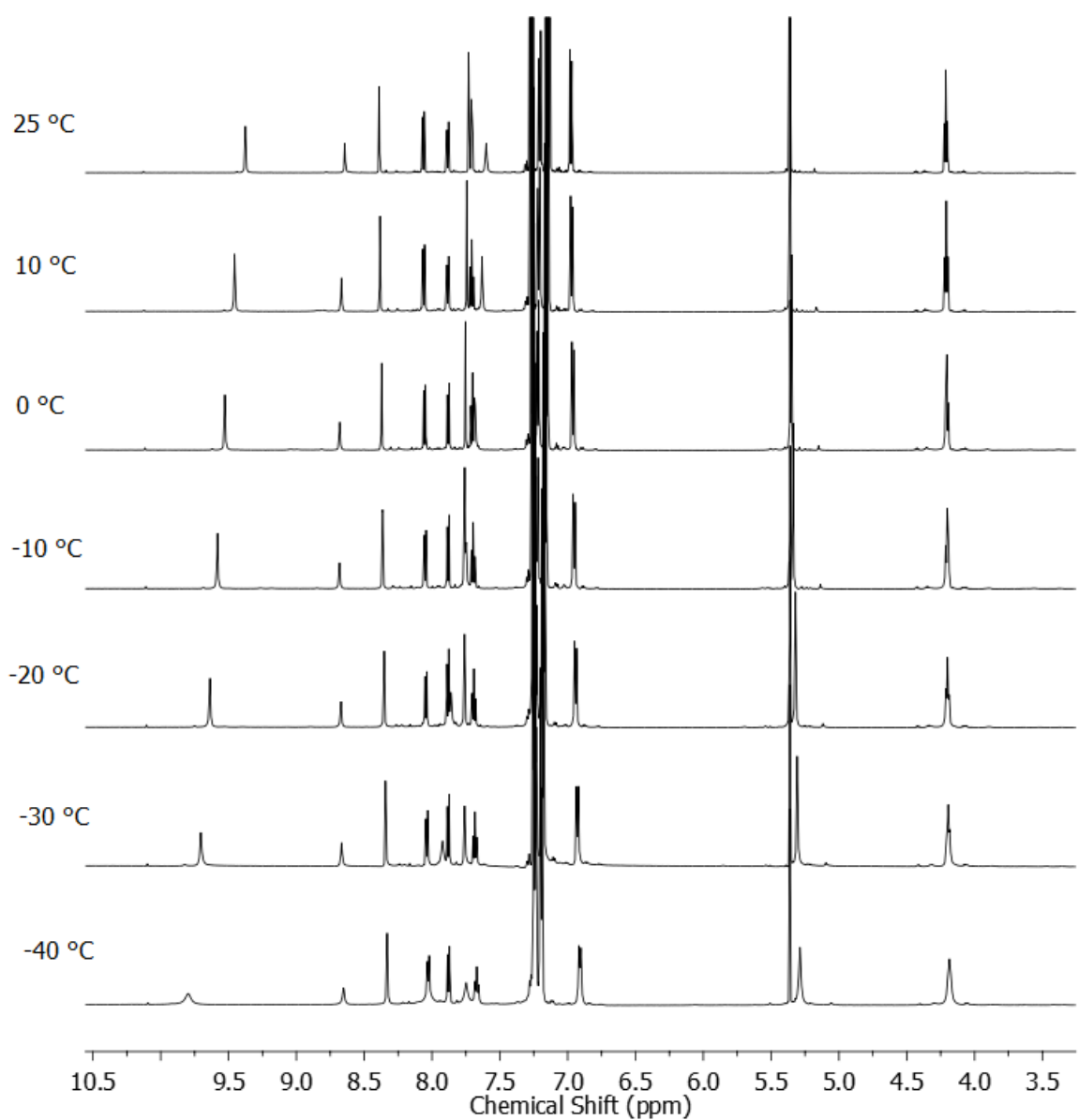

**Figure S14**. ¹H-NMR of rotaxane **1** at different temperatures (600 MHz in CD$_2$Cl$_2$).

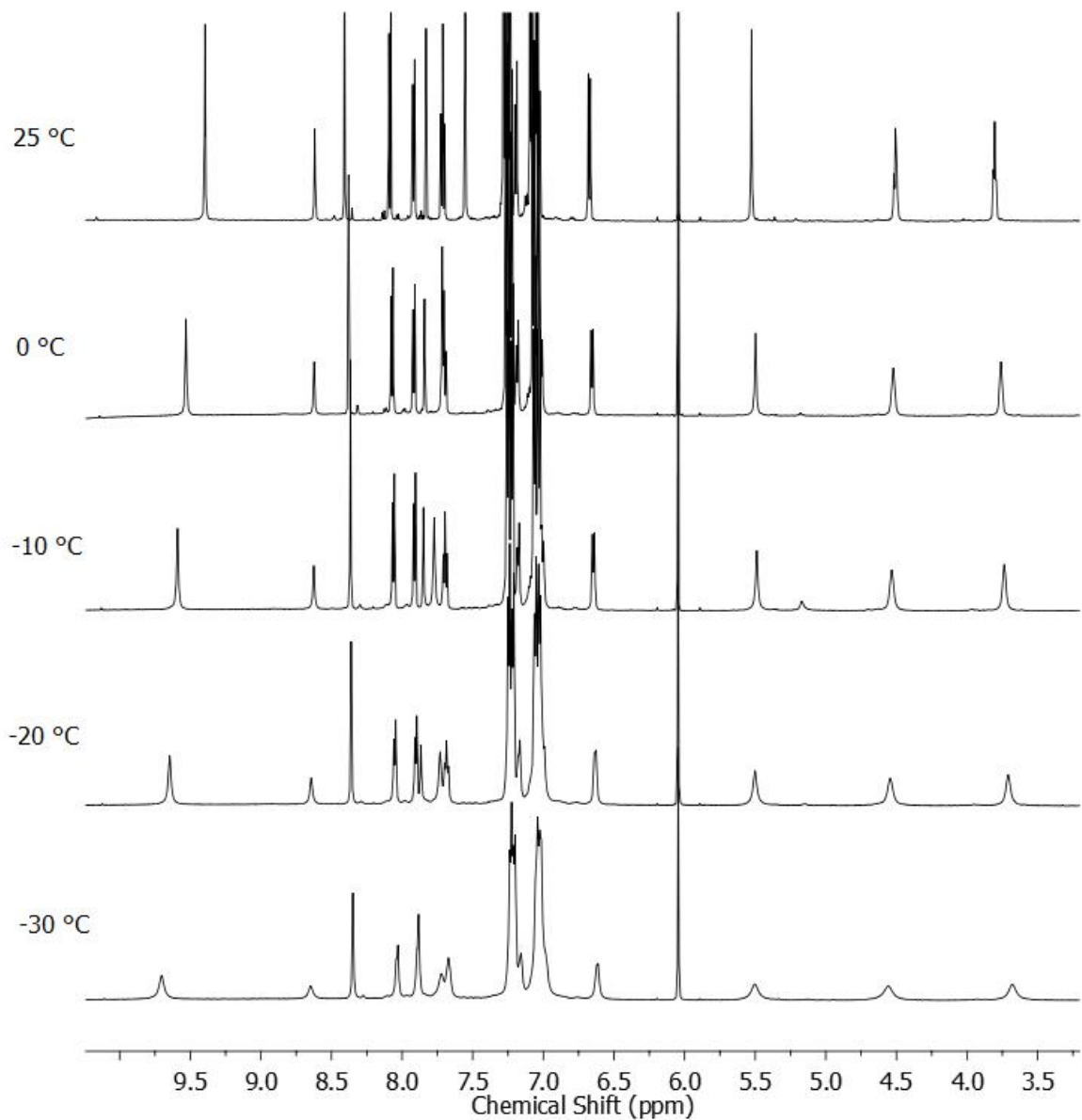

**Figure S15**. $^1$H-NMR of rotaxane **7** at different temperatures (600 MHz in TCE-$d_2$).

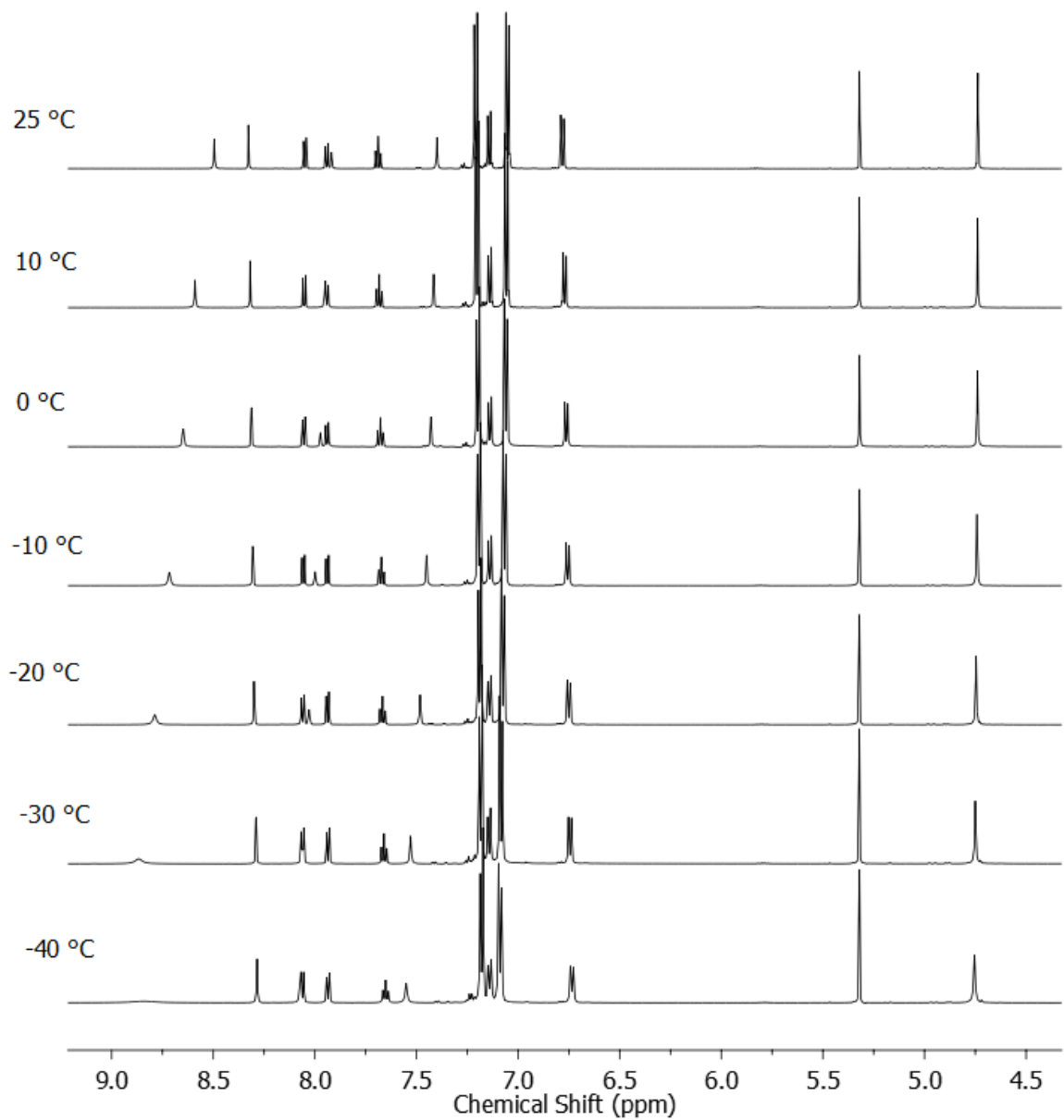

**Figure S16.** $^1$H-NMR of rotaxane **8** at different temperatures (600 MHz in CD$_2$Cl$_2$).

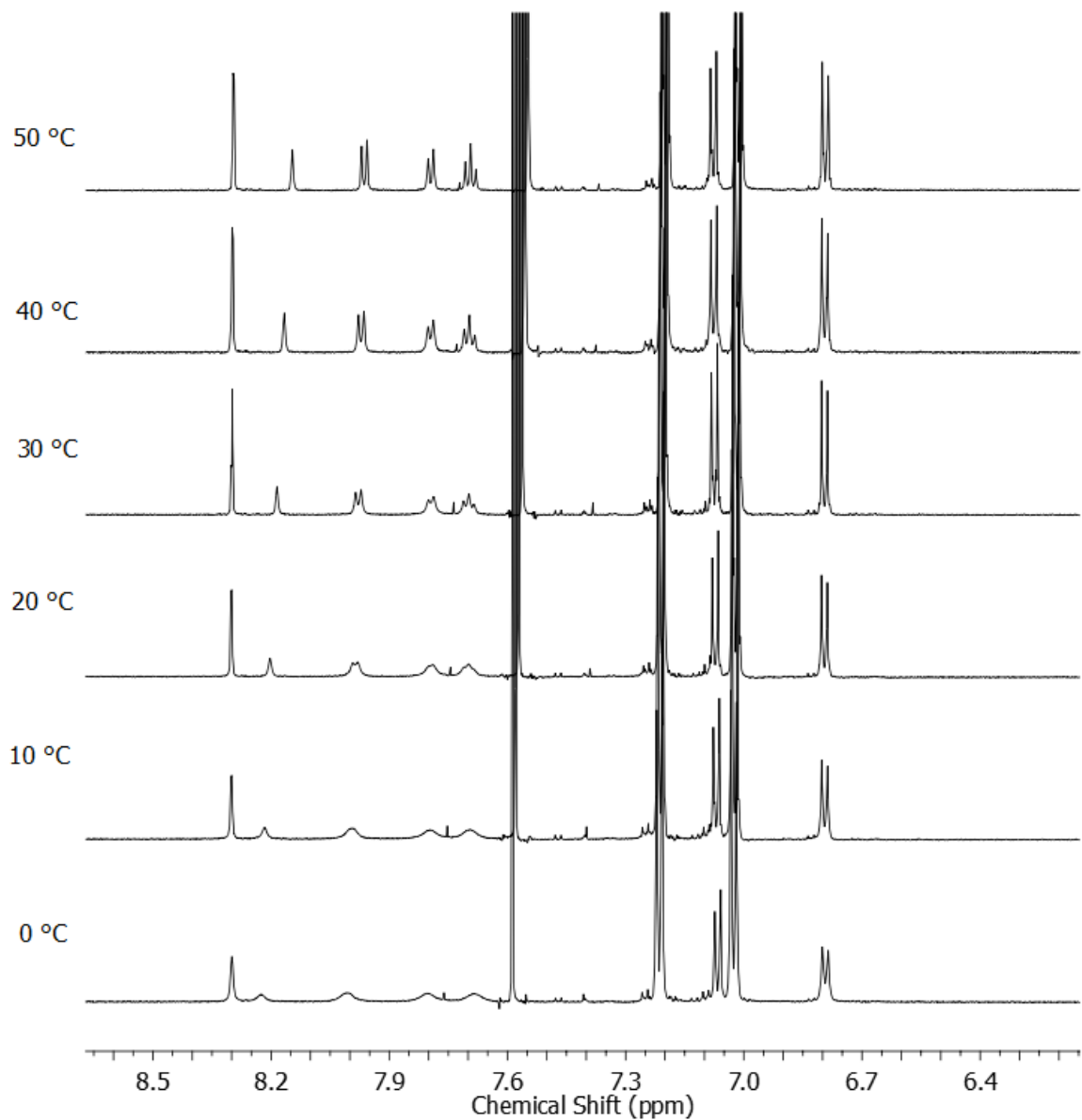

**Figure S17**. $^1$H-NMR of rotaxane **9** at different temperatures (600 MHz in CDCl$_3$/MeOD/D$_2$O 45:45:10).

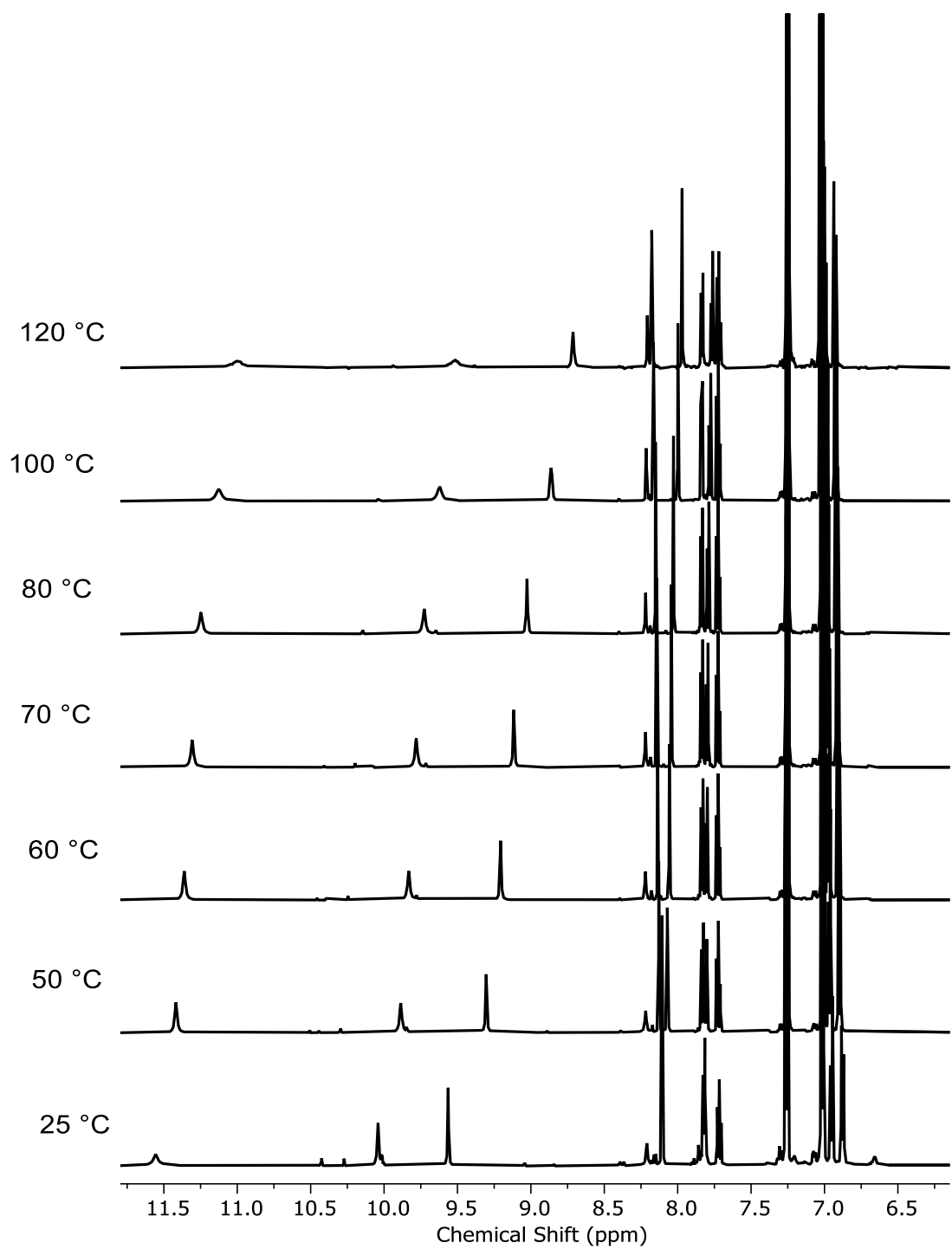

**Figure S18**. $^1$H-NMR of rotaxane **9** at different temperatures (600 MHz in DMSO-$d_6$).

# Molecular Modelling

Structures were minimized with the MMFF force field were carried out with the Spartan '18 software.[S1] The isophthalamide motif in the macrocycle was constrained to be planar.

Molecular modelling allowed estimating the shuttling process in rotaxane **1** using representative conformations of the macrocycle positioned on the barbiturate station, the hydrocarbon chain and the triazole station (Figure S19).

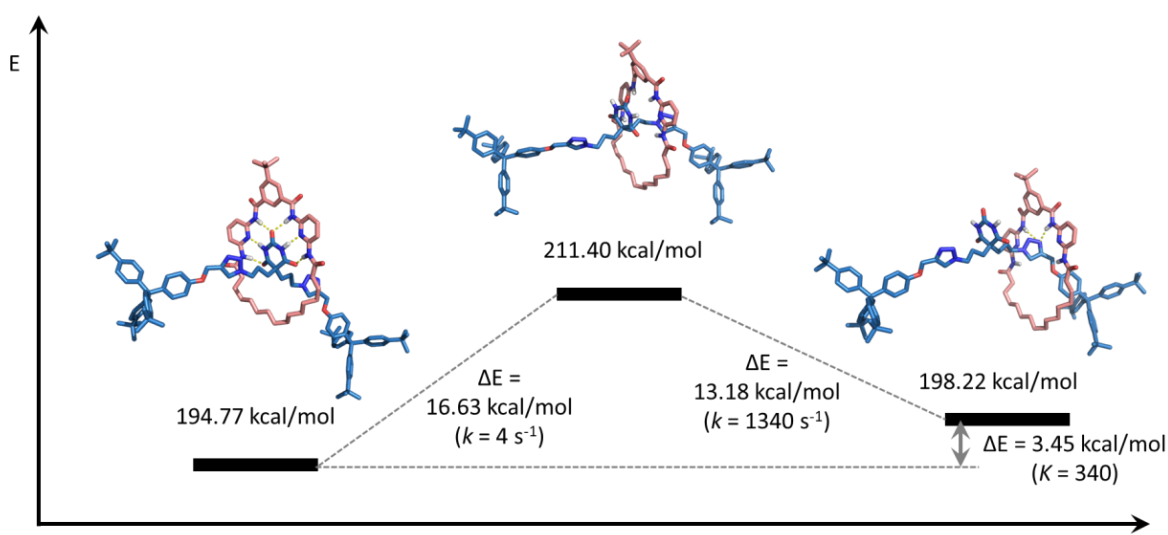

**Figure S19**. Estimated shuttling rate for rotaxane **1** based on molecular modelling energies using representative conformations (MMFF, Spartan '18 software). Energy barriers have been converted to kinetic constants using the Eyring equation $k = k_b T/h \exp(-\Delta E^{\ddagger}/(R T))$ and the equilibrium constant has been obtained using the standard equation $K = \exp(-\Delta E/(R T))$.

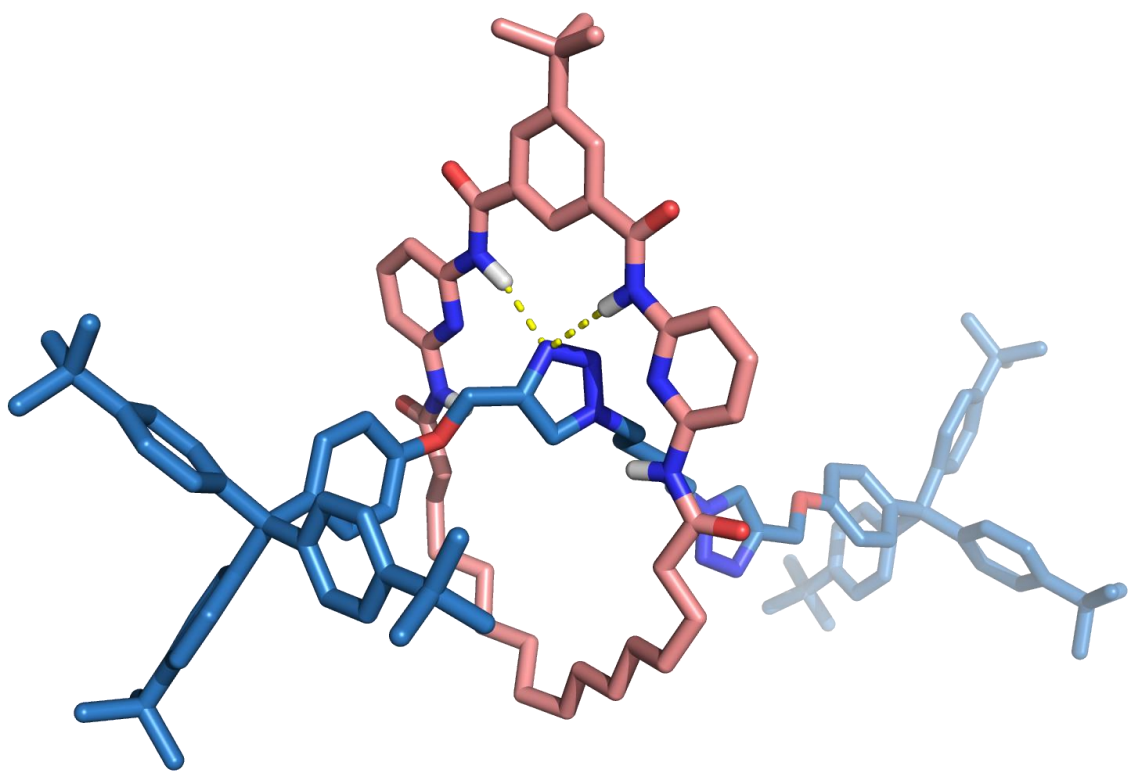

**Figure S20**. Optimized structure of rotaxane **1** (MMFF, Spartan '18 software).

XYZ coordinates of rotaxane **1**

| | | | | | | | | |
|---|---|---|---|---|---|---|---|---|
| C | -23.442950 | 29.329658 | -2.370120 | | H | -22.661258 | 23.094377 | 9.240255 |
| C | -23.076807 | 27.440212 | -0.274551 | | H | -23.781722 | 23.080252 | 10.599499 |
| C | -23.591329 | 27.955870 | -2.584131 | | C | -22.041437 | 21.810921 | 10.847250 |
| C | -23.077493 | 29.725520 | -1.077062 | | H | -20.956682 | 21.907832 | 10.717069 |
| C | -22.909017 | 28.807506 | -0.019883 | | H | -22.228553 | 21.682303 | 11.919482 |
| C | -23.417897 | 27.003626 | -1.560991 | | C | -21.877703 | 17.525108 | 6.810999 |
| H | -23.843048 | 27.590755 | -3.577573 | | H | -20.874692 | 17.083974 | 6.844661 |
| H | -22.925308 | 30.784325 | -0.870858 | | H | -22.594059 | 16.739245 | 7.076722 |
| H | -22.950849 | 26.723928 | 0.517946 | | C | -22.538430 | 20.588848 | 10.071946 |
| C | -23.574762 | 25.570145 | -1.975425 | | H | -22.615369 | 20.867407 | 9.016721 |
| C | -22.613200 | 29.404859 | 1.325206 | | H | -23.546941 | 20.316082 | 10.403987 |
| O | -23.915382 | 25.357446 | -3.137238 | | C | -21.962113 | 18.681105 | 7.805656 |
| O | -22.488837 | 30.626800 | 1.374591 | | H | -22.902706 | 19.219441 | 7.637400 |
| N | -22.497118 | 28.565959 | 2.413659 | | H | -21.144515 | 19.382943 | 7.597137 |
| H | -22.677771 | 27.572636 | 2.342908 | | C | -21.591704 | 19.394518 | 10.223048 |
| N | -23.363436 | 24.586030 | -1.036107 | | H | -21.624061 | 19.029213 | 11.256373 |
| H | -23.114916 | 24.802590 | -0.080462 | | C | -21.899550 | 18.233438 | 9.266907 |
| C | -23.421784 | 23.200928 | -1.257242 | | H | -22.851976 | 17.767340 | 9.545806 |
| C | -23.459987 | 20.465235 | -1.461729 | | H | -20.560579 | 19.721872 | 10.038562 |
| C | -23.792493 | 22.616475 | -2.461500 | | H | -21.121681 | 17.470426 | 9.390522 |
| N | -23.055028 | 22.464281 | -0.185681 | | C | -22.373226 | 23.608254 | 3.745813 |
| C | -23.073281 | 21.117650 | -0.302782 | | H | -21.986238 | 22.828631 | 4.385078 |
| C | -23.822082 | 21.230771 | -2.559610 | | C | -21.783285 | 24.701148 | 3.147076 |
| H | -24.082595 | 23.191772 | -3.331294 | | N | -22.734519 | 25.384464 | 2.435589 |
| H | -24.123698 | 20.754079 | -3.488226 | | N | -23.897403 | 24.771259 | 2.595355 |
| H | -23.473919 | 19.386260 | -1.550248 | | N | -23.686295 | 23.714934 | 3.401061 |
| C | -22.284215 | 28.959842 | 3.747392 | | C | -20.372068 | 25.168431 | 3.171654 |
| C | -21.990579 | 29.520584 | 6.416305 | | H | -20.350230 | 26.254516 | 3.025522 |
| N | -22.284344 | 27.935906 | 4.631044 | | H | -19.850412 | 24.683147 | 2.335687 |
| C | -22.100443 | 30.276855 | 4.147696 | | C | -24.789018 | 22.894355 | 3.858464 |
| C | -21.953795 | 30.559192 | 5.499451 | | H | -24.419306 | 21.865296 | 3.935305 |
| C | -22.157061 | 28.227409 | 5.946775 | | H | -25.566736 | 22.917522 | 3.087112 |
| H | -22.079883 | 31.107261 | 3.454346 | | O | -19.745433 | 24.854942 | 4.422348 |
| H | -21.819295 | 31.583862 | 5.834092 | | C | -18.380074 | 24.803646 | 4.357458 |
| H | -21.870805 | 29.750467 | 7.467587 | | C | -15.566484 | 24.474663 | 4.298910 |
| N | -22.647169 | 20.462830 | 0.852144 | | C | -17.593923 | 25.642035 | 3.570858 |
| H | -22.333237 | 21.110427 | 1.560700 | | C | -17.774916 | 23.828966 | 5.142658 |
| N | -22.188708 | 27.107363 | 6.781798 | | C | -16.390667 | 23.650323 | 5.096481 |
| H | -22.222047 | 26.239427 | 6.264796 | | C | -16.206522 | 25.464166 | 3.525796 |
| C | -22.284171 | 27.093464 | 8.157076 | | H | -18.031390 | 26.433281 | 2.969641 |
| C | -22.652913 | 19.110743 | 1.109391 | | H | -18.381461 | 23.173906 | 5.761576 |
| O | -23.048621 | 18.232069 | 0.352716 | | H | -15.962047 | 22.831434 | 5.671473 |
| O | -22.415083 | 28.075399 | 8.879409 | | H | -15.638808 | 26.096527 | 2.846020 |
| C | -22.085981 | 18.798556 | 2.483365 | | C | -25.322922 | 23.386801 | 5.202877 |
| H | -22.460426 | 19.552910 | 3.184600 | | H | -25.644887 | 24.431601 | 5.108250 |
| H | -20.995311 | 18.888296 | 2.427676 | | H | -24.516015 | 23.376728 | 5.947390 |
| C | -22.197454 | 25.685474 | 8.721227 | | C | -26.487244 | 22.533658 | 5.707633 |
| H | -22.898043 | 25.048571 | 8.169442 | | H | -26.149587 | 21.500968 | 5.857597 |
| H | -21.179357 | 25.322551 | 8.536108 | | H | -27.282584 | 22.509734 | 4.953192 |
| C | -22.511211 | 25.632331 | 10.221387 | | O | -33.912294 | 21.556218 | 10.503280 |
| H | -23.589397 | 25.765905 | 10.372942 | | C | -35.167657 | 21.153593 | 10.886312 |
| H | -22.020207 | 26.466506 | 10.737505 | | C | -37.843002 | 20.553761 | 11.670395 |
| C | -22.480572 | 17.399416 | 2.957552 | | C | -35.553550 | 19.839549 | 11.135079 |
| H | -23.575325 | 17.335148 | 3.007403 | | C | -36.110378 | 22.169883 | 11.048784 |
| H | -22.163757 | 16.651824 | 2.219810 | | C | -37.423818 | 21.879313 | 11.452954 |
| C | -21.894515 | 17.012911 | 4.318958 | | C | -36.860055 | 19.550114 | 11.539168 |
| H | -20.809822 | 16.880732 | 4.224801 | | H | -34.868620 | 19.003275 | 11.044378 |
| H | -22.309123 | 16.040049 | 4.608906 | | H | -35.819920 | 23.204607 | 10.882784 |
| C | -22.043926 | 24.337833 | 10.892042 | | H | -38.107452 | 22.710407 | 11.613962 |
| H | -20.955697 | 24.259143 | 10.777748 | | H | -37.093501 | 18.511322 | 11.767457 |
| H | -22.242399 | 24.402503 | 11.968828 | | C | -39.285781 | 20.192302 | 12.212720 |
| C | -22.183443 | 18.045613 | 5.405127 | | C | -40.457129 | 21.090799 | 11.652526 |
| H | -21.581163 | 18.942883 | 5.216583 | | C | -42.720807 | 22.516684 | 10.597696 |
| H | -23.237129 | 18.348402 | 5.359844 | | C | -41.678024 | 21.179922 | 12.352595 |
| C | -22.717976 | 23.085536 | 10.333723 | | C | -40.419324 | 21.685855 | 10.382542 |
| C | -41.520399 | 22.392520 | 9.875342 | | C | -43.915120 | 23.281708 | 10.003170 |
| C | -42.775312 | 21.887933 | 11.846539 | | C | -45.134660 | 23.370223 | 10.949768 |
| H | -41.801425 | 20.675892 | 13.310198 | | H | -44.879428 | 23.873243 | 11.889838 |
| H | -39.543721 | 21.588118 | 9.745222 | | H | -45.532978 | 22.377642 | 11.190809 |
| H | -41.428188 | 22.830015 | 8.883267 | | H | -45.951798 | 23.938970 | 10.489871 |
| H | -43.674033 | 21.909871 | 12.456501 | | C | -44.386838 | 22.576272 | 8.713964 |
| C | -39.732410 | 18.746804 | 11.745248 | | H | -43.608597 | 22.560378 | 7.943095 |
| C | -40.615570 | 16.190595 | 10.773887 | | H | -45.258000 | 23.082839 | 8.282276 |
| C | -39.501890 | 18.330960 | 10.417484 | | H | -44.671828 | 21.536541 | 8.914762 |
| C | -40.484253 | 17.880003 | 12.551999 | | C | -43.496082 | 24.729239 | 9.668389 |
| C | -40.898297 | 16.624595 | 12.081754 | | H | -42.708978 | 24.768184 | 8.907695 |
| C | -39.917783 | 17.079365 | 9.948824 | | H | -43.120108 | 25.246448 | 10.559118 |
| H | -38.990503 | 18.988284 | 9.715928 | | H | -44.343757 | 25.306309 | 9.280571 |
| H | -40.777196 | 18.166524 | 13.559432 | | C | -38.390710 | 20.885308 | 18.098679 |
| H | -41.464013 | 15.990344 | 12.761387 | | C | -39.000317 | 22.177243 | 18.690525 |
| H | -39.684406 | 16.835140 | 8.916406 | | H | -38.841488 | 22.229110 | 19.774421 |
| C | -39.105224 | 20.352895 | 13.774884 | | H | -40.081627 | 22.226473 | 18.516218 |
| C | -38.640024 | 20.719905 | 16.589909 | | H | -38.544031 | 23.076422 | 18.261733 |
| C | -38.535510 | 19.340010 | 14.564735 | | C | -39.006289 | 19.704125 | 18.882213 |
| C | -39.355877 | 21.580349 | 14.417033 | | H | -38.888569 | 19.842970 | 19.963488 |
| C | -39.146194 | 21.753543 | 15.792303 | | H | -38.533435 | 18.747689 | 18.635378 |
| C | -38.318752 | 19.516751 | 15.938934 | | H | -40.078273 | 19.608201 | 18.673127 |
| H | -38.232351 | 18.394435 | 14.120077 | | C | -36.870459 | 20.925292 | 18.354417 |
| H | -39.713036 | 22.439715 | 13.852056 | | H | -36.652056 | 21.058654 | 19.420351 |
| H | -39.378648 | 22.729665 | 16.209019 | | H | -36.399841 | 21.753942 | 17.811848 |
| H | -37.872135 | 18.692379 | 16.490624 | | H | -36.376706 | 20.001128 | 18.033529 |
| C | -14.024497 | 24.148600 | 4.136741 | | C | -14.513236 | 19.674434 | 0.175593 |
| C | -14.063943 | 22.960844 | 3.088792 | | C | -14.933265 | 20.124169 | -1.242336 |
| C | -14.372403 | 20.834626 | 1.175634 | | H | -14.181147 | 20.774985 | -1.702727 |
| C | -14.273090 | 23.216012 | 1.719239 | | H | -15.887597 | 20.663674 | -1.224929 |
| C | -14.067677 | 21.611781 | 3.480958 | | H | -15.057756 | 19.262812 | -1.909611 |
| C | -14.193013 | 20.576438 | 2.544662 | | C | -15.581437 | 18.666121 | 0.656984 |
| C | -14.409147 | 22.179354 | 0.784931 | | H | -15.298738 | 18.172591 | 1.592869 |
| H | -14.349744 | 24.239424 | 1.355363 | | H | -15.735721 | 17.870885 | -0.081960 |
| H | -13.991554 | 21.332449 | 4.529470 | | H | -16.545944 | 19.161007 | 0.820910 |
| H | -14.157614 | 19.554370 | 2.915389 | | C | -13.155072 | 18.953714 | 0.056092 |
| H | -14.555421 | 22.464403 | -0.253307 | | H | -12.827521 | 18.536037 | 1.014909 |
| C | -13.328917 | 23.763222 | 5.500081 | | H | -12.371954 | 19.640550 | -0.286702 |
| C | -11.982499 | 23.142582 | 7.961612 | | H | -13.208306 | 18.125136 | -0.659553 |
| C | -12.141375 | 23.010389 | 5.520862 | | C | -11.264681 | 22.753851 | 9.264744 |
| C | -13.774425 | 24.267037 | 6.734918 | | C | -9.776655 | 23.161494 | 9.196867 |
| C | -13.126712 | 23.950398 | 7.937710 | | H | -9.261593 | 22.933867 | 10.137573 |
| C | -11.498988 | 22.689077 | 6.723074 | | H | -9.670466 | 24.236518 | 9.009587 |
| H | -11.693718 | 22.659031 | 4.592878 | | H | -9.235809 | 22.633410 | 8.404240 |
| H | -14.635514 | 24.930240 | 6.788430 | | C | -11.364436 | 21.225835 | 9.458546 |
| H | -13.546479 | 24.365538 | 8.849948 | | H | -10.883871 | 20.913120 | 10.392827 |
| H | -10.601871 | 22.075990 | 6.669951 | | H | -10.879407 | 20.674702 | 8.645112 |
| C | -13.169149 | 25.381720 | 3.649119 | | H | -12.410980 | 20.900344 | 9.496735 |
| C | -11.529501 | 27.622365 | 2.902065 | | C | -11.858779 | 23.419109 | 10.526981 |
| C | -13.355541 | 26.658972 | 4.211460 | | H | -11.307709 | 23.121292 | 11.427129 |
| C | -12.095106 | 25.241403 | 2.754696 | | H | -12.904835 | 23.130768 | 10.682909 |
| C | -11.310793 | 26.338685 | 2.374718 | | H | -11.811338 | 24.512650 | 10.465101 |
| C | -12.564662 | 27.754803 | 3.836505 | | C | -10.647849 | 28.802576 | 2.460031 |
| H | -14.121651 | 26.821023 | 4.967841 | | C | -9.160054 | 28.480865 | 2.722638 |
| H | -11.834284 | 24.269663 | 2.340840 | | H | -8.519202 | 29.329392 | 2.455424 |
| H | -10.509612 | 26.164716 | 1.659303 | | H | -8.810168 | 27.623067 | 2.138205 |
| H | -12.781908 | 28.708157 | 4.310426 | | H | -8.986768 | 28.251930 | 3.780779 |
| C | -41.083025 | 14.807768 | 10.290515 | | C | -10.853379 | 29.054174 | 0.951080 |
| C | -42.620886 | 14.717336 | 10.383235 | | H | -10.257215 | 29.907619 | 0.607522 |
| H | -42.979695 | 14.810218 | 11.414031 | | H | -11.904795 | 29.269927 | 0.726570 |
| H | -43.098487 | 15.510321 | 9.795468 | | H | -10.559001 | 28.190288 | 0.344784 |
| C | -40.452426 | 13.710591 | 11.174252 | | C | -10.961185 | 30.125574 | 3.194849 |
| H | -40.764832 | 13.791649 | 12.221055 | | H | -10.302954 | 30.933172 | 2.852395 |
| H | -39.357522 | 13.767414 | 11.152114 | | H | -10.817169 | 30.028335 | 4.277300 |
| H | -40.741095 | 12.710836 | 10.829183 | | H | -11.991361 | 30.454302 | 3.015257 |

```
C   -31.125007   22.087995    9.390707        H   -42.983897   13.755484   10.002633
H   -31.654503   22.841642    8.827237        C   -40.694413   14.493197    8.827174
C   -31.551742   21.030601   10.166785        H   -39.607258   14.508185    8.685951
N   -30.451824   20.366010   10.653919        H   -41.139466   15.210880    8.127980
N   -29.364715   20.979934   10.213023        H   -41.043776   13.496994    8.530415
N   -29.766923   22.011288    9.453010        C   -23.642134   30.359141   -3.492629
C   -32.915122   20.524476   10.478472        C   -24.436785   31.585850   -2.990074
H   -32.871305   20.034865   11.459974        H   -23.907637   32.138349   -2.206786
H   -33.176064   19.785708    9.710185        H   -25.409928   31.284586   -2.584587
C   -27.042116   23.094463    7.019889        H   -24.617927   32.298462   -3.802821
H   -26.240696   23.134663    7.768286        C   -24.426034   29.789328   -4.697697
H   -27.385703   24.123660    6.860243        H   -23.882092   28.978182   -5.195084
C   -28.195702   22.248752    7.560022        H   -24.601769   30.557665   -5.458621
H   -27.835627   21.232952    7.766518        H   -25.403397   29.401405   -4.385554
H   -28.981684   22.154594    6.800371        C   -22.260098   30.821332   -3.991538
C   -28.784306   22.868566    8.828303        H   -21.673760   29.978288   -4.376324
H   -28.002759   23.058333    9.573560        H   -21.674790   31.289176   -3.191533
H   -29.269334   23.824954    8.603115        H   -22.357924   31.554668   -4.800260
```

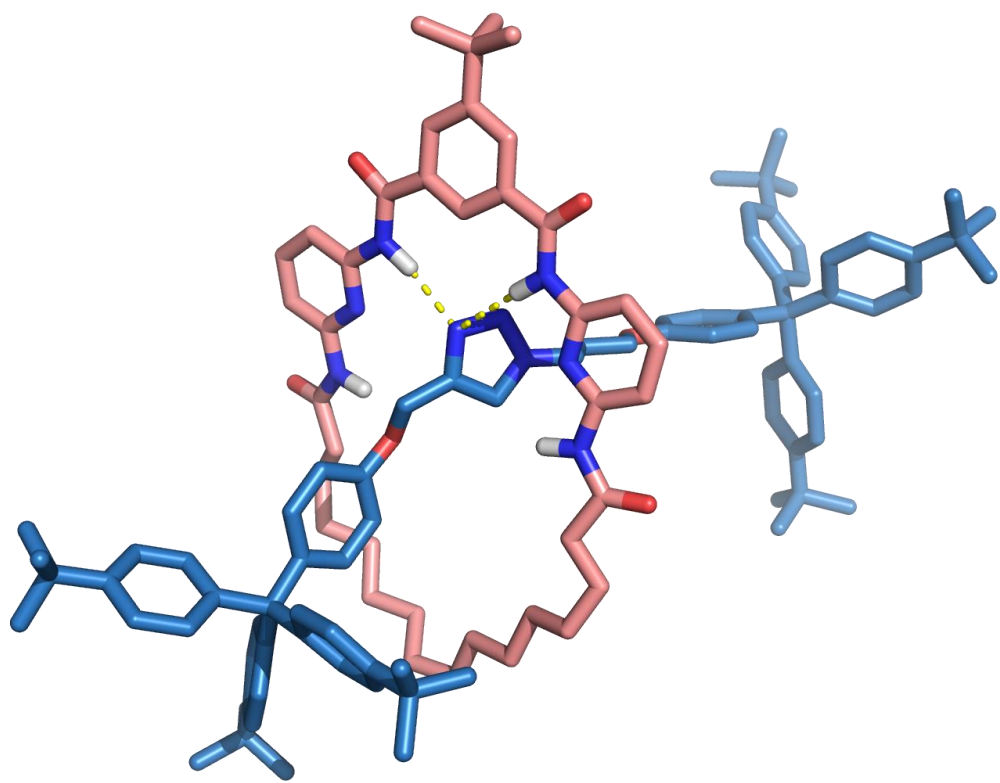

**Figure S21**. Optimized structure of rotaxane **7** (MMFF, Spartan '18 software).

XYZ coordinates of rotaxane **7**

```
C   -27.968866   29.037021   -0.716749        C   -27.117145   29.531556    0.285572
C   -26.188884   27.341770    0.708190        C   -26.229715   28.707349    1.008004
C   -27.908168   27.660713   -0.970508        C   -27.031740   26.801739   -0.273395
```

| | | | | | | | | |
|---|---|---|---|---|---|---|---|---|
| H | -28.550660 | 27.223106 | -1.732492 | | H | -18.426787 | 18.037651 | 4.897798 |
| H | -27.150020 | 30.594976 | 0.524934 | | H | -19.646309 | 17.284470 | 5.934975 |
| H | -25.490219 | 26.709380 | 1.224099 | | C | -18.991476 | 21.086365 | 8.906210 |
| C | -27.064444 | 25.348469 | -0.664181 | | H | -19.670408 | 21.291904 | 8.073998 |
| C | -25.439906 | 29.368109 | 2.104166 | | H | -19.576610 | 20.568238 | 9.674641 |
| O | -27.862971 | 25.019711 | -1.538917 | | C | -19.188963 | 19.343051 | 6.439873 |
| O | -25.571092 | 30.585684 | 2.215283 | | H | -20.176770 | 19.631986 | 6.817980 |
| N | -24.610990 | 28.594795 | 2.898278 | | H | -18.771411 | 20.214512 | 5.921363 |
| H | -24.594276 | 27.585050 | 2.820004 | | C | -17.839819 | 20.194239 | 8.427947 |
| N | -26.211591 | 24.473368 | -0.022469 | | H | -17.253088 | 19.860930 | 9.292286 |
| H | -25.612411 | 24.781465 | 0.734200 | | C | -18.294640 | 18.967828 | 7.622744 |
| C | -26.053522 | 23.101842 | -0.287261 | | H | -18.831760 | 18.273699 | 8.279916 |
| C | -25.565394 | 20.422993 | -0.650617 | | H | -17.158861 | 20.779306 | 7.797449 |
| C | -26.806342 | 22.397506 | -1.215561 | | H | -17.402475 | 18.441866 | 7.262194 |
| N | -25.080223 | 22.506837 | 0.443102 | | C | -23.929126 | 23.465954 | 3.838042 |
| C | -24.839504 | 21.189644 | 0.245031 | | H | -23.550621 | 22.578462 | 4.322663 |
| C | -26.563652 | 21.039831 | -1.391387 | | C | -23.320345 | 24.617653 | 3.384340 |
| H | -27.586686 | 22.856767 | -1.808848 | | N | -24.272038 | 25.440631 | 2.843261 |
| H | -27.144651 | 20.469595 | -2.111262 | | N | -25.450578 | 24.847603 | 2.960973 |
| H | -25.371491 | 19.369878 | -0.812474 | | N | -25.246322 | 23.656562 | 3.550759 |
| C | -23.853156 | 29.045566 | 4.000510 | | C | -21.887573 | 25.011256 | 3.391701 |
| C | -22.426155 | 29.722019 | 6.247666 | | H | -21.799716 | 26.081042 | 3.170956 |
| N | -23.203502 | 28.068372 | 4.682111 | | H | -21.383146 | 24.442076 | 2.599166 |
| C | -23.782226 | 30.377828 | 4.387521 | | C | -26.351248 | 22.750627 | 3.777856 |
| C | -23.059469 | 30.718722 | 5.522864 | | H | -25.954790 | 21.730224 | 3.724140 |
| C | -22.520306 | 28.411688 | 5.803898 | | H | -27.058923 | 22.877707 | 2.950596 |
| H | -24.275242 | 31.175954 | 3.848679 | | O | -21.293860 | 24.750627 | 4.672387 |
| H | -22.998320 | 31.754852 | 5.843387 | | C | -19.937117 | 24.583951 | 4.624944 |
| H | -21.863299 | 30.002795 | 7.129010 | | C | -17.146209 | 24.101269 | 4.686657 |
| N | -23.779541 | 20.694727 | 1.009540 | | C | -19.081627 | 25.335637 | 3.825149 |
| H | -23.312303 | 21.419870 | 1.535628 | | C | -19.415100 | 23.605211 | 5.463000 |
| N | -21.893265 | 27.344164 | 6.460216 | | C | -18.041142 | 23.356944 | 5.483305 |
| H | -21.975397 | 26.463673 | 5.966363 | | C | -17.706536 | 25.076259 | 3.835803 |
| C | -21.293690 | 27.363227 | 7.704424 | | H | -19.453957 | 26.124164 | 3.177883 |
| C | -23.347538 | 19.389935 | 1.114500 | | H | -20.077681 | 23.010193 | 6.085271 |
| O | -23.865435 | 18.415609 | 0.580953 | | H | -17.677536 | 22.548864 | 6.113886 |
| O | -21.279211 | 28.311272 | 8.482513 | | H | -17.079185 | 25.641301 | 3.148675 |
| C | -22.105375 | 19.268574 | 1.984052 | | C | -27.022525 | 23.003725 | 5.119411 |
| H | -22.232391 | 19.923675 | 2.853141 | | H | -27.367538 | 24.043679 | 5.187892 |
| H | -21.249239 | 19.627161 | 1.401356 | | H | -26.300487 | 22.882786 | 5.937524 |
| C | -20.597568 | 26.051428 | 8.029135 | | C | -28.202193 | 22.063395 | 5.336740 |
| H | -21.293413 | 25.229378 | 7.829900 | | H | -27.855054 | 21.023951 | 5.297549 |
| H | -19.746559 | 25.966359 | 7.344504 | | H | -28.956835 | 22.237332 | 4.560122 |
| C | -20.098209 | 25.979701 | 9.477736 | | O | -28.743778 | 22.347174 | 6.625184 |
| H | -20.950461 | 25.841608 | 10.153994 | | C | -29.856682 | 21.649019 | 7.001622 |
| H | -19.630574 | 26.931700 | 9.757798 | | C | -32.200458 | 20.388219 | 8.008065 |
| C | -21.865117 | 17.828373 | 2.443207 | | C | -30.469404 | 20.627868 | 6.281983 |
| H | -22.725391 | 17.496908 | 3.039062 | | C | -30.383285 | 22.012247 | 8.239387 |
| H | -21.821005 | 17.160821 | 1.573607 | | C | -31.525475 | 21.382301 | 8.744284 |
| C | -20.585094 | 17.632788 | 3.263461 | | C | -31.613965 | 19.995415 | 6.787575 |
| H | -19.713369 | 17.751801 | 2.610289 | | H | -30.080784 | 20.283199 | 5.329926 |
| H | -20.565805 | 16.599809 | 3.631351 | | H | -29.896561 | 22.787730 | 8.826048 |
| C | -19.061182 | 24.874327 | 9.703157 | | H | -31.876057 | 21.676918 | 9.731539 |
| H | -18.213519 | 25.050480 | 9.029405 | | H | -32.034309 | 19.174201 | 6.209947 |
| H | -18.670205 | 24.954418 | 10.724786 | | C | -33.454864 | 19.634392 | 8.611580 |
| C | -20.471657 | 18.604031 | 4.434892 | | C | -34.423235 | 20.582355 | 9.425645 |
| H | -20.266359 | 19.610129 | 4.048821 | | C | -36.311007 | 22.249724 | 10.812341 |
| H | -21.428752 | 18.651409 | 4.969380 | | C | -35.273970 | 20.070345 | 10.423367 |
| C | -19.613453 | 23.463883 | 9.492315 | | C | -34.596691 | 21.935653 | 9.088754 |
| H | -20.171712 | 23.425182 | 8.552306 | | C | -35.504208 | 22.751353 | 9.777378 |
| H | -20.319807 | 23.212380 | 10.292007 | | C | -36.183171 | 20.887216 | 11.110893 |
| C | -18.486581 | 22.427172 | 9.442411 | | H | -35.251427 | 19.012001 | 10.677569 |
| H | -17.701076 | 22.799258 | 8.778775 | | H | -34.037522 | 22.383281 | 8.269864 |
| H | -18.037955 | 22.297965 | 10.433864 | | H | -35.578473 | 23.794095 | 9.475373 |
| C | -19.371129 | 18.215188 | 5.425515 | | H | -36.793384 | 20.414555 | 11.875536 |
| C | -34.418734 | 19.046786 | 7.504601 | | H | -37.848700 | 24.348720 | 9.760872 |
| C | -36.289039 | 18.078934 | 5.547359 | | H | -39.059060 | 24.378037 | 11.041033 |
| C | -34.713778 | 19.773735 | 6.335847 | | H | -38.868551 | 22.932105 | 10.036647 |
| C | -35.143807 | 17.859566 | 7.703466 | | C | -36.526072 | 24.345417 | 12.194411 |
| C | -36.041622 | 17.379868 | 6.740663 | | H | -36.016453 | 24.969764 | 11.452396 |
| C | -35.613592 | 19.293585 | 5.373514 | | H | -35.765288 | 23.976872 | 12.892767 |
| H | -34.252849 | 20.743867 | 6.158477 | | H | -37.202577 | 25.001922 | 12.754128 |
| H | -35.036547 | 17.287915 | 8.623034 | | C | -30.673613 | 15.472418 | 11.904937 |
| H | -36.560398 | 16.447008 | 6.951847 | | C | -30.499073 | 15.833286 | 13.397588 |
| H | -35.775233 | 19.911600 | 4.494632 | | H | -29.994018 | 15.026730 | 13.942646 |
| C | -32.773646 | 18.521884 | 9.504778 | | H | -31.465583 | 15.998685 | 13.887839 |
| C | -31.394718 | 16.554659 | 11.084307 | | H | -29.891435 | 16.736517 | 13.526717 |
| C | -32.311353 | 17.317582 | 8.944130 | | C | -31.475418 | 14.153712 | 11.847658 |
| C | -32.457562 | 18.745773 | 10.856750 | | H | -31.000859 | 13.375693 | 12.457123 |
| C | -31.799976 | 17.778897 | 11.631500 | | H | -31.552175 | 13.758306 | 10.828933 |
| C | -31.654323 | 16.352756 | 9.718585 | | H | -32.495652 | 14.295886 | 12.223453 |
| H | -32.447971 | 17.111845 | 7.884199 | | C | -29.264037 | 15.235350 | 11.322207 |
| H | -32.706889 | 19.690376 | 11.336591 | | H | -28.714380 | 14.489504 | 11.908243 |
| H | -31.607207 | 18.026337 | 12.671847 | | H | -28.675692 | 16.160632 | 11.323796 |
| H | -31.335746 | 15.436974 | 9.225044 | | H | -29.299085 | 14.870195 | 10.289874 |
| C | -15.612133 | 23.711576 | 4.612299 | | C | -16.007092 | 19.326299 | 0.547516 |
| C | -15.636959 | 22.547132 | 3.540645 | | C | -15.954502 | 19.799997 | -0.922454 |
| C | -15.889976 | 20.466509 | 1.572806 | | H | -15.007676 | 20.302914 | -1.151255 |
| C | -15.616568 | 22.824277 | 2.161224 | | H | -16.773276 | 20.491773 | -1.152434 |
| C | -15.856574 | 21.207120 | 3.907853 | | H | -16.043756 | 18.953432 | -1.613849 |
| C | -15.961206 | 20.191756 | 2.948629 | | C | -17.347917 | 18.585225 | 0.739497 |
| C | -15.723547 | 21.807179 | 1.201760 | | H | -17.415345 | 18.102254 | 1.720178 |
| H | -15.529695 | 23.849087 | 1.804217 | | H | -17.476565 | 17.798031 | -0.012582 |
| H | -15.965713 | 20.925824 | 4.953455 | | H | -18.196920 | 19.273527 | 0.651118 |
| H | -16.110766 | 19.173605 | 3.302264 | | C | -14.844465 | 18.331426 | 0.750978 |
| H | -15.682857 | 22.105678 | 0.157689 | | H | -14.867014 | 17.866330 | 1.742754 |
| C | -15.018645 | 23.269981 | 6.005978 | | H | -13.874298 | 18.831154 | 0.644403 |
| C | -13.865435 | 22.554552 | 8.538241 | | H | -14.884757 | 17.520353 | 0.014532 |
| C | -13.941341 | 22.371787 | 6.096444 | | C | -13.256539 | 22.116954 | 9.880292 |
| C | -15.437717 | 23.869866 | 7.206421 | | C | -11.758991 | 22.487717 | 9.914723 |
| C | -14.883167 | 23.511374 | 8.443364 | | H | -11.305773 | 22.213165 | 10.874421 |
| C | -13.393320 | 22.008581 | 7.333262 | | H | -11.616120 | 23.565774 | 9.774473 |
| H | -13.501957 | 21.939487 | 5.199490 | | H | -11.188169 | 21.976285 | 9.131839 |
| H | -16.207240 | 24.639355 | 7.202898 | | C | -13.407167 | 20.588539 | 10.045224 |
| H | -15.276202 | 24.010254 | 9.325157 | | H | -13.016326 | 20.254955 | 11.013668 |
| H | -12.574705 | 21.291786 | 7.334823 | | H | -12.864226 | 20.031998 | 9.273536 |
| C | -14.675107 | 24.916303 | 4.212234 | | H | -14.460209 | 20.287743 | 9.988725 |
| C | -12.890102 | 27.093418 | 3.629970 | | C | -13.923221 | 22.772515 | 11.111007 |
| C | -14.905708 | 26.216554 | 4.699186 | | H | -13.468755 | 22.415897 | 12.043176 |
| C | -13.485474 | 24.719630 | 3.490446 | | H | -14.993067 | 22.538045 | 11.162282 |
| C | -12.625785 | 25.784971 | 3.191593 | | H | -13.812658 | 23.863051 | 11.097768 |
| C | -14.043828 | 27.281694 | 4.401287 | | C | -11.922414 | 28.237694 | 3.284539 |
| H | -15.762019 | 26.424861 | 5.337962 | | C | -10.530493 | 27.937958 | 3.881088 |
| H | -13.194547 | 23.725699 | 3.156327 | | H | -9.829016 | 28.755583 | 3.677886 |
| H | -11.728310 | 25.566425 | 2.616410 | | H | -10.088971 | 27.025989 | 3.464488 |
| H | -14.301664 | 28.256196 | 4.807107 | | H | -10.586726 | 27.809042 | 4.968520 |
| C | -37.282789 | 17.521885 | 4.514492 | | C | -11.800289 | 28.369463 | 1.750867 |
| C | -38.684563 | 17.411116 | 5.151521 | | H | -11.142590 | 29.202687 | 1.476820 |
| H | -38.700352 | 16.717777 | 5.999615 | | H | -12.779088 | 28.552080 | 1.291605 |
| H | -39.029998 | 18.385499 | 5.516968 | | H | -11.383340 | 27.467879 | 1.288915 |
| C | -36.822719 | 16.122071 | 4.052252 | | C | -12.366061 | 29.615729 | 3.825895 |
| H | -36.803119 | 15.398960 | 4.874938 | | H | -11.652032 | 30.399224 | 3.544831 |
| H | -35.814311 | 16.160118 | 3.623317 | | H | -12.430396 | 29.618150 | 4.920258 |
| H | -37.495605 | 15.717439 | 3.287111 | | H | -13.343043 | 29.910772 | 3.425529 |
| C | -37.296616 | 23.175556 | 11.544493 | | H | -39.420930 | 17.046881 | 4.425603 |
| C | -38.093576 | 22.475367 | 12.668364 | | C | -37.422189 | 18.396114 | 3.247655 |
| H | -37.429533 | 22.065975 | 13.438594 | | H | -36.464609 | 18.503931 | 2.724929 |
| H | -38.708937 | 21.656696 | 12.277397 | | H | -37.795247 | 19.398604 | 3.487622 |
| H | -38.772723 | 23.177175 | 13.167281 | | H | -38.129891 | 17.952190 | 2.537180 |
| C | -38.322757 | 23.739876 | 10.538559 | | C | -28.915575 | 29.987639 | -1.470378 |

S28

| | | | | | | |
|---|---|---|---|---|---|---|
| C | -29.888475 | 30.650227 | -0.471191 | H | -28.742606 | 31.754310 | -2.757545 |
| H | -29.366525 | 31.263395 | 0.271559 | H | -27.368302 | 30.644079 | -2.877477 |
| H | -30.468159 | 29.895838 | 0.073920 | C | -29.774593 | 29.290752 | -2.549908 |
| H | -30.598100 | 31.307499 | -0.987258 | H | -29.151543 | 28.818740 | -3.318559 |
| C | -28.093255 | 31.083945 | -2.182302 | H | -30.426759 | 30.009499 | -3.060606 |
| H | -27.534765 | 31.708481 | -1.476558 | H | -30.423292 | 28.521329 | -2.115221 |

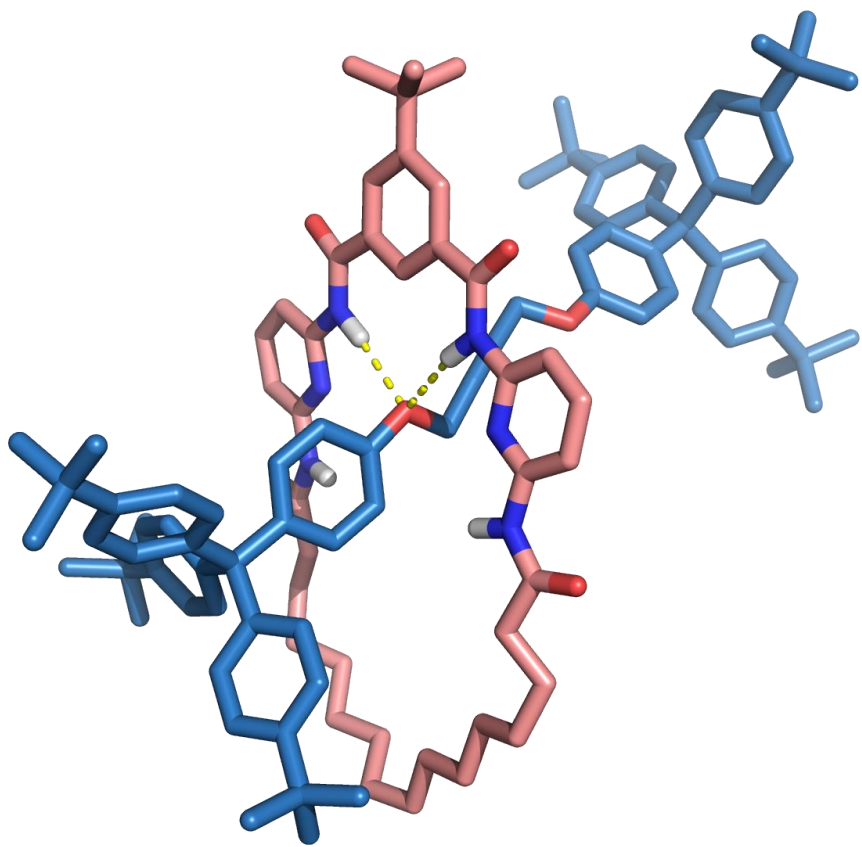

**Figure S22**. Optimized structure of rotaxane **8** (MMFF, Spartan '18 software).

XYZ coordinates of rotaxane **8**

| | | | | | | | |
|---|---|---|---|---|---|---|---|
| C | -29.239414 | 29.079263 | 2.760120 | H | -24.666625 | 27.621511 | 4.348747 |
| C | -26.910685 | 27.516101 | 2.945959 | N | -26.875674 | 25.020011 | 1.321974 |
| C | -29.141491 | 27.879683 | 2.067984 | H | -26.037552 | 25.270143 | 1.825149 |
| C | -28.179009 | 29.491469 | 3.555923 | C | -26.714490 | 23.822254 | 0.599245 |
| C | -27.007622 | 28.718204 | 3.670253 | C | -26.194221 | 21.494286 | -0.758863 |
| C | -27.984733 | 27.080617 | 2.149729 | C | -27.712706 | 23.250761 | -0.179891 |
| H | -29.984520 | 27.570801 | 1.449862 | N | -25.481383 | 23.271144 | 0.702758 |
| H | -28.279465 | 30.428577 | 4.103320 | C | -25.237569 | 22.123047 | 0.022613 |
| H | -26.019498 | 26.914540 | 3.017149 | C | -27.450345 | 22.069528 | -0.860609 |
| C | -27.994715 | 25.828186 | 1.322297 | H | -28.703011 | 23.676009 | -0.276117 |
| C | -25.967560 | 29.241638 | 4.619083 | H | -28.220837 | 21.604162 | -1.468920 |
| O | -29.034871 | 25.576277 | 0.717812 | H | -25.999398 | 20.580214 | -1.305194 |
| O | -26.218455 | 30.303824 | 5.183873 | C | -23.765961 | 28.802601 | 5.699148 |
| N | -24.819644 | 28.503701 | 4.815510 | C | -21.646786 | 29.176275 | 7.388165 |

| | | | | | | | | |
|---|---|---|---|---|---|---|---|---|
| N | -22.782228 | 27.879654 | 5.706128 | | C | -21.531942 | 27.073502 | 1.312682 |
| C | -23.721674 | 29.937594 | 6.503902 | | C | -20.602378 | 25.010533 | 2.116344 |
| C | -22.638487 | 30.161211 | 7.366621 | | H | -21.838467 | 23.831495 | 3.398253 |
| C | -21.746745 | 28.069958 | 6.552595 | | H | -23.513279 | 27.545381 | 1.984732 |
| H | -24.529034 | 30.656224 | 6.487624 | | H | -21.461097 | 28.020551 | 0.780717 |
| H | -20.791637 | 29.285281 | 8.040965 | | H | -19.779479 | 24.303974 | 2.214502 |
| N | -23.930785 | 21.645683 | 0.164876 | | O | -27.750252 | 23.835497 | 7.399985 |
| H | -23.356279 | 22.259230 | 0.725127 | | C | -28.580211 | 23.772612 | 8.482941 |
| N | -20.769463 | 27.072648 | 6.484319 | | C | -30.214732 | 23.345433 | 10.771938 |
| H | -20.933303 | 26.420416 | 5.730137 | | C | -29.182776 | 24.850222 | 9.123800 |
| C | -19.691150 | 26.887581 | 7.323819 | | C | -28.835281 | 22.482944 | 8.945957 |
| C | -23.408516 | 20.463883 | -0.317116 | | C | -29.653312 | 22.268669 | 10.060092 |
| O | -24.015392 | 19.612745 | -0.957909 | | C | -29.997474 | 24.636699 | 10.244612 |
| O | -19.459392 | 27.515452 | 8.351147 | | H | -29.057273 | 25.870387 | 8.776819 |
| C | -21.930609 | 20.313028 | -0.000112 | | H | -28.400425 | 21.629961 | 8.430120 |
| H | -21.731383 | 20.778117 | 0.971664 | | H | -29.851223 | 21.239750 | 10.352710 |
| H | -21.369251 | 20.859669 | -0.766291 | | H | -30.475208 | 25.503222 | 10.699164 |
| C | -18.777638 | 25.776173 | 6.834981 | | C | -31.189545 | 23.117659 | 11.998353 |
| H | -19.376705 | 24.863914 | 6.736766 | | C | -30.877040 | 21.798216 | 12.812004 |
| H | -18.415403 | 26.060952 | 5.841398 | | C | -30.289520 | 19.491652 | 14.425615 |
| C | -17.592080 | 25.527377 | 7.774090 | | C | -31.891919 | 21.057498 | 13.443973 |
| H | -17.945907 | 25.021658 | 8.680868 | | C | -29.554187 | 21.397831 | 13.073279 |
| H | -17.166634 | 26.484829 | 8.099691 | | C | -29.269718 | 20.264120 | 13.845516 |
| C | -21.501685 | 18.842675 | 0.025942 | | C | -31.606106 | 19.922324 | 14.217001 |
| H | -22.031516 | 18.335964 | 0.842942 | | H | -32.934266 | 21.358999 | 13.359584 |
| H | -21.813066 | 18.338715 | -0.897327 | | H | -28.714845 | 21.972647 | 12.686676 |
| C | -19.993812 | 18.629466 | 0.199427 | | H | -28.224452 | 20.003083 | 13.997346 |
| H | -19.484106 | 18.896200 | -0.732828 | | H | -32.449984 | 19.399526 | 14.658777 |
| H | -19.809409 | 17.561218 | 0.364920 | | C | -31.058464 | 24.232573 | 13.111240 |
| C | -16.464432 | 24.714582 | 7.130975 | | C | -30.783102 | 26.133335 | 15.250612 |
| H | -16.068814 | 25.275520 | 6.277643 | | C | -29.836221 | 24.874627 | 13.379302 |
| H | -15.641743 | 24.616677 | 7.849571 | | C | -32.120172 | 24.513082 | 13.990065 |
| C | -19.409614 | 19.443653 | 1.348090 | | C | -31.989856 | 25.451739 | 15.021533 |
| H | -19.440153 | 20.508271 | 1.085484 | | C | -29.706732 | 25.814377 | 14.412605 |
| H | -20.031287 | 19.313240 | 2.242904 | | H | -28.946986 | 24.642463 | 12.796574 |
| C | -16.896927 | 23.325854 | 6.668635 | | H | -33.070708 | 23.991591 | 13.894172 |
| H | -17.796035 | 23.409444 | 6.048784 | | H | -32.854369 | 25.629178 | 15.657847 |
| H | -17.159012 | 22.706447 | 7.534410 | | H | -28.730526 | 26.272781 | 14.545113 |
| C | -15.800743 | 22.641276 | 5.845085 | | C | -32.604637 | 23.096377 | 11.294526 |
| H | -15.459656 | 23.328386 | 5.060975 | | C | -35.096351 | 23.067293 | 9.861141 |
| H | -14.934729 | 22.410029 | 6.475539 | | C | -33.364433 | 24.264989 | 11.115613 |
| C | -17.966936 | 19.063130 | 1.686442 | | C | -33.090434 | 21.930513 | 10.673226 |
| H | -17.342855 | 19.082685 | 0.786289 | | C | -34.313670 | 21.912653 | 9.988043 |
| H | -17.936045 | 18.044541 | 2.089833 | | C | -34.586928 | 24.245863 | 10.431491 |
| C | -16.344558 | 21.365574 | 5.201160 | | H | -33.012552 | 25.222629 | 11.493549 |
| H | -17.333717 | 21.597360 | 4.795380 | | H | -32.512616 | 21.008201 | 10.702131 |
| H | -16.484444 | 20.589956 | 5.963222 | | H | -34.620786 | 20.967905 | 9.547975 |
| C | -17.431457 | 20.058857 | 2.713047 | | H | -35.131099 | 25.183603 | 10.339777 |
| H | -18.153245 | 20.115428 | 3.536244 | | C | -19.087933 | 26.499950 | 0.597123 |
| H | -17.385211 | 21.053949 | 2.252194 | | C | -18.277136 | 27.211073 | 1.755110 |
| C | -15.437771 | 20.844742 | 4.082247 | | C | -16.887042 | 28.527525 | 3.900414 |
| H | -14.483240 | 20.513021 | 4.507608 | | C | -17.397398 | 26.515055 | 2.601622 |
| C | -16.059108 | 19.692470 | 3.277290 | | C | -18.511954 | 28.562663 | 2.070620 |
| H | -16.150362 | 18.804778 | 3.914341 | | C | -17.828100 | 29.204786 | 3.109417 |
| H | -15.206568 | 21.663822 | 3.389476 | | C | -16.698914 | 27.165113 | 3.631613 |
| H | -15.377337 | 19.432189 | 2.458933 | | H | -17.229567 | 25.447427 | 2.475556 |
| H | -30.138929 | 29.685022 | 2.684804 | | H | -19.237914 | 29.145670 | 1.506365 |
| C | -24.096148 | 24.544930 | 4.463430 | | H | -18.048817 | 30.255229 | 3.288750 |
| H | -24.149815 | 23.549458 | 4.008413 | | H | -16.007190 | 26.565820 | 4.216780 |
| H | -23.215812 | 24.606821 | 5.115012 | | C | -19.261036 | 27.415532 | -0.680821 |
| O | -24.036532 | 25.535197 | 3.435318 | | C | -19.523474 | 28.987936 | -3.072817 |
| C | -22.843387 | 25.654892 | 2.776872 | | C | -18.285617 | 28.356183 | -1.054633 |
| C | -20.458539 | 26.161601 | 1.313798 | | C | -20.327341 | 27.226074 | -1.579487 |
| C | -21.786460 | 24.748142 | 2.819872 | | C | -20.464264 | 28.005502 | -2.737597 |
| C | -22.711046 | 26.813135 | 2.015483 | | C | -18.422875 | 29.132966 | -2.212164 |
| H | -17.388569 | 28.498678 | -0.455339 | | H | -16.617311 | 30.296599 | 6.887939 |
| H | -21.075539 | 26.457209 | -1.396280 | | H | -17.690273 | 28.942203 | 6.507999 |
| H | -21.325774 | 27.807655 | -3.369657 | | C | -15.383449 | 30.489026 | 4.407347 |
| H | -17.638285 | 29.853174 | -2.435336 | | H | -16.067547 | 31.223606 | 3.968972 |
| C | -18.363248 | 25.228459 | 0.002431 | | H | -14.689976 | 30.172496 | 3.619168 |
| C | -17.043467 | 23.007735 | -1.260783 | | H | -14.800646 | 31.014670 | 5.172915 |
| C | -19.080968 | 24.126859 | -0.498120 | | C | -19.644392 | 29.866080 | -4.329602 |
| C | -16.973183 | 25.219189 | -0.212047 | | C | -18.450506 | 29.591046 | -5.268381 |
| C | -16.329156 | 24.129170 | -0.809628 | | H | -18.526551 | 30.182576 | -6.188246 |
| C | -18.435043 | 23.035514 | -1.099089 | | H | -18.408922 | 28.533312 | -5.554263 |
| H | -20.167899 | 24.107762 | -0.456789 | | H | -17.491442 | 29.843297 | -4.802660 |
| H | -16.363064 | 26.076583 | 0.066592 | | C | -19.640977 | 31.357459 | -3.927156 |
| H | -15.249919 | 24.185310 | -0.937015 | | H | -19.765958 | 32.003921 | -4.803775 |
| H | -19.060877 | 22.222825 | -1.457322 | | H | -18.704533 | 31.655133 | -3.442898 |
| C | -30.674919 | 27.154362 | 16.395384 | | H | -20.457422 | 31.579629 | -3.229740 |
| C | -30.935990 | 26.449513 | 17.743809 | | C | -20.935558 | 29.618833 | -5.141981 |
| H | -31.944929 | 26.026426 | 17.801282 | | H | -20.984198 | 30.278403 | -6.016815 |
| H | -30.226038 | 25.629445 | 17.904374 | | H | -21.832304 | 29.811502 | -4.541505 |
| C | -31.720580 | 28.273101 | 16.196093 | | H | -20.988173 | 28.589096 | -5.514108 |
| H | -32.747166 | 27.891989 | 16.226546 | | C | -16.301169 | 21.839618 | -1.931222 |
| H | -31.583559 | 28.773314 | 15.230016 | | C | -15.586039 | 22.340691 | -3.204393 |
| H | -31.639154 | 29.034613 | 16.980576 | | H | -15.076486 | 21.518769 | -3.720832 |
| C | -29.940487 | 18.253029 | 15.267478 | | H | -14.827251 | 23.098998 | -2.982276 |
| C | -31.175465 | 17.496678 | 15.807350 | | H | -16.299311 | 22.785742 | -3.908356 |
| H | -31.819361 | 17.143518 | 14.993408 | | C | -15.254686 | 21.260289 | -0.954631 |
| H | -31.778182 | 18.128246 | 16.470359 | | H | -14.743382 | 20.394226 | -1.390856 |
| H | -30.876535 | 16.615824 | 16.388261 | | H | -15.725003 | 20.934452 | -0.019805 |
| C | -29.093760 | 18.677545 | 16.486431 | | H | -14.480839 | 21.990879 | -0.694997 |
| H | -28.140390 | 19.129321 | 16.191071 | | C | -17.224120 | 20.676327 | -2.354134 |
| H | -28.857520 | 17.816633 | 17.122750 | | H | -16.651206 | 19.861989 | -2.813535 |
| H | -29.628274 | 19.410755 | 17.102060 | | H | -17.971574 | 20.997672 | -3.088753 |
| C | -29.133310 | 17.252194 | 14.412029 | | H | -17.751139 | 20.254534 | -1.492752 |
| H | -28.177306 | 17.669076 | 14.076889 | | C | -27.333592 | 25.128477 | 6.966644 |
| H | -29.694514 | 16.955214 | 13.518070 | | H | -28.172416 | 25.645408 | 6.487217 |
| H | -28.903747 | 16.342482 | 14.979218 | | H | -26.966811 | 25.723549 | 7.811622 |
| C | -36.440846 | 23.085593 | 9.115158 | | C | -25.291458 | 24.785339 | 5.279708 |
| C | -36.873099 | 21.704224 | 8.573430 | | C | -26.231939 | 24.959255 | 6.007315 |
| H | -37.841397 | 21.765684 | 8.062268 | | H | -30.832987 | 27.148780 | 18.581831 |
| H | -36.979967 | 20.970204 | 9.380660 | | C | -29.291929 | 27.837077 | 16.494557 |
| H | -36.151851 | 21.311149 | 7.847457 | | H | -28.495107 | 27.109597 | 16.688461 |
| C | -37.557186 | 23.570525 | 10.065633 | | H | -29.268342 | 28.564535 | 17.314887 |
| H | -38.533764 | 23.557398 | 9.567588 | | H | -29.043235 | 28.379240 | 15.574761 |
| H | -37.389997 | 24.595802 | 10.413373 | | C | -22.582311 | 31.406402 | 8.268805 |
| H | -37.626222 | 22.929258 | 10.952378 | | C | -23.487724 | 31.161336 | 9.491617 |
| C | -36.349850 | 24.043573 | 7.908071 | | H | -23.157546 | 30.284423 | 10.061248 |
| H | -37.286328 | 24.051368 | 7.338170 | | H | -24.528631 | 30.986162 | 9.195715 |
| H | -35.546249 | 23.742835 | 7.225357 | | H | -23.478055 | 32.021853 | 10.170479 |
| H | -36.151718 | 25.076852 | 8.213591 | | C | -23.073201 | 32.672298 | 7.527587 |
| C | -16.129963 | 29.277901 | 5.009548 | | H | -22.516599 | 32.825711 | 6.595714 |
| C | -15.074675 | 28.417488 | 5.738495 | | H | -22.941854 | 33.568630 | 8.145589 |
| H | -14.323239 | 28.024756 | 5.043714 | | H | -24.138536 | 32.624818 | 7.278517 |
| H | -15.535470 | 27.571758 | 6.258186 | | C | -21.157782 | 31.725663 | 8.780215 |
| H | -14.540944 | 29.003530 | 6.496310 | | H | -20.451704 | 31.836472 | 7.948791 |
| C | -17.132160 | 29.777263 | 6.071317 | | H | -20.776535 | 30.946848 | 9.450057 |
| H | -17.863558 | 30.477978 | 5.653852 | | H | -21.142892 | 32.661029 | 9.352563 |

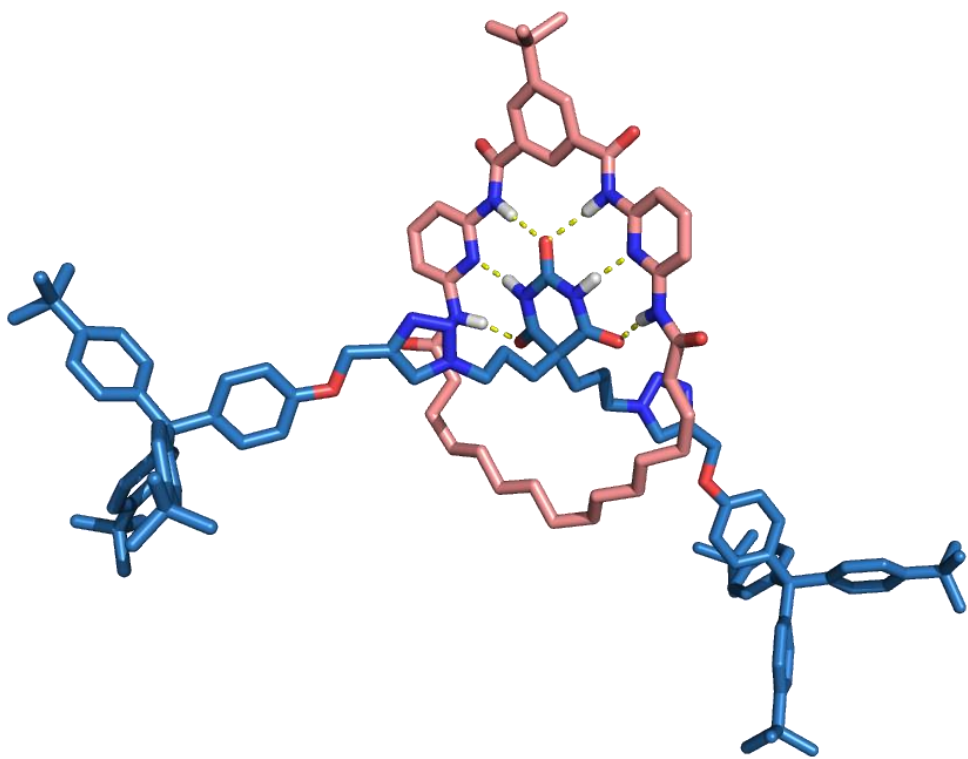

**Figure S23**. Optimized structure of rotaxane **9** with the macrocycle placed in the barbiturate station (MMFF, Spartan '18 software).

XYZ coordinates of rotaxane **9** with the macrocycle placed in the barbiturate station

| | | | | | | | |
|---|---|---|---|---|---|---|---|
| C | -8.957918 | 6.585806 | -5.696027 | C | -6.738777 | -1.050243 | -5.630255 |
| C | -6.406187 | 5.554168 | -4.971946 | H | -7.802277 | 0.672900 | -6.266551 |
| C | -8.610383 | 5.249894 | -5.941940 | H | -7.431893 | -1.735406 | -6.110892 |
| C | -8.005071 | 7.372689 | -5.025448 | H | -5.497823 | -2.607156 | -4.870947 |
| C | -6.744832 | 6.869827 | -4.647464 | C | -4.122543 | 8.000485 | -2.131395 |
| C | -7.353105 | 4.724192 | -5.578880 | C | -2.575877 | 9.300382 | -0.257238 |
| H | -9.311594 | 4.583618 | -6.441910 | N | -3.122241 | 7.306703 | -1.520405 |
| H | -8.233144 | 8.415319 | -4.801804 | C | -4.384336 | 9.332613 | -1.827242 |
| H | -5.400108 | 5.198040 | -4.805932 | C | -3.609109 | 9.984439 | -0.876192 |
| C | -7.108085 | 3.284409 | -5.900105 | C | -2.349835 | 7.980571 | -0.612164 |
| C | -5.830150 | 7.811191 | -3.932121 | H | -5.196586 | 9.887308 | -2.281516 |
| O | -7.727572 | 2.810079 | -6.848314 | H | -3.808703 | 11.022615 | -0.624785 |
| O | -5.974253 | 9.009133 | -4.160193 | H | -1.962149 | 9.823467 | 0.466246 |
| N | -4.909262 | 7.270250 | -3.050355 | N | -3.632770 | -1.036894 | -3.657323 |
| H | -4.775110 | 6.267779 | -3.014993 | H | -2.993043 | -0.280224 | -3.437535 |
| N | -6.223548 | 2.586484 | -5.095584 | N | -1.286637 | 7.243537 | -0.084276 |
| H | -5.665822 | 3.082828 | -4.412597 | H | -1.140804 | 6.346217 | -0.535768 |
| C | -6.041402 | 1.185774 | -5.094963 | C | -0.386213 | 7.580014 | 0.910622 |
| C | -5.653007 | -1.536018 | -4.920402 | C | -3.247672 | -2.307769 | -3.271768 |
| C | -6.931978 | 0.322322 | -5.724792 | O | -3.875184 | -3.345591 | -3.439392 |
| N | -4.971318 | 0.726378 | -4.388890 | O | -0.373357 | 8.609611 | 1.572333 |
| C | -4.786773 | -0.628651 | -4.332363 | C | -1.899422 | -2.269423 | -2.566487 |

| | | | | | | | | |
|---|---|---|---|---|---|---|---|---|
| H | -1.956917 | -1.471085 | -1.818551 | | H | 0.289753 | 2.601252 | -3.524400 |
| H | -1.140661 | -2.004246 | -3.312118 | | H | 1.190353 | 3.700241 | -2.480111 |
| C | 0.618055 | 6.454801 | 1.116268 | | C | 0.194313 | 2.116130 | -1.393314 |
| H | 0.035811 | 5.554758 | 1.342821 | | H | 0.284262 | 1.022050 | -1.385467 |
| H | 1.144111 | 6.307160 | 0.165195 | | C | 13.245170 | 2.207151 | -3.844305 |
| C | 1.642861 | 6.671178 | 2.226236 | | C | 13.468255 | 1.235895 | -5.071793 |
| H | 1.143158 | 6.650408 | 3.202017 | | C | 13.734522 | -0.533669 | -7.321612 |
| H | 2.094730 | 7.665328 | 2.125473 | | C | 13.529749 | 1.718881 | -6.391592 |
| C | -1.501696 | -3.560753 | -1.855113 | | C | 13.461802 | -0.161449 | -4.915405 |
| H | -2.219101 | -3.758682 | -1.048881 | | C | 13.609635 | -1.022192 | -6.010559 |
| H | -1.570055 | -4.410768 | -2.544573 | | C | 13.676863 | 0.856169 | -7.487645 |
| C | -0.079610 | -3.513648 | -1.264909 | | H | 13.455626 | 2.785643 | -6.594839 |
| H | 0.645563 | -3.603834 | -2.083321 | | H | 13.332583 | -0.610357 | -3.932530 |
| H | 0.062512 | -4.387815 | -0.618642 | | H | 13.613954 | -2.092900 | -5.817097 |
| C | 2.763411 | 5.617235 | 2.186167 | | H | 13.733313 | 1.312026 | -8.472343 |
| H | 3.293893 | 5.706485 | 1.229840 | | C | 13.883320 | 1.668566 | -2.500377 |
| H | 3.493230 | 5.844861 | 2.972377 | | C | 15.129517 | 0.823093 | -0.050297 |
| C | 0.215601 | -2.231091 | -0.482436 | | C | 15.102763 | 0.967957 | -2.495991 |
| H | 0.191257 | -1.383002 | -1.175924 | | C | 13.348199 | 2.001761 | -1.242819 |
| H | -0.569350 | -2.061765 | 0.264856 | | C | 13.947097 | 1.574555 | -0.048127 |
| C | 2.264901 | 4.177630 | 2.364361 | | C | 15.699948 | 0.541673 | -1.302611 |
| H | 1.439590 | 3.978097 | 1.673759 | | H | 15.621272 | 0.749655 | -3.427627 |
| H | 1.873089 | 4.046125 | 3.379834 | | H | 12.453779 | 2.617117 | -1.166999 |
| C | 3.363770 | 3.145103 | 2.088547 | | H | 13.461873 | 1.863851 | 0.879996 |
| H | 3.795044 | 3.338299 | 1.098528 | | H | 16.636589 | -0.007608 | -1.372519 |
| H | 4.172294 | 3.245840 | 2.821605 | | C | 13.962952 | 3.604709 | -4.013277 |
| C | 1.579784 | -2.239624 | 0.215255 | | C | 15.357507 | 6.114240 | -4.164781 |
| H | 2.352364 | -2.606663 | -0.470599 | | C | 13.469224 | 4.765920 | -3.391366 |
| H | 1.552576 | -2.917576 | 1.075795 | | C | 15.210868 | 3.717631 | -4.651337 |
| C | 2.794622 | 1.721006 | 2.125337 | | C | 15.880774 | 4.944481 | -4.739958 |
| H | 1.872195 | 1.700877 | 1.536078 | | C | 14.141558 | 5.993838 | -3.479455 |
| H | 2.524372 | 1.456066 | 3.154245 | | H | 12.551788 | 4.733434 | -2.806837 |
| C | 1.934098 | -0.817718 | 0.662607 | | H | 15.696419 | 2.843850 | -5.081579 |
| H | 1.104164 | -0.411086 | 1.252997 | | H | 16.836661 | 4.963696 | -5.259382 |
| H | 2.034578 | -0.190791 | -0.230306 | | H | 13.686446 | 6.844616 | -2.979835 |
| C | 3.783933 | 0.694673 | 1.564876 | | C | 13.902230 | -1.508693 | -8.498831 |
| H | 4.688174 | 0.683845 | 2.185245 | | C | 14.045974 | -0.812005 | -9.870774 |
| C | 3.224405 | -0.734795 | 1.479609 | | H | 14.918978 | -0.149410 | -9.896944 |
| H | 3.042783 | -1.122362 | 2.488970 | | H | 13.158323 | -0.217698 | -10.116859 |
| H | 4.096702 | 1.005651 | 0.559822 | | H | 14.173759 | -1.546270 | -10.675260 |
| H | 3.991512 | -1.373930 | 1.025870 | | C | 12.672041 | -2.437456 | -8.582926 |
| C | 4.599940 | 2.372822 | -3.241984 | | H | 12.558253 | -3.056387 | -7.686151 |
| H | 4.977698 | 1.804405 | -2.404844 | | H | 12.750624 | -3.121059 | -9.436375 |
| C | 5.232288 | 3.081446 | -4.240386 | | H | 11.748320 | -1.858964 | -8.702579 |
| N | 4.278597 | 3.604451 | -5.079701 | | C | 15.170792 | -2.363977 | -8.288259 |
| N | 3.083156 | 3.214797 | -4.648205 | | H | 15.108836 | -2.983177 | -7.386660 |
| N | 3.277319 | 2.475649 | -3.545434 | | H | 16.061506 | -1.731634 | -8.192644 |
| C | 6.689331 | 3.294571 | -4.474750 | | H | 15.333682 | -3.044077 | -9.132575 |
| H | 6.946424 | 4.309182 | -4.147903 | | C | 15.813249 | 0.337875 | 1.238946 |
| H | 6.893067 | 3.186960 | -5.546813 | | C | 17.215558 | 0.973322 | 1.351869 |
| C | 2.190547 | 1.807501 | -2.870618 | | H | 17.714119 | 0.668435 | 2.279445 |
| H | 2.595774 | 1.463473 | -1.911318 | | H | 17.154026 | 2.068066 | 1.350432 |
| H | 1.947107 | 0.916767 | -3.461595 | | H | 17.870720 | 0.680050 | 0.524197 |
| O | 7.461500 | 2.336419 | -3.740392 | | C | 15.954036 | -1.199876 | 1.207869 |
| C | 8.821370 | 2.401960 | -3.829919 | | H | 16.405669 | -1.574318 | 2.134066 |
| C | 11.666359 | 2.315485 | -3.790574 | | H | 16.588169 | -1.542274 | 0.382826 |
| C | 9.559321 | 3.316752 | -4.572611 | | H | 14.976396 | -1.683204 | 1.094328 |
| C | 9.495700 | 1.423920 | -3.100193 | | C | 15.040847 | 0.697042 | 2.528446 |
| C | 10.892823 | 1.367334 | -3.094015 | | H | 15.558798 | 0.317980 | 3.417708 |
| C | 10.961841 | 3.262119 | -4.564979 | | H | 14.034871 | 0.261356 | 2.530143 |
| H | 9.084698 | 4.078195 | -5.182604 | | H | 14.943663 | 1.781863 | 2.652486 |
| H | 8.929837 | 0.683429 | -2.539939 | | C | 16.123755 | 7.442579 | -4.278968 |
| H | 11.366686 | 0.557005 | -2.543744 | | C | 17.510487 | 7.301535 | -3.614935 |
| H | 11.500045 | 3.970176 | -5.193091 | | H | 18.064495 | 8.246834 | -3.652107 |
| C | 0.926857 | 2.647672 | -2.634480 | | H | 18.131945 | 6.546794 | -4.109261 |
| H | 17.415955 | 7.011389 | -2.561771 | | C | -7.541296 | -9.598485 | 4.800540 |
| C | 16.308582 | 7.808277 | -5.767751 | | C | -8.256432 | -10.346039 | 6.955578 |
| H | 16.822516 | 8.770400 | -5.877644 | | H | -9.404086 | -8.972873 | 8.125064 |
| H | 15.341003 | 7.886642 | -6.277599 | | H | -8.114073 | -7.619995 | 4.234302 |
| H | 16.904840 | 7.063885 | -6.306779 | | H | -7.062561 | -9.741169 | 3.835625 |
| C | 15.410979 | 8.634732 | -3.601527 | | H | -8.346692 | -11.108405 | 7.726617 |
| H | 15.989842 | 9.558890 | -3.717821 | | C | -10.950915 | -6.958506 | 5.945725 |
| H | 15.282465 | 8.469756 | -2.525397 | | C | -13.539823 | -7.635942 | 4.895727 |
| H | 14.422633 | 8.815755 | -4.039872 | | C | -11.253565 | -6.829559 | 4.577726 |
| C | -3.648580 | 3.648523 | -2.383035 | | C | -11.955916 | -7.507623 | 6.760955 |
| O | -4.594856 | 4.174587 | -2.961623 | | C | -13.222927 | -7.820186 | 6.251852 |
| N | -2.794629 | 4.377329 | -1.616349 | | C | -12.521817 | -7.144357 | 4.068573 |
| H | -2.951331 | 5.386586 | -1.626131 | | H | -10.495960 | -6.492404 | 3.872756 |
| N | -3.358045 | 2.331844 | -2.555866 | | H | -11.766930 | -7.717170 | 7.811970 |
| H | -3.927647 | 1.846796 | -3.251227 | | H | -13.959442 | -8.228481 | 6.940854 |
| C | -2.132412 | 1.787958 | -2.298216 | | H | -12.675999 | -7.004590 | 3.002138 |
| C | -1.535824 | 3.967420 | -1.282899 | | C | -9.311918 | -4.718769 | 12.161535 |
| O | -1.764491 | 0.774839 | -2.887248 | | C | -10.533392 | -3.879129 | 12.599085 |
| O | -0.670709 | 4.788665 | -0.994975 | | H | -11.464941 | -4.451824 | 12.521380 |
| C | -1.309281 | 2.474225 | -1.229320 | | H | -10.638285 | -2.972072 | 11.992460 |
| H | 0.750329 | 2.469412 | -0.513942 | | H | -10.438113 | -3.557467 | 13.643184 |
| C | -5.045134 | -0.178738 | 2.887262 | | C | -8.059810 | -3.833486 | 12.339973 |
| H | -4.343312 | -0.961615 | 3.134613 | | H | -7.132120 | -4.385866 | 12.154212 |
| C | -6.414503 | -0.086905 | 3.020239 | | H | -7.998189 | -3.437022 | 13.360219 |
| N | -6.822611 | 1.138633 | 2.553682 | | H | -8.078160 | -2.980589 | 11.651115 |
| N | -5.749420 | 1.808686 | 2.151114 | | C | -9.221986 | -5.925114 | 13.121407 |
| N | -4.684693 | 1.016894 | 2.353554 | | H | -8.317090 | -6.520135 | 12.956452 |
| C | -7.388347 | -1.073236 | 3.565920 | | H | -10.082510 | -6.593160 | 12.996838 |
| H | -8.053239 | -1.389806 | 2.753290 | | H | -9.202705 | -5.597439 | 14.167443 |
| H | -7.972253 | -0.582651 | 4.353970 | | C | -6.922178 | -11.994312 | 5.533538 |
| C | -3.350640 | 1.467638 | 2.030768 | | C | -8.006229 | -13.092478 | 5.571486 |
| H | -2.661991 | 0.655534 | 2.292093 | | H | -7.573667 | -14.081552 | 5.380499 |
| H | -3.127932 | 2.318144 | 2.685146 | | H | -8.776497 | -12.912858 | 4.812067 |
| O | -6.707228 | -2.210691 | 4.105762 | | H | -8.508964 | -13.144471 | 6.543473 |
| C | -7.465392 | -3.210456 | 4.641399 | | C | -5.883329 | -12.266788 | 6.643235 |
| C | -8.780332 | -5.459287 | 5.780672 | | H | -5.377864 | -13.226617 | 6.484546 |
| C | -8.851684 | -3.228974 | 4.751296 | | H | -6.339739 | -12.306868 | 7.638286 |
| C | -6.732753 | -4.289934 | 5.131109 | | H | -5.114977 | -11.484776 | 6.664703 |
| C | -7.375384 | -5.388617 | 5.707448 | | C | -6.185874 | -12.122551 | 4.180629 |
| C | -9.497411 | -4.331381 | 5.329884 | | H | -5.728476 | -13.113348 | 4.072032 |
| H | -9.467213 | -2.401397 | 4.414607 | | H | -5.381336 | -11.383628 | 4.087917 |
| H | -5.646922 | -4.274307 | 5.077949 | | H | -6.869940 | -11.990860 | 3.334196 |
| H | -6.755595 | -6.186786 | 6.111022 | | C | -14.941175 | -7.992113 | 4.372411 |
| H | -10.579954 | -4.286375 | 5.436057 | | C | -15.218748 | -9.492261 | 4.609217 |
| C | -3.186725 | 1.857015 | 0.556487 | | H | -16.199399 | -9.780169 | 4.212606 |
| H | -3.697749 | 2.809735 | 0.387470 | | H | -15.215404 | -9.751388 | 5.673647 |
| H | -3.671929 | 1.104405 | -0.076209 | | H | -14.463327 | -10.115844 | 4.116499 |
| C | -1.701625 | 1.956791 | 0.193210 | | C | -16.001687 | -7.154951 | 5.119952 |
| H | -1.277928 | 0.950867 | 0.326600 | | H | -17.009172 | -7.364395 | 4.741820 |
| H | -1.208427 | 2.594122 | 0.941303 | | H | -15.815452 | -6.081440 | 4.996504 |
| C | -9.499803 | -6.672105 | 6.500041 | | H | -16.013484 | -7.364959 | 6.195093 |
| C | -9.486776 | -6.209407 | 8.011935 | | C | -15.130434 | -7.724289 | 2.862164 |
| C | -9.388967 | -5.217310 | 10.708864 | | H | -16.145598 | -7.984426 | 2.538999 |
| C | -10.488287 | -5.367448 | 8.529813 | | H | -14.438063 | -8.320601 | 2.256503 |
| C | -8.397091 | -6.480405 | 8.857883 | | H | -14.974993 | -6.667058 | 2.617451 |
| C | -8.358758 | -6.010645 | 10.177173 | | C | -10.292403 | 7.200827 | -6.154222 |
| C | -10.448830 | -4.897379 | 9.850632 | | C | -11.259539 | 6.185208 | -6.804078 |
| H | -11.320501 | -5.048500 | 7.904804 | | H | -10.829510 | 5.736683 | -7.707237 |
| H | -7.544432 | -7.054343 | 8.500578 | | H | -12.197155 | 6.668701 | -7.104068 |
| H | -7.491782 | -6.268302 | 10.782031 | | H | -11.519069 | 5.376697 | -6.110838 |
| H | -11.269301 | -4.263242 | 10.175115 | | C | -11.034684 | 7.829220 | -4.955150 |
| C | -8.772215 | -8.055093 | 6.259703 | | H | -10.474317 | 8.655276 | -4.504467 |
| C | -7.572450 | -10.619220 | 5.759106 | | H | -11.216676 | 7.085819 | -4.170346 |
| C | -8.856788 | -9.102632 | 7.193415 | | H | -12.005905 | 8.235612 | -5.261750 |
| C | -8.141395 | -8.353504 | 5.037926 | | C | -10.006252 | 8.296845 | -7.204021 |

| | | | | | | | |
|---|---|---|---|---|---|---|---|
| H | -9.462513 | 7.887348 | -8.063914 | H | -10.936560 | 8.739572 | -7.578513 |
| H | -9.402393 | 9.113859 | -6.793541 | | | | |

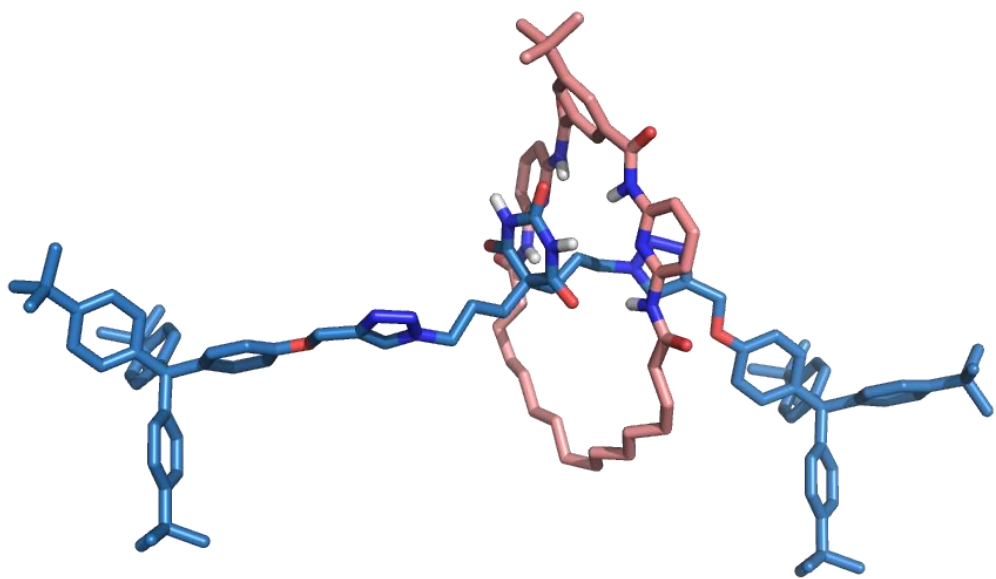

**Figure S24**. Optimized structure of rotaxane **9** with the with the macrocycle placed in the hydrocarbon chain (MMFF, Spartan '18 software).

XYZ coordinates of rotaxane **9** with the with the macrocycle placed in the hydrocarbon chain

| | | | | | | | |
|---|---|---|---|---|---|---|---|
| C | -5.244847 | 6.197032 | -6.794522 | C | -1.872816 | -1.858492 | -6.902132 |
| C | -2.656722 | 5.155815 | -6.169007 | C | -2.711987 | 0.139009 | -7.935594 |
| C | -4.837816 | 4.931580 | -7.237637 | N | -1.745141 | 0.247060 | -5.740225 |
| C | -4.316658 | 6.929470 | -6.036155 | C | -1.523944 | -1.089702 | -5.804773 |
| C | -3.042555 | 6.423888 | -5.720974 | C | -2.474084 | -1.230133 | -7.984298 |
| C | -3.565844 | 4.409921 | -6.927615 | H | -3.186450 | 0.609917 | -8.787597 |
| H | -5.501072 | 4.316705 | -7.843626 | H | -2.758290 | -1.803661 | -8.862425 |
| H | -4.577899 | 7.924385 | -5.674483 | H | -1.685723 | -2.924253 | -6.951850 |
| H | -1.652148 | 4.786716 | -5.986561 | C | -0.541081 | 7.309727 | -3.035870 |
| C | -3.251772 | 3.059342 | -7.470983 | C | 1.148901 | 8.296135 | -1.119661 |
| C | -2.151840 | 7.317156 | -4.929654 | N | 0.032652 | 6.440432 | -2.170519 |
| O | -3.659201 | 2.786943 | -8.593866 | C | -0.330066 | 8.678704 | -2.966804 |
| O | -2.216504 | 8.522415 | -5.143856 | C | 0.519086 | 9.177327 | -1.987467 |
| N | -1.370449 | 6.697872 | -3.983082 | C | 0.891073 | 6.939948 | -1.245666 |
| H | -1.417548 | 5.698501 | -3.872447 | H | -0.796299 | 9.378112 | -3.649326 |
| N | -2.582676 | 2.213023 | -6.616816 | H | 0.701105 | 10.245760 | -1.911752 |
| H | -2.310902 | 2.534476 | -5.702281 | H | 1.829122 | 8.695457 | -0.377657 |
| C | -2.344121 | 0.843005 | -6.798577 | N | -0.872185 | -1.599845 | -4.680551 |

S35

| | | | |
|---|---|---|---|
| H | -0.692435 | -0.889542 | -3.989599 |
| N | 1.514893 | 5.964661 | -0.463680 |
| H | 1.197310 | 5.027727 | -0.677360 |
| C | 2.438342 | 6.143244 | 0.544413 |
| C | -0.541185 | -2.908819 | -4.400039 |
| O | -0.869736 | -3.894155 | -5.048467 |
| O | 2.780334 | 7.215635 | 1.029813 |
| C | 0.382700 | -2.999378 | -3.194916 |
| H | 0.086181 | -2.242635 | -2.458975 |
| H | 1.387629 | -2.747331 | -3.555700 |
| C | 3.049151 | 4.822114 | 0.987443 |
| H | 2.235113 | 4.139178 | 1.254663 |
| H | 3.582064 | 4.404701 | 0.123984 |
| C | 4.014697 | 4.974559 | 2.167644 |
| H | 3.441214 | 5.161987 | 3.083620 |
| H | 4.655154 | 5.852695 | 2.018482 |
| C | 0.397346 | -4.375881 | -2.528539 |
| H | -0.536953 | -4.504869 | -1.968027 |
| H | 0.414161 | -5.172263 | -3.281995 |
| C | 1.590037 | -4.589860 | -1.584163 |
| H | 2.493965 | -4.741330 | -2.187439 |
| H | 1.430246 | -5.515457 | -1.018445 |
| C | 4.931112 | 3.762068 | 2.365833 |
| H | 5.566643 | 3.651903 | 1.478190 |
| H | 5.602663 | 3.960328 | 3.210132 |
| C | 1.826143 | -3.426041 | -0.623031 |
| H | 2.081621 | -2.530814 | -1.202262 |
| H | 0.903959 | -3.201009 | -0.073613 |
| C | 4.172942 | 2.459833 | 2.620089 |
| H | 3.452956 | 2.293687 | 1.811403 |
| H | 3.600689 | 2.534085 | 3.552209 |
| C | 5.120681 | 1.257936 | 2.684298 |
| H | 5.784881 | 1.279537 | 1.811592 |
| H | 5.754669 | 1.318392 | 3.576329 |
| C | 2.955005 | -3.701929 | 0.375744 |
| H | 3.808136 | -4.166166 | -0.132862 |
| H | 2.608400 | -4.401813 | 1.144470 |
| C | 4.327057 | -0.051200 | 2.687181 |
| H | 3.547430 | 0.026720 | 1.923191 |
| H | 3.818727 | -0.181473 | 3.649653 |
| C | 3.402981 | -2.384848 | 1.008961 |
| H | 2.523424 | -1.869897 | 1.414304 |
| H | 3.817531 | -1.745679 | 0.218750 |
| C | 5.219271 | -1.261913 | 2.398268 |
| H | 5.900250 | -1.425016 | 3.242081 |
| C | 4.439801 | -2.556609 | 2.121037 |
| H | 3.939738 | -2.892775 | 3.036939 |
| H | 5.849051 | -1.050073 | 1.524842 |
| H | 5.159654 | -3.336593 | 1.845527 |
| C | 3.229570 | 2.483046 | -3.453325 |
| H | 3.605168 | 1.670971 | -2.848597 |
| C | 3.849982 | 3.570078 | -4.033554 |
| N | 2.915675 | 4.292912 | -4.733778 |
| N | 1.747215 | 3.674221 | -4.627603 |
| N | 1.940793 | 2.584074 | -3.867388 |
| C | 5.285784 | 3.964814 | -4.035022 |
| H | 5.367406 | 5.037408 | -3.823897 |
| H | 5.676614 | 3.754601 | -5.038420 |
| C | 0.864575 | 1.642329 | -3.658616 |
| H | 1.269765 | 0.783699 | -3.111884 |
| H | 0.578449 | 1.291034 | -4.655718 |
| O | 6.021453 | 3.233099 | -3.047961 |
| C | 7.384671 | 3.283519 | -3.117486 |
| C | 10.227224 | 3.197800 | -3.030570 |
| C | 8.138031 | 4.082526 | -3.971980 |
| C | 8.043327 | 2.424650 | -2.239779 |
| C | 9.440044 | 2.366680 | -2.208595 |
| C | 9.538715 | 4.021083 | -3.944501 |
| H | 7.677912 | 4.761013 | -4.682925 |
| H | 7.467617 | 1.778965 | -1.581431 |
| H | 9.906271 | 1.651366 | -1.533861 |
| H | 10.085219 | 4.628228 | -4.664108 |
| C | -0.343267 | 2.256995 | -2.944729 |
| H | -1.214210 | 2.144980 | -3.599114 |
| H | -0.204827 | 3.333448 | -2.793429 |
| C | -0.632090 | 1.576766 | -1.593174 |
| H | -0.736555 | 0.496552 | -1.760977 |
| C | 11.806136 | 3.083361 | -3.032256 |
| C | 12.052391 | 1.872542 | -4.019041 |
| C | 12.383148 | -0.330257 | -5.837448 |
| C | 12.080544 | 2.059195 | -5.413599 |
| C | 12.106868 | 0.543600 | -3.564359 |
| C | 12.285675 | -0.527344 | -4.449993 |
| C | 12.259402 | 0.986389 | -6.299586 |
| H | 11.959455 | 3.053487 | -5.840097 |
| H | 12.004253 | 0.311991 | -2.506184 |
| H | 12.339267 | -1.529385 | -4.029457 |
| H | 12.289937 | 1.219830 | -7.360357 |
| C | 12.412562 | 2.847913 | -1.591010 |
| C | 13.598756 | 2.547520 | 1.010014 |
| C | 13.658562 | 2.219323 | -1.416958 |
| C | 11.817617 | 3.390725 | -0.437950 |
| C | 12.387786 | 3.229028 | 0.833088 |
| C | 14.227438 | 2.058612 | -0.147158 |
| H | 14.217922 | 1.850719 | -2.274770 |
| H | 10.897778 | 3.968180 | -0.506445 |
| H | 11.856737 | 3.668375 | 1.673030 |
| H | 15.187911 | 1.551727 | -0.082327 |
| C | 12.530054 | 4.414366 | -3.483159 |
| C | 13.915288 | 6.841653 | -4.158354 |
| C | 12.056218 | 5.676737 | -3.079856 |
| C | 13.759154 | 4.395836 | -4.164756 |
| C | 14.424563 | 5.580014 | -4.508121 |
| C | 12.722637 | 6.861515 | -3.424220 |
| H | 11.156560 | 5.761274 | -2.472973 |
| H | 14.234793 | 3.454161 | -4.431192 |
| H | 15.366269 | 5.494400 | -5.046395 |
| H | 12.281649 | 7.794528 | -3.084262 |
| C | 12.595226 | -1.528935 | -6.777378 |
| C | 12.715866 | -1.139890 | -8.268216 |
| H | 13.559415 | -0.461296 | -8.440569 |
| H | 11.804674 | -0.652221 | -8.633605 |
| H | 12.879671 | -2.024509 | -8.895303 |
| C | 11.404871 | -2.503583 | -6.652054 |
| H | 11.311955 | -2.915562 | -5.641199 |
| H | 11.516178 | -3.353197 | -7.335838 |
| H | 10.458844 | -2.003936 | -6.891874 |
| C | 13.897352 | -2.266426 | -6.394484 |
| H | 13.857481 | -2.680325 | -5.380995 |
| H | 14.760786 | -1.592154 | -6.440911 |
| H | 14.091656 | -3.104366 | -7.074164 |
| C | 14.251950 | 2.352683 | 2.388257 |
| C | 15.618081 | 3.070481 | 2.418151 |
| H | 16.093309 | 2.974924 | 3.401312 |
| H | 15.506553 | 4.140014 | 2.203940 |
| H | 16.317512 | 2.659357 | 1.681912 |
| C | 14.462885 | 0.846612 | 2.657506 |
| H | 14.892682 | 0.679330 | 3.652018 |
| H | 15.144789 | 0.386811 | 1.933900 |
| H | 13.513378 | 0.300491 | 2.608192 |

| | | | | | | | | |
|---|---|---|---|---|---|---|---|---|
| C | 13.413392 | 2.910195 | 3.560725 | | H | -11.495424 | -5.041921 | 8.697484 |
| H | 13.911384 | 2.733715 | 4.521601 | | H | -8.815095 | -8.017200 | 7.085394 |
| H | 12.428322 | 2.432129 | 3.614444 | | H | -7.765641 | -8.236885 | 9.247853 |
| H | 13.263562 | 3.992467 | 3.471835 | | H | -10.451034 | -5.257106 | 10.850113 |
| C | 14.673764 | 8.120125 | -4.551709 | | C | -10.980236 | -7.726984 | 5.420194 |
| C | 16.062004 | 8.124057 | -3.877016 | | C | -11.116112 | -10.123131 | 3.836927 |
| H | 16.614579 | 9.040400 | -4.115035 | | C | -11.167013 | -8.990102 | 6.009023 |
| H | 16.682080 | 7.281349 | -4.202037 | | C | -10.945112 | -7.690046 | 4.014307 |
| H | 15.970649 | 8.063313 | -2.786027 | | C | -10.994464 | -8.860395 | 3.242612 |
| C | 14.853700 | 8.170641 | -6.084813 | | C | -11.214789 | -10.158890 | 5.237805 |
| H | 15.358004 | 9.093861 | -6.393393 | | H | -11.290236 | -9.088331 | 7.085729 |
| H | 13.884971 | 8.132437 | -6.597011 | | H | -10.890014 | -6.740196 | 3.486095 |
| H | 15.457405 | 7.337168 | -6.460054 | | H | -10.947700 | -8.744100 | 2.163777 |
| C | 13.956082 | 9.424240 | -4.136074 | | H | -11.345596 | -11.104898 | 5.758338 |
| H | 14.525203 | 10.306525 | -4.452679 | | C | -12.526837 | -6.057825 | 6.353148 |
| H | 13.838980 | 9.492388 | -3.048278 | | C | -15.357086 | -5.556566 | 6.392012 |
| H | 12.962023 | 9.499738 | -4.592092 | | C | -13.164037 | -5.326349 | 5.334525 |
| C | -3.499755 | 4.189411 | -2.051657 | | C | -13.358087 | -6.600254 | 7.349100 |
| O | -3.960014 | 5.030498 | -2.814519 | | C | -14.734263 | -6.340376 | 7.377314 |
| N | -2.517139 | 4.487559 | -1.161357 | | C | -14.542448 | -5.067986 | 5.362217 |
| H | -2.237976 | 5.461616 | -1.100549 | | H | -12.596894 | -4.951550 | 4.484653 |
| N | -3.885737 | 2.888787 | -2.121084 | | H | -12.950220 | -7.249771 | 8.120994 |
| H | -4.610364 | 2.683449 | -2.797924 | | H | -15.316399 | -6.778614 | 8.184695 |
| C | -3.097036 | 1.834708 | -1.774142 | | H | -14.953388 | -4.484483 | 4.543431 |
| C | -1.664098 | 3.589759 | -0.591808 | | C | -8.324589 | -6.948916 | 11.624239 |
| O | -3.332076 | 0.707132 | -2.205122 | | C | -8.916745 | -6.057578 | 12.740205 |
| O | -0.694108 | 3.967008 | 0.066617 | | H | -9.974978 | -6.283992 | 12.917992 |
| C | -1.909639 | 2.120801 | -0.878488 | | H | -8.832306 | -4.993617 | 12.494147 |
| H | 0.242471 | 1.706366 | -0.941620 | | H | -8.389969 | -6.211520 | 13.688241 |
| C | -5.413100 | -0.406140 | 3.424583 | | C | -6.832660 | -6.574899 | 11.485439 |
| H | -4.912153 | -1.361951 | 3.380902 | | H | -6.303676 | -7.235097 | 10.788496 |
| C | -6.677581 | -0.040313 | 3.836217 | | H | -6.316887 | -6.646398 | 12.448566 |
| N | -6.811183 | 1.319432 | 3.693301 | | H | -6.716512 | -5.549091 | 11.119042 |
| N | -5.671832 | 1.802077 | 3.211458 | | C | -8.438806 | -8.412726 | 12.102938 |
| N | -4.834512 | 0.765108 | 3.046748 | | H | -7.945805 | -9.112172 | 11.418695 |
| C | -7.787409 | -0.873154 | 4.381581 | | H | -9.488280 | -8.718734 | 12.186974 |
| H | -8.724310 | -0.573321 | 3.896955 | | H | -7.973197 | -8.544369 | 13.085571 |
| H | -7.852352 | -0.684310 | 5.460229 | | C | -11.170578 | -11.426274 | 3.022627 |
| C | -3.492902 | 0.963744 | 2.542665 | | C | -12.526156 | -12.127269 | 3.256226 |
| H | -3.002162 | -0.016334 | 2.538331 | | H | -12.605450 | -13.042490 | 2.659540 |
| H | -2.966016 | 1.599643 | 3.262824 | | H | -13.361502 | -11.473539 | 2.977392 |
| O | -7.551018 | -2.265403 | 4.142919 | | H | -12.668224 | -12.413387 | 4.303770 |
| C | -8.436409 | -3.158504 | 4.675387 | | C | -10.029275 | -12.367595 | 3.467360 |
| C | -10.110261 | -5.242916 | 5.654444 | | H | -10.025689 | -13.289730 | 2.875955 |
| C | -9.604241 | -2.853723 | 5.367181 | | H | -10.119826 | -12.662069 | 4.518399 |
| C | -8.087889 | -4.496106 | 4.496062 | | H | -9.051337 | -11.887604 | 3.343738 |
| C | -8.898126 | -5.521843 | 4.993286 | | C | -11.018291 | -11.217541 | 1.498596 |
| C | -10.418535 | -3.882475 | 5.864945 | | H | -11.051292 | -12.173271 | 0.963895 |
| H | -9.913274 | -1.830297 | 5.552406 | | H | -10.062631 | -10.741470 | 1.251551 |
| H | -7.166054 | -4.748487 | 3.977998 | | H | -11.825486 | -10.595595 | 1.093776 |
| H | -8.556552 | -6.546929 | 4.865277 | | C | -16.869893 | -5.290010 | 6.457187 |
| H | -11.299598 | -3.598821 | 6.438309 | | C | -17.636669 | -6.627270 | 6.369239 |
| C | -3.461351 | 1.585877 | 1.143129 | | H | -18.719285 | -6.463104 | 6.381200 |
| H | -3.689747 | 2.655796 | 1.215803 | | H | -17.403422 | -7.292872 | 7.207042 |
| H | -4.248772 | 1.133765 | 0.527451 | | H | -17.389959 | -7.162757 | 5.444395 |
| C | -2.095800 | 1.369844 | 0.468611 | | C | -17.219378 | -4.591949 | 7.790375 |
| H | -1.973006 | 0.290928 | 0.302400 | | H | -18.287970 | -4.357790 | 7.844777 |
| H | -1.303249 | 1.664462 | 1.168940 | | H | -16.665230 | -3.652366 | 7.900372 |
| C | -10.986441 | -6.402414 | 6.283871 | | H | -16.984044 | -5.215344 | 8.659677 |
| C | -10.322596 | -6.580539 | 7.707966 | | C | -17.394202 | -4.384321 | 5.319674 |
| C | -9.025994 | -6.792399 | 10.265001 | | H | -18.471763 | -4.212317 | 5.417421 |
| C | -10.704500 | -5.782248 | 8.801973 | | H | -17.229166 | -4.836852 | 4.334603 |
| C | -9.221042 | -7.431016 | 7.907285 | | H | -16.907429 | -3.402618 | 5.330472 |
| C | -8.601504 | -7.546862 | 9.158501 | | C | -6.627144 | 6.792551 | -7.107890 |
| C | -10.083487 | -5.898223 | 10.054245 | | C | -7.535399 | 5.857881 | -7.937579 |

| | | | | | | | |
|---|---|---|---|---|---|---|---|
| H | -7.091994 | 5.625648 | -8.912838 | H | -8.375807 | 7.488973 | -5.983075 |
| H | -8.509729 | 6.322724 | -8.129590 | C | -6.458342 | 8.100107 | -7.911116 |
| H | -7.726010 | 4.913156 | -7.415200 | H | -5.906277 | 7.922669 | -8.841672 |
| C | -7.370389 | 7.097951 | -5.788929 | H | -5.913974 | 8.865719 | -7.347397 |
| H | -6.848649 | 7.845961 | -5.181852 | H | -7.431369 | 8.529219 | -8.176601 |
| H | -7.475719 | 6.193996 | -5.177317 | | | | |

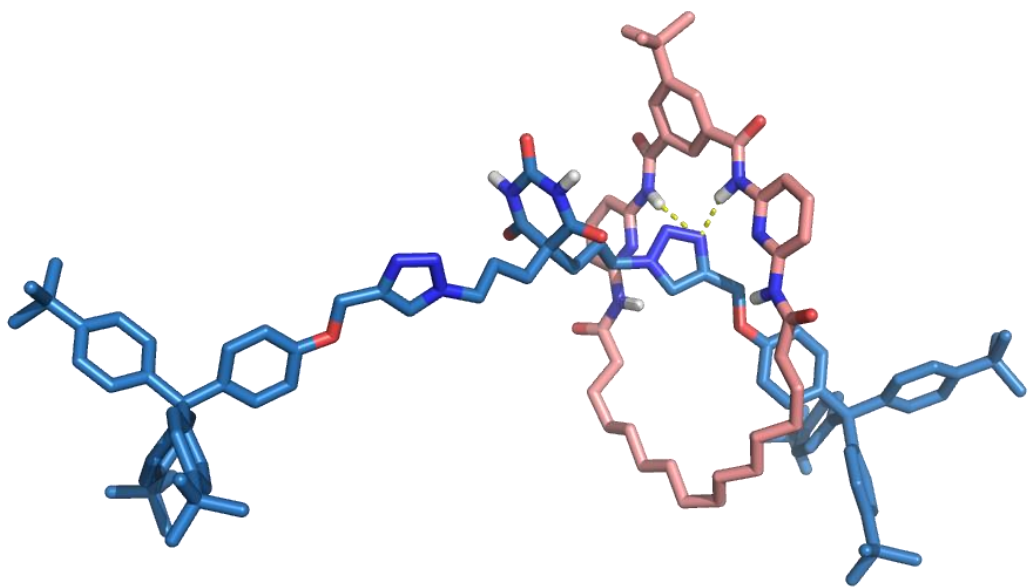

**Figure S25**. Optimized structure of rotaxane **9** with the with the macrocycle placed in the triazole station (MMFF, Spartan '18 software).

XYZ coordinates of rotaxane **9** with the with the macrocycle placed in the triazole station

| | | | | | | | |
|---|---|---|---|---|---|---|---|
| C | -2.448861 | 7.664295 | -7.430590 | O | 0.606549 | 9.436417 | -5.274864 |
| C | -0.095959 | 6.246351 | -6.666594 | N | 1.261196 | 7.431415 | -4.300511 |
| C | -2.181942 | 6.374302 | -7.909311 | H | 1.202966 | 6.417726 | -4.335014 |
| C | -1.516566 | 8.207750 | -6.530286 | N | -0.183801 | 3.370642 | -7.348589 |
| C | -0.356034 | 7.512128 | -6.139761 | H | 0.225004 | 3.634532 | -6.455812 |
| C | -1.025631 | 5.661373 | -7.530491 | C | -0.075349 | 1.993936 | -7.604298 |
| H | -2.867207 | 5.891559 | -8.604014 | C | 0.147135 | -0.732017 | -7.850609 |
| H | -1.679231 | 9.206859 | -6.124858 | C | -0.699016 | 1.357850 | -8.668683 |
| H | 0.838641 | 5.745920 | -6.446969 | N | 0.654095 | 1.312291 | -6.689889 |
| C | -0.852140 | 4.300074 | -8.116355 | C | 0.759067 | -0.030262 | -6.825604 |
| C | 0.563092 | 8.211978 | -5.198991 | C | -0.588680 | -0.022733 | -8.788959 |
| O | -1.349688 | 4.095029 | -9.219510 | H | -1.288529 | 1.889388 | -9.404947 |

| | | | | | | | | |
|---|---|---|---|---|---|---|---|---|
| H | -1.075108 | -0.540729 | -9.610952 | | N | 0.554278 | 3.907403 | -4.417263 |
| H | 0.239592 | -1.805529 | -7.960642 | | N | 0.639723 | 2.834662 | -3.608648 |
| C | 2.032674 | 7.894714 | -3.219243 | | C | 4.121127 | 3.821703 | -3.922055 |
| C | 3.511056 | 8.605500 | -1.018059 | | H | 4.305698 | 4.898900 | -3.841896 |
| N | 2.671051 | 6.924961 | -2.520876 | | H | 4.492422 | 3.456463 | -4.887982 |
| C | 2.113212 | 9.233651 | -2.858749 | | C | -0.524576 | 2.029431 | -3.310482 |
| C | 2.859142 | 9.591211 | -1.743133 | | H | -0.176102 | 1.172377 | -2.722682 |
| C | 3.398299 | 7.288931 | -1.436268 | | H | -0.890899 | 1.638528 | -4.266386 |
| H | 1.600262 | 10.018985 | -3.398893 | | O | 4.788677 | 3.147540 | -2.845462 |
| H | 2.927815 | 10.632264 | -1.440303 | | C | 6.151668 | 3.065730 | -2.928455 |
| H | 4.094919 | 8.898547 | -0.154695 | | C | 8.980696 | 2.756345 | -2.853395 |
| N | 1.563232 | -0.628436 | -5.852946 | | C | 6.965442 | 3.834391 | -3.755327 |
| H | 2.020734 | 0.050881 | -5.261696 | | C | 6.741094 | 2.122192 | -2.088460 |
| N | 4.046127 | 6.224722 | -0.800853 | | C | 8.130508 | 1.952216 | -2.067476 |
| H | 3.895757 | 5.338781 | -1.263791 | | C | 8.357252 | 3.664026 | -3.733542 |
| C | 4.803076 | 6.262574 | 0.351781 | | H | 6.560158 | 4.579676 | -4.432624 |
| C | 1.736727 | -1.974376 | -5.610563 | | H | 6.119214 | 1.498183 | -1.451358 |
| O | 1.148649 | -2.889859 | -6.174064 | | H | 8.541530 | 1.175897 | -1.424471 |
| O | 4.973150 | 7.237439 | 1.074942 | | H | 8.950439 | 4.259842 | -4.425592 |
| C | 2.791232 | -2.225825 | -4.545388 | | C | -1.644335 | 2.791341 | -2.590406 |
| H | 2.680935 | -1.467087 | -3.762066 | | H | -2.472042 | 2.930386 | -3.295330 |
| H | 3.771489 | -2.094741 | -5.018249 | | H | -1.314145 | 3.793660 | -2.293337 |
| C | 5.447085 | 4.918624 | 0.645905 | | C | -2.124586 | 2.027611 | -1.344121 |
| H | 4.652378 | 4.166004 | 0.693358 | | H | -2.149345 | 0.953364 | -1.573063 |
| H | 6.105424 | 4.678587 | -0.196701 | | C | 10.550857 | 2.542825 | -2.850218 |
| C | 6.253976 | 4.920079 | 1.948549 | | C | 10.741273 | 1.375342 | -3.898551 |
| H | 5.570563 | 4.970876 | 2.804931 | | C | 11.077237 | -0.732075 | -5.823910 |
| H | 6.879252 | 5.820399 | 1.995781 | | C | 10.811880 | 1.636591 | -5.279494 |
| C | 2.686034 | -3.622598 | -3.928443 | | C | 10.731557 | 0.022743 | -3.517563 |
| H | 1.753099 | -3.687019 | -3.354019 | | C | 10.916563 | -1.002636 | -4.454673 |
| H | 2.613317 | -4.382426 | -4.716158 | | C | 10.994043 | 0.609680 | -6.217238 |
| C | 3.866428 | -3.990053 | -3.020626 | | H | 10.739138 | 2.656601 | -5.653154 |
| H | 4.757987 | -4.148908 | -3.639684 | | H | 10.590564 | -0.262520 | -2.477014 |
| H | 3.646780 | -4.946257 | -2.530662 | | H | 10.937690 | -2.027020 | -4.088270 |
| C | 7.175735 | 3.705014 | 2.084226 | | H | 11.072230 | 0.900303 | -7.261387 |
| H | 7.873549 | 3.701299 | 1.239661 | | C | 11.128081 | 2.195960 | -1.419258 |
| H | 7.780429 | 3.814140 | 2.992573 | | C | 12.254916 | 1.678622 | 1.174457 |
| C | 4.170288 | -2.926462 | -1.968927 | | C | 12.328521 | 1.479641 | -1.265968 |
| H | 4.506925 | -2.011823 | -2.471473 | | C | 10.561161 | 2.722442 | -0.245964 |
| H | 3.255472 | -2.674305 | -1.418461 | | C | 11.099500 | 2.454838 | 1.020834 |
| C | 6.422223 | 2.377703 | 2.138713 | | C | 12.868345 | 1.214825 | -0.000759 |
| H | 5.728118 | 2.315432 | 1.294039 | | H | 12.876390 | 1.122365 | -2.135977 |
| H | 5.820160 | 2.325609 | 3.053503 | | H | 9.689420 | 3.370530 | -0.296142 |
| C | 7.376778 | 1.182482 | 2.072235 | | H | 10.590118 | 2.888372 | 1.877106 |
| H | 8.062375 | 1.321145 | 1.228637 | | H | 13.795516 | 0.647396 | 0.046101 |
| H | 7.986664 | 1.125768 | 2.980980 | | C | 11.361362 | 3.849725 | -3.223213 |
| C | 5.249954 | -3.367620 | -0.976923 | | C | 12.901971 | 6.216830 | -3.764378 |
| H | 6.103365 | -3.800206 | -1.512018 | | C | 10.959254 | 5.117974 | -2.764833 |
| H | 4.850119 | -4.140973 | -0.311096 | | C | 12.598437 | 3.788774 | -3.888690 |
| C | 6.590690 | -0.115285 | 1.876835 | | C | 13.339059 | 4.944577 | -4.167930 |
| H | 5.817357 | 0.071085 | 1.125617 | | C | 11.702087 | 6.274680 | -3.043760 |
| H | 6.073745 | -0.386419 | 2.804798 | | H | 10.057830 | 5.229626 | -2.165303 |
| C | 5.715797 | -2.155999 | -0.171677 | | H | 13.021369 | 2.833757 | -4.194025 |
| H | 4.836949 | -1.672831 | 0.272325 | | H | 14.280908 | 4.827073 | -4.700043 |
| H | 6.169996 | -1.432801 | -0.861230 | | H | 11.313978 | 7.216250 | -2.664749 |
| C | 7.490189 | -1.267353 | 1.421431 | | C | 11.330119 | -1.878824 | -6.817267 |
| H | 8.155392 | -1.559401 | 2.242669 | | C | 11.484735 | -1.413211 | -8.282926 |
| C | 6.713197 | -2.500316 | 0.936119 | | H | 12.327733 | -0.722121 | -8.400148 |
| H | 6.180947 | -2.959898 | 1.777325 | | H | 10.579366 | -0.912349 | -8.644849 |
| H | 8.135360 | -0.926086 | 0.601768 | | H | 11.671010 | -2.263519 | -8.950062 |
| H | 7.436138 | -3.239989 | 0.571968 | | C | 10.155323 | -2.878737 | -6.775581 |
| C | 1.928023 | 2.597309 | -3.236784 | | H | 10.045151 | -3.348189 | -5.791983 |
| H | 2.219587 | 1.781954 | -2.591500 | | H | 10.299485 | -3.687612 | -7.501520 |
| C | 2.651808 | 3.576300 | -3.883776 | | H | 9.206695 | -2.382747 | -7.012283 |
| N | 1.785236 | 4.365650 | -4.598326 | | C | 12.634747 | -2.612118 | -6.434154 |

```
H    12.575860   -3.071386   -5.441239              H    -5.528377    2.745823    1.204009
H    13.487532   -1.922676   -6.427259              H    -5.866616    1.190075    0.442973
H    12.861250   -3.414796   -7.145756              C    -3.757675    1.674108    0.520504
C    12.871309    1.366084    2.547831              H    -3.502669    0.617416    0.358281
C    14.247041    2.055909    2.658080              H    -3.054315    2.043886    1.279721
H    14.703790    1.873017    3.637681              C   -11.884747   -6.742904    6.847070
H    14.156173    3.141104    2.529705              C   -11.583566   -6.454365    8.372384
H    14.951248    1.694078    1.900545              C   -10.951576   -5.798238   11.099073
C    13.053139   -0.159341    2.711805              C   -12.369108   -5.558817    9.120937
H    13.447861   -0.405814    3.704530              C   -10.429427   -6.952827    9.002010
H    13.753588   -0.576408    1.980215              C   -10.132139   -6.646631   10.336177
H    12.098541   -0.685352    2.592076              C   -12.069921   -5.251869   10.456329
C    12.012211    1.847367    3.739140              H   -13.231150   -5.069286    8.671434
H    12.474996    1.575108    4.695342              H    -9.728665   -7.584200    8.459295
H    11.012678    1.397054    3.721635              H    -9.231013   -7.078398   10.766759
H    11.895794    2.937311    3.740910              H   -12.736152   -4.561852   10.966832
C    13.740956    7.463506   -4.091170              C   -11.427387   -8.182625    6.380516
C    15.127001    7.345230   -3.421888              C   -10.729564  -10.824963    5.488261
H    15.735060    8.237443   -3.611855              C   -11.523907   -9.297389    7.231781
H    15.694511    6.485255   -3.794277              C   -11.051898   -8.441007    5.049519
H    15.032202    7.231043   -2.335507              C   -10.695080   -9.727201    4.618815
C    13.924081    7.581815   -5.620176              C   -11.166516  -10.581574    6.800986
H    14.488034    8.484762   -5.882109              H   -11.891049   -9.188471    8.250444
H    12.955039    7.633899   -6.130619              H   -11.039745   -7.640510    4.312211
H    14.471535    6.730776   -6.039871              H   -10.402615   -9.837441    3.578212
C    13.107508    8.786763   -3.605556              H   -11.249651  -11.397514    7.515964
H    13.733764    9.646239   -3.873470              C   -13.427354   -6.757103    6.506416
H    12.990629    8.802680   -2.515708              C   -16.214910   -6.939858    5.827653
H    12.122355    8.951507   -4.057388              C   -13.902934   -6.414629    5.227589
C    -5.672275    4.208332   -1.558088              C   -14.374902   -7.265846    7.412141
O    -6.492605    4.994163   -2.018779              C   -15.735300   -7.337930    7.086472
N    -4.678933    4.621682   -0.726953              C   -15.265309   -6.488002    4.902077
H    -4.655559    5.616010   -0.540284              H   -13.214104   -6.095382    4.447808
N    -5.667276    2.896954   -1.915384              H   -14.068334   -7.630941    8.390463
H    -6.353434    2.643090   -2.614596              H   -16.416671   -7.729477    7.839002
C    -4.583758    2.073046   -1.820979              H   -15.550623   -6.193439    3.895890
C    -3.521158    3.940632   -0.484839              C   -10.593607   -5.484708   12.561267
O    -4.513499    1.056525   -2.508604              C   -11.606396   -4.554393   13.266332
O    -2.553791    4.513363    0.011964              H   -12.610891   -4.993064   13.285858
C    -3.521784    2.457329   -0.805256              H   -11.669592   -3.577739   12.772492
H    -1.365823    2.151651   -0.559238              H   -11.314483   -4.369350   14.307103
C    -7.003605   -0.486204    3.322268              C    -9.215924   -4.790701   12.620077
H    -6.418586   -1.393394    3.287100              H    -8.413104   -5.430904   12.238160
C    -8.296574   -0.231805    3.728126              H    -8.951195   -4.524420   13.650049
N    -8.550974    1.109430    3.576357              H    -9.212651   -3.869239   12.025795
N    -7.453723    1.690887    3.106909              C   -10.537082   -6.795564   13.375656
N    -6.525627    0.732540    2.956221              H    -9.761112   -7.477587   13.011211
C    -9.332657   -1.156884    4.267144              H   -11.493319   -7.330154   13.329410
H   -10.165541   -1.197223    3.554961              H   -10.317220   -6.595937   14.430993
H    -9.683650   -0.756070    5.225945              C   -10.339677  -12.247099    5.052344
C    -5.194151    1.051873    2.493624              C   -11.559062  -13.183560    5.188960
H    -4.624116    0.115538    2.478551              H   -11.316835  -14.199048    4.854339
H    -4.739099    1.707347    3.244995              H   -12.401272  -12.825250    4.585185
O    -8.798820   -2.472107    4.457198              H   -11.907000  -13.261661    6.224845
C    -9.618313   -3.413035    5.011409              C    -9.194680  -12.770279    5.947077
C   -11.093810   -5.573527    6.130197              H    -8.873485  -13.770465    5.633378
C   -10.954259   -3.243293    5.359671              H    -9.490870  -12.845498    6.999065
C    -9.009759   -4.644534    5.247035              H    -8.321641  -12.108704    5.897434
C    -9.727117   -5.701881    5.814724              C    -9.850989  -12.339759    3.589077
C   -11.675092   -4.304177    5.927402              H    -9.572462  -13.368372    3.330122
H   -11.470520   -2.299318    5.219932              H    -8.967580  -11.713312    3.418767
H    -7.958933   -4.782425    5.004071              H   -10.630111  -12.031133    2.882383
H    -9.191908   -6.626690    6.020396              C   -17.715429   -7.035190    5.505164
H   -12.704911   -4.114930    6.225754              C   -18.175357   -8.505303    5.607928
C    -5.173046    1.713992    1.111462              H   -19.235958   -8.606601    5.349614
```

| | | | | | | | |
|---|---|---|---|---|---|---|---|
| H | -18.051971 | -8.908120 | 6.619243 | H | -4.073696 | 7.491717 | -9.786168 |
| H | -17.603284 | -9.146279 | 4.926621 | H | -5.461684 | 8.363168 | -9.135786 |
| C | -18.518619 | -6.180366 | 6.509643 | H | -4.999075 | 6.814186 | -8.430464 |
| H | -19.590170 | -6.201985 | 6.278913 | C | -4.533704 | 8.829310 | -6.622970 |
| H | -18.193137 | -5.133571 | 6.485379 | H | -3.988892 | 9.453696 | -5.906480 |
| H | -18.406553 | -6.535752 | 7.539818 | H | -4.852046 | 7.922900 | -6.094619 |
| C | -18.082357 | -6.537713 | 4.088782 | H | -5.435635 | 9.383214 | -6.908862 |
| H | -19.161787 | -6.613819 | 3.910695 | C | -3.221181 | 9.785359 | -8.546505 |
| H | -17.586544 | -7.130585 | 3.311342 | H | -2.585326 | 9.572658 | -9.414226 |
| H | -17.804867 | -5.486777 | 3.946241 | H | -2.649065 | 10.430123 | -7.870365 |
| C | -3.680352 | 8.479684 | -7.860730 | H | -4.079080 | 10.370781 | -8.897514 |
| C | -4.601609 | 7.741811 | -8.858423 | | | | |